\documentclass{MICE}

\usepackage{amsmath}					
\usepackage{subcaption}					
\usepackage{mathtools}					
\usepackage[colorlinks]{hyperref}		

\numberwithin{equation}{section}
\numberwithin{figure}{section}
\numberwithin{table}{section}

\begin{document}

\begin{titlepage}
\begin{center}

\includegraphics[height=4cm]{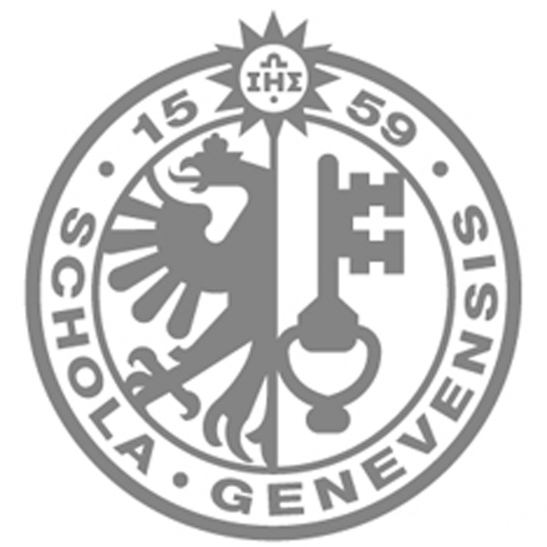}~\\[1cm]

{\LARGE Universit\'e of Gen\`eve \\[0.5cm]
 D\'epartement de Physique Nucl\'eaire et Corpusculaire}\\[2cm]

{\huge Master Thesis}\\[1cm]

{\huge \bfseries Electron-Muon Ranger:\\hardware characterization}

\vfill
\begin{minipage}{0.45\textwidth}
\begin{flushleft} \large
\emph{Author:}\\
Fran\c cois \textsc{DRIELSMA}
\end{flushleft}
\end{minipage}
\begin{minipage}{0.45\textwidth}
\begin{flushright} \large
\emph{Supervisor:} \\
Prof.~Alain \textsc{BLONDEL} \\[0.2cm]
\emph{Assistants:} \\
Ruslan \textsc{ASFANDIYAROV} \\
Dr.~Yordan \textsc{KARADZHOV}
\end{flushright}
\end{minipage}

\vfill
{\large June 19, 2014}
\end{center}
\end{titlepage}

\begin{abstract}
	The Electron-Muon Ranger (EMR) is a fully-active tracking-calorimeter in charge of the electron background rejection downstream of the cooling channel at the international Muon Ionization Cooling Experiment. It consists of 2832 plastic scintillator bars segmented in 48 planes in an X--Y arrangement and uses particle range as its main variable to tag muons and discriminate electrons. An array of analyses were conducted to characterize the hardware of the EMR and determine whether the detector performs to specifications. The clear fibres coming from the bars were shown to transmit the desired amount of light, and only four dead channels were identified in the electronics. Two channels had indubitably been mismatched during assembly and the DAQ channel map was subsequently corrected. The level of crosstalk is within acceptable values for the type of multi-anode photomultiplier used with an average of $0.20\pm0.03$\,\% probability of occurrence in adjacent channels and a mean amplitude equivalent to $4.5\pm0.1$\,\% of the primary signal intensity. The efficiency of the signal acquisition, defined as the probability of recording a signal in a plane when a particle goes through it in beam conditions, reached $99.73\pm0.02$\,\%.
\end{abstract}

\footnotesize
\tableofcontents
\normalsize

\section*{Preamble}
\addcontentsline{toc}{section}{Preamble}
Over eighty years have passed since Pauli first postulated the existence of the neutrino in order to \emph{save the law of conservation of energy} in his famous open letter to the \emph{radioactive} people of the T\"ubingen meeting. A little over 25 years later, Cowan and Reines confirmed its existence and a new era of particle physics was underway. The initial concern was to determine its characteristics and classify its role in the mainstream theories. In the 70s and 80s, the neutrino was used to investigate the nature of the weak interaction and probe the nucleon structure. In the last three decades, the focus has shifted back to its roots: the nature of the three generations of neutrinos, their mass scale and their flavour mixing. This field of research stays constantly in motion as much uncertainty remains as to how the neutrino fits into the Standard Model.

Despite the continuous international effort, several parameters of this \emph{ghost particle} are still unknown today. A fraction of its oscillation parameters has been measured but its leptonic CP violating phase and its mass eigenstates ordering remain elusive. The Neutrino Factory (NuFact) collaboration, which studies the feasibility and design of such a facility since the beginning of the 21$^\text{st}$ century, has proposed to answer most of them in a single state-of-the-art experiment. The idea behind this machine involves a very high luminosity collimated muon storage ring as a source of intense and exquisitely well understood neutrino beams, a near and a far detector to observe the neutrino oscillations from different baselines and beam energies.

One of the main challenges associated with such a device is the development of an operational muon cooling channel. The muon beam, produced from the decay of pions, cannot fit the acceptance of an accelerator without reducing its emittance. The international Muon Ionization Cooling Experiment (MICE) collaboration intends to experimentally demonstrate that the use of ionization cooling as a front-end to the Neutrino Factory cooling channel is a viable solution. It aims to observe a 10\,\% reduction in the muon beam emittance with a short section of a cooling channel. The experiment is currently ongoing and should yield its first results in the upcoming years.

MICE has used several detector technologies in order to measure the effect of ionization cooling on a muon beam. These detectors are necessary to sample the beam emittance at the entrance and at the exit of the channel and to reject the particle background. The latter task is challenging and requires a calorimeter to discriminate the electron background downstream the cooling channel, a key element of which is the Electron-Muon Ranger (EMR). The EMR is a 1.5\,tons fully-active tracking-calorimeter composed of 2832 scintillating bars divided in 48 planes. It uses particle range reconstruction as a powerful tool to identify the muon tracks.

Following the implementation of the EMR in the MICE beam line in October 2013, this master thesis intends to investigate the performance of the EMR hardware in working conditions. A brief summary of the motivations behind this experiment, its design choices and the conception of the EMR is presented first. The analyses are introduced chronologically, starting with the tests performed during construction and ending with channel mismatch, crosstalk, misalignment and efficiency investigations of the whole detector.
\newpage

\section{Current status of neutrino physics}
Neutrinos are spin $1/2$, electrically neutral particles with a very light mass (of order $\sim10^6$ times smaller than the electron mass\,\cite{bib:nu_mass}) and are the most abundant particles in the universe. The three known neutrino flavour eigenstates ($\nu_e$, $\nu_\mu$ and $\nu_\tau$) interact weakly with matter via the weak gauge bosons $W^\pm, Z^0$. They stem from various sources, both natural and artificial.

The strongest nearby neutrino source is the Sun. The frequent fusion reactions in which hydrogen is transformed into helium ($2 \text{H}\rightarrow\prescript{4}{}{\text{He}}$) produce $\sim2\times10^{39}$ electron neutrinos per second, i.e. a $6\times10^{10}\,\nu/\text{cm}^{2}/\text{s}$ flux at the Earth's surface\,\cite{bib:nu_sun_flux}. Of order $10^{58}$ neutrinos can be produced in a few seconds in Supernovae explosions\,\cite{bib:nu_sn_flux} while relic neutrinos, the so-called Cosmic neutrino Background (C$\nu$B), fill the Universe with a density of $\sim100\,\nu/\text{cm}^3$\,\cite{bib:cnub}.

Neutrinos are also produced on Earth in the surrounding atmosphere and in its crust. Hadronic showers induced by the interaction of cosmic rays with the high layer of the atmosphere generate a flux of typically $\sim1\,\nu/\text{cm}^2/\text{s}$ at the Earth's surface\,\cite{bib:nu_atm} while geoneutrinos are produced through the $\beta$-decay of radioactive nuclei (mainly $^{238}$U, $^{232}$Th and $^{40}$K) with a flux of $\sim10^7\,\nu/\text{cm}^2/\text{s}$\,\cite{bib:nu_geo}.

In the last century, man-made machines have been added to the spectrum of neutrino sources. Electron antineutrinos are generated with a typical rate of the order of $10^{20}\,\nu/\text{s}$ in nuclear reactors\,\cite{bib:nu_reactor} while fluxes of $10^6\,\nu_\mu/\text{cm}^2/\text{s}$ can currently be produced with conventional accelerators by dumping a high energy proton beam on a target and letting the mesons produced (mostly $\pi^\pm,\,K^\pm$) decay\,\cite{bib:nu_acc}.

A short summary of the current status of neutrino physics is presented from the historical, experimental and theoretical point of view in this section.

\subsection{History and major experiments}
\begin{figure}[!htb]
	\begin{minipage}[b]{.45\textwidth}
		\centering
		\includegraphics[width=\textwidth]{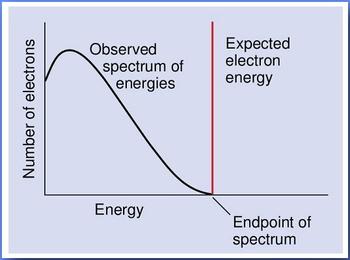}
		\caption{Expected electron energy spectrum for a two body final state (red line) as opposed to the observed spectrum (smooth black curve).\\ \\}
		\label{fig:beta_spectrum}    
	\end{minipage}
	\hfill
	\begin{minipage}[b]{.45\textwidth}
		\centering
		\includegraphics[width=.9\textwidth]{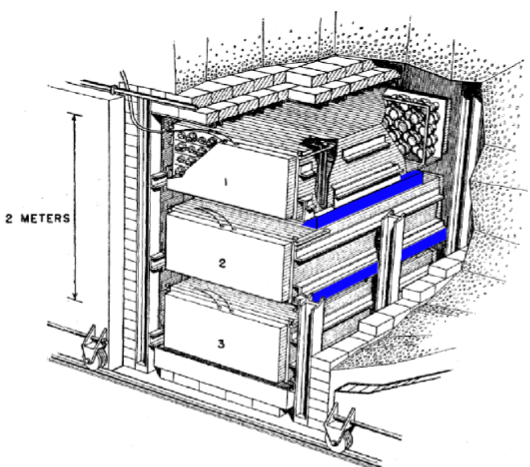}
		\caption{Schematic layout of the detector used by Cowans and Reines to detect the neutrino. It consisted of two large plastic tanks (in light blue) filled with 200 liters of water mixed with CdCl$_2$, sandwiched between three 1.4\,t liquid scintillator detectors.}
		\label{fig:savannah}
	\end{minipage}
\end{figure}

The first notable event in the history of neutrino can be identified as Lise Meitner and Otto Hahn's discovery, in 1911\,\cite{bib:beta_spectrum}, of the continuous $\beta$-decay energy spectrum. It could not be explained with the law of conservation of energy for a two body final state ($\prescript{n}{m}{\text{X}}\rightarrow\prescript{n}{m+1}{\text{X}}+e^-$) as it seemed that some energy was lost\,\cite{bib:beta_decay}, as shown in figure\,\ref{fig:beta_spectrum}. Almost two decades later, Wolfgang Pauli, in his famous open \emph{neutrino letter} to the meeting in T\" ubingen, postulated the existence of a new particle, that he called \emph{neutron} at the time, to solve the $\beta$-decay puzzle\,\cite{bib:pauli_letter}.

Three years later, Enrico Fermi published the first account of his theory on $\beta$-decay which culminated in his famous weak interaction theory in 1934\,\cite{bib:fermi_theory}. The following two decades were characterized by continuous efforts to prove the existence of the neutrino and to determine its physical properties. Pauli will have to wait until 1956, two years before his death, to see his hunch confirmed when Clyde L. Cowans and Frederick Reines saw the first neutrino event by detecting the inverse $\beta$-decay process at the Savannah River reactor\,\cite{bib:nu_discovery} using the detector sketched in figure\,\ref{fig:savannah}.

A few years later, Maurice Goldhaber showed that neutrinos involved in the weak interaction are left-handed particles\,\cite{bib:nu_helicity}. In 1962, the first evidence of the existence of different neutrino generations came from the $\nu_\mu$ observation at the Brookhaven National Laboratory\,\cite{bib:nu_generations}, completed much later by the discovery of the third generation, $\nu_\tau$, at the Fermi National Laboratory in 2001 \cite{bib:nu_tau}. These experiments consolidated the idea of neutrino oscillations, first proposed by Bruno Pontecorvo in 1957\,\cite{bib:nu_osc} using the same formalism as for the quark mixing, and subsequently developed by Ziro Maki, Masami Nakagawa and Shoichi Sakata in 1962\,\cite{bib:nu_osc_mns}.

\begin{figure}[!htb]  
	\begin{minipage}[b]{.45\textwidth}
		\centering
		\includegraphics[width=\textwidth]{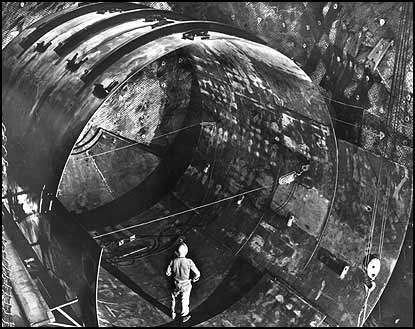}
		\caption{Picture of the mine that hosted the Homestake neutrino detector and tanker of perchloroethylene. \\}
		\label{fig:homestake}
	\end{minipage}
	\hfill
	\begin{minipage}[b]{.45\textwidth}
		\centering
		\includegraphics[width=.65\textwidth]{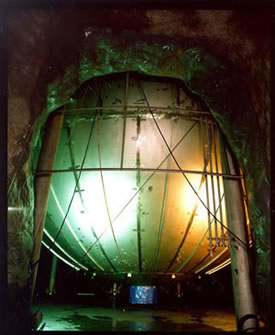}
		\caption{Picture of the cave harbouring KamLAND and its main spherical liquid scintillator vessel. The inner surface of the sphere is covered by 1879 50\,cm PTMs.}
		\label{fig:kamland}
	\end{minipage}
\end{figure}

The experimental neutrino physics era expanded in 1968 when John Bahcall and Ray Davis first showed a non-negligible deficit in the solar neutrino flux measurement with the Homestake chlorine-based detector, pictured in figure\,\ref{fig:homestake}\,\cite{bib:nu_bahcall}. Neutrino oscillations was proposed as a possible solution to the so-called \emph{solar neutrino puzzle}. In the following twenty years, the main solar neutrino experiments (Homestake, SAGE\,\cite{bib:sage} and GALLEX\,\cite{bib:gallex}) confirmed the solar neutrino flux deficit observed by Davis and Bahcall, but it was only in 2002 that the puzzle was solved by the Sudbury Neutrino Observatory (SNO) experiment \cite{bib:sno}. Its unique ability to detect the three neutrino flavours and distinguish the $\nu_e$ from the other generations provided the scientific community with the final piece of evidence of solar neutrino oscillations.

At the same time, neutrino oscillations had also been observed in the atmospheric neutrino field: an upward-downward $\nu_\mu$ asymmetry was observed without a corresponding deficit in $\nu_e$. The oscillation between the $\nu_\mu$ and $\nu_\tau$ was proposed as an explanation to this deficit of muon neutrinos\,\cite{bib:nu_atm_osc}.

Additional experiments dedicated to the determination of the oscillation parameters were built in recent years, exploiting nuclear reactor neutrino production. The KamLAND experiment (figure\,\ref{fig:kamland}) provided several confirmations of both the solar and atmospheric neutrino oscillation results\,\cite{bib:kamland}.

As described in more details in section \ref{subsec:mix}, the six relevant parameters of three-flavour oscillations are three mixing angles ($\theta_{12}$, $\theta_{13}$ and $\theta_{23}$), a CP violating phase $\delta$ and two mass-squared differences $\Delta m_{21}^2$ and $|\Delta m_{31}^2|$ , with $\Delta m_{ji}^2 = m_j^2 - m_i^2$. The oscillation parameters obtained from the two most recent global fits to the world neutrino data (e.g. T2K \cite{bib:t2k}, Double Chooz \cite{bib:double_chooz}) are summarised in table\,\ref{tab:osc_param} \cite{bib:global_fits}.

\begin{table}[!htb]
\centering
\begin{tabular}{c|c|c|c|c}
Parameter & Best Fit & $1\sigma$ CI & $2\sigma$ CI & $3\sigma$ CI \\
\hline

$\Delta m_{21}^2$ [$10^{-5}$eV$^2$] & 7.62 & $7.43-7.81$ & $7.27-8.01$ & $7.12-8.20$ \\[.2cm]

$|\Delta m_{31}^2|$ [$10^{-3}$eV$^2$] & 
\begin{tabular}{@{}c@{}}2.55 \\ 2.43\end{tabular} & \begin{tabular}{@{}c@{}}$2.46-2.61$ \\ $2.37-2.50$ \end{tabular} & 
\begin{tabular}{@{}c@{}}$2.38-2.68$ \\ $2.29-2.58$ \end{tabular} & 
\begin{tabular}{@{}c@{}}$2.31-2.74$ \\ $2.21-2.64$ \end{tabular} \\[.4cm]

$\sin^2\theta_{12}$ & 0.320 & $0.303-0.336$ & $0.29-0.35$ & $0.27-037$ \\[.2cm]

$\sin^2\theta_{23}$ & 
\begin{tabular}{@{}c@{}}0.613 \\ 0.600 \end{tabular} & \begin{tabular}{@{}c@{}}$0.573-0.625$ \\ $0.569-0.626$ \end{tabular} & 
\begin{tabular}{@{}c@{}}$0.38-0.66$ \\ $0.39-0.65$ \end{tabular} & 
\begin{tabular}{@{}c@{}}$0.36-0.68$ \\ $0.37-0.67$ \end{tabular} \\[.4cm]

$\sin^2\theta_{13}$ & 
\begin{tabular}{@{}c@{}}0.0246 \\ 0.0250 \end{tabular} & \begin{tabular}{@{}c@{}}$0.0218-0.0275$ \\ $0.0223-0.0276$ \end{tabular} & 
\begin{tabular}{@{}c@{}}$0.019-0.030$ \\ $0.020-0.030$ \end{tabular} & 
$0.017-0.033$ \\[.4cm]

$\delta$ & 
\begin{tabular}{@{}c@{}} $0.80\pi$ \\ $-0.03\pi$ \end{tabular} &
$0-2\pi$ & $0-2\pi$ & $0-2\pi$ \\
\end{tabular}
\caption{Neutrino oscillation parameters summary. For $\Delta_{31}^2$, $\sin^2\theta_{23}$, $\sin^2\theta_{13}$ and $\delta$, the upper (resp. lower) row corresponds to normal (resp. inverted) neutrino mass hierarchy.}
\label{tab:osc_param}
\end{table}

\subsection{Mixing}
\label{subsec:mix}

The fact that neutrinos have masses implies that there is a spectrum of neutrino mass eigenstates $\nu_i,\,i = 1, 2, 3$, each with a mass $m_i$. Mixing among the three known neutrino flavours is in turn possible: in the $W^\pm$ decays to the particular charged lepton $l_\alpha$ (resp. antilepton $\bar{l}_\alpha$), $\alpha = e, \mu, \tau$, the accompanying neutrino flavour eigenstate is not a single $\nu_i$ , but a mixture of the different $\nu_i$s.

The neutrino mixing is described by the $3\times3$ Pontecorvo-Maki-Nakagawa-Sakata unitary matrix $U_\text{PMNS}$, which is analogous to the CKM matrix in the quark sector\,\cite{bib:PDG,bib:nu_mix},
\begin{equation}
U_{\text{PMNS}}= 
\begin{bmatrix}
c_{12} c_{13} & s_{12} c_{13} & s_{13} e^{-i\delta} \\
- s_{12} c_{23} - c_{12} s_{23} s_{13} e^{i \delta} & c_{12} c_{23} - s_{12} s_{23} s_{13} e^{i \delta} & s_{23} c_{13}\\
s_{12} s_{23} - c_{12} c_{23} s_{13} e^{i \delta} & - c_{12} s_{23} - s_{12} c_{23} s_{13} e^{i \delta} & c_{23} c_{13}
\end{bmatrix},
\label{eq:pmns}
\end{equation}
where $c_{ij} = \cos\theta_{ij}$ and  $s_{ij} = \sin\theta_{ij}$ with $i, j = 1, 2, 3$. The phase $\delta$ is non-zero only if the neutrino oscillation violates CP symmetry. The $U_{\text{PMNS}}$ matrix relates the mass eigenstates ($\nu_1$, $\nu_2$, $\nu_3$) to the light-neutrino flavour eigenstates ($\nu_e$, $\nu_\mu$, $\nu_\tau$).

Using the Dirac formalism, a neutrino of flavour $\alpha$ can be expressed as a superposition of the three mass eigenstates and vice versa through the following formulas
\begin{equation}
\left|\nu_\alpha\right\rangle=\sum_iU^*_{\alpha i}\left|\nu_i\right\rangle,
\end{equation}
\begin{equation}
\left|\nu_i\right\rangle=\sum_\alpha U_{\alpha i}\left|\nu_\alpha\right\rangle.
\end{equation}
Here, $U_{\alpha i}$ corresponds to the $U_\text{PMNS}$ matrix element and denotes the probability amplitude of the $W^+$ decay to produce the specific combination $l_\alpha + \nu_i$. The fraction of flavour $\alpha$ in $\nu_i$ is $|U_{\alpha i}|^2$.

\begin{figure}[!htb]
  \centering
  \includegraphics[width=.5\textwidth]{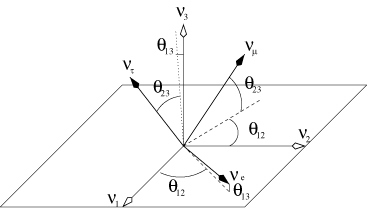}
  \caption{Rotation of the neutrino mass eigenstates $(\nu_1 , \nu_2 , \nu_3)$ into the flavour eigenstates $(\nu_e , \nu_\mu , \nu_\tau)$ as stated in equation \ref{eq:pmns}. The locations of the Euler angles ($\theta_{12}$, $\theta_{13}$ and $\theta_{23}$) are indicated.}
  \label{fig:nu_mix}
\end{figure}

The relationship between the weak and the mass eigenstates through the mixing angles ($\theta_{12}$, $\theta_{13}$ and $\theta_{23}$), that arises from equation \ref{eq:pmns}, are presented in figure\,\ref{fig:nu_mix}.

\subsection{Oscillation}
The oscillation probability $P(\nu_\alpha\rightarrow\nu_\beta)$, that indicates the probability of finding a neutrino created in a given flavour state to be in another one, can be derived using an efficient and simple approach that contains all the essential quantum physics. The reasoning is summarised briefly in this section and developed extensively in \cite{bib:nu_physics}.

\begin{figure}[!htb]
  \centering
  \includegraphics[width=.75\textwidth]{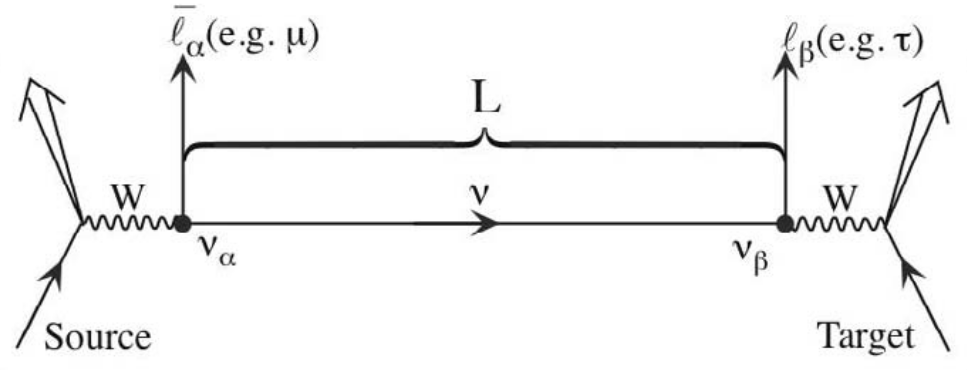}
  \caption{Neutrino oscillation scheme. A neutrino of flavour $\alpha$ travels from its source over a distance $L$ and produces a charged lepton $l_\beta$ in its interaction with the detector. The $\nu_\alpha$ has oscillated into a $\nu_\beta$ while travelling from the source to the target.}
  \label{fig:oscillation}
\end{figure}

A typical oscillation is represented schematically in figure\,\ref{fig:oscillation}. A neutrino source produces a neutrino of flavour $\alpha$ together with the corresponding charged antilepton $\bar{l}_\alpha$. It travels a distance $L$ to a detector where it interacts with a target and produces another charged lepton $l_\beta$. At the time of its interaction in the detector, the neutrino is a $\nu_\beta$. If $\alpha \neq \beta$, the neutrino has changed from a $\nu_\alpha$ to a $\nu_\beta$ while travelling from the source to the detector.

Since  $\left| \nu_{i} \right\rangle$ are mass eigenstates, their propagation can be described in natural units by plane wave solutions of the Schr\" odinger equation of the form:
\begin{equation}
|\nu_{i}(t)\rangle = e^{ -i ( E_{i} t - \vec{p}_{i} \cdot \vec{x}) }|\nu_{i}(0)\rangle,
\end{equation}
with $t$ the propagation time, $\vec{p_i}$ the three-momentum and $\vec{x}$ the position. $E_i$ is the energy of the mass eigenstate $i$ and, in the ultrarelativistic limit $|\vec{p_i}|=p_i\gg m$, reads
\begin{equation}
E_{i} = \sqrt{p_{i}^2 + m_{i}^2 }\simeq p_{i} + \frac{m_{i}^2}{2 p_{i}} \simeq E + \frac{m_{i}^2}{2 E},
\end{equation}
with $E$ is the total energy of the particle. This limit applies to all currently observed neutrinos. Since their masses are of order eV and their energies are of order at least MeV, the Lorentz factor $\gamma$ is greater than $10^6$ in all cases. Using also $t\simeq L$ (holds for $v\simeq c$), where $L$ is the distance travelled, and dropping the phase factors, the wave function becomes:
\begin{equation}
|\nu_{i}(L)\rangle = e^{ -i m_{i}^2 L/2E }|\nu_{i}(0)\rangle.
\end{equation}

As the flavour eigenstates are a linear combination of the mass eigenstates with evolving parameters, it is possible to observe a neutrino change its flavour during its propagation. The probability of a neutrino originally of flavour $\alpha$ to be later observed at target with flavour $\beta$ is
\begin{equation}
P_{\alpha\rightarrow\beta}=\left|\left\langle \nu_{\beta}|\nu_{\alpha}(t)\right\rangle \right|^{2}=\left|\sum_{i}U_{\alpha i}^{*}U_{\beta i}e^{ -i m_{i}^2 L/2E }\right|^{2},
\end{equation}
which can be more conveniently written as
\begin{equation}
\begin{matrix}P_{\alpha\rightarrow\beta}=\delta_{\alpha\beta} & - & 4{\displaystyle \sum_{i>j}{\rm Re}(U_{\alpha i}^{*}U_{\beta i}U_{\alpha j}U_{\beta j}^{*}})\sin^{2}(\frac{\Delta m_{ij}^{2}L}{4E})\\ & + & {\displaystyle 2\sum_{i>j}{\rm Im}(U_{\alpha i}^{*}U_{\beta i}U_{\alpha j}U_{\beta j}^{*})\sin(}\frac{\Delta m_{ij}^{2}L}{2E}).\end{matrix}
\label{eq:prob}
\end{equation}
Several characteristics of the neutrinos can be drawn from this formula:
\begin{itemize}
\item if neutrinos were massless, i.e. $\Delta m_{ij}^2 = 0,\,\forall i,j$, equation \ref{eq:prob} would become $P_{\alpha\rightarrow\beta} = \delta_{\alpha\beta}$. The observation that neutrinos do change flavour implies non-degenerate neutrino masses, and in particular at least two non-zero masses;

\item the probability depends on the quantity $L/E$, with the distance $L$ referred to as baseline. Experiments are classified as either Short BaseLine (SBL) or Long BaseLine (LBL);

\item there are two fundamental ways to detect neutrino flavour oscillations: \emph{appearance} and \emph{disappearance}. In a beam of neutrinos which are initially of flavour $\alpha$, the observation of neutrinos of a new flavour $\beta$ or of a $\nu_\alpha$ flux reduction are two equivalent approaches;

\item the neutrino oscillation probability depends only on the neutrino squared mass splittings $\Delta m_{ij}^2$ and not on the absolute neutrino masses. Oscillation experiments can determine the neutrino squared-mass spectral patterns, but not how far above zero the entire spectrum lies. Two mass orderings are still possible today: the situation where $m_3 > m_2 > m_1$, known as direct (or normal) hierarchy and the inverse hierarchy given as $m_2 > m_1 > m_3$.
\end{itemize}

\begin{figure}[!htb]
  \centering
  \includegraphics[width=.5\textwidth]{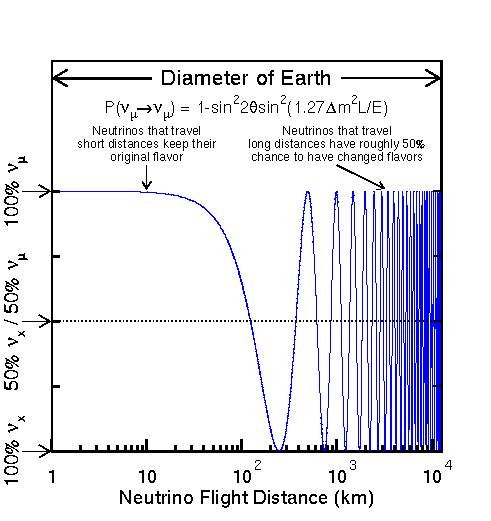}
  \caption{Survival probability of a $\nu_\mu$ as a function of the flight distance $L$. Provided an appropriate choice of baseline, it is possible to observe a lower flux of $\nu_\mu$s than expected (disappearance).}
  \label{fig:mu_osc}
\end{figure}

Both the results from solar and atmospheric neutrinos have shown that a simplified two-flavour approximation can be an accurate description for several sets of data. In the simple case of two neutrino mixing between $\nu_\alpha,\,\nu_\beta$ and $\nu_i,\,\nu_j$, there is only one squared-mass difference $\Delta m_{ij}^2 \equiv \Delta m^2 = m_i^2-m_j^2$ and the mixing matrix $U$ can be parametrized in terms of a single mixing angle $\theta$:
\begin{equation}
U = \begin{pmatrix} \cos\theta & \sin\theta \\ -\sin\theta & \cos\theta \end{pmatrix}.
\end{equation}
The resulting survival probability of a given flavour can be written as
\begin{equation}
P_{\alpha\rightarrow\alpha} = 1-\sin^{2}(2\theta) \, \sin^{2}\left( 1.267 \frac{\Delta m^2 L}{E} \frac{\rm GeV}{\rm eV^{2}\,\rm km}\right),
\end{equation}
with $\sin^2 2\theta$ is the oscillation amplitude. The survival probability of a 1\,GeV muon neutrino is represented in figure\,\ref{fig:mu_osc} as a function of flight distance $L$.

\section{International Muon Ionization Cooling Experiment}
Particle physics experiments based on muon acceleration have been the object of increasing interest in the last 20 years. A Neutrino Factory\,\cite{bib:nu_fact}, based on a high energy muon storage ring, is the ultimate tool to study the neutrino mixing parameters summarised in table\,\ref{tab:osc_param}. This type of facility also provides the best chance of discovering and studying accurately the leptonic CP violation. The potential outcome of a Neutrino Factory goes beyond its own results as it would pave the way to a brand new line of high brilliance muon accelerators by taking the first step towards the Muon Collider\,\cite{bib:mu_collider}, potential candidate of choice for multi-TeV lepton-antilepton collisions.

A Neutrino Factory could be built using accessible technologies, with a performance matching the requirements of an exciting physics program. Cost estimates are quite high (\$1.9 billion in US Feasibility Study II\,\cite{bib:nu_fact_feas}), and several techniques considered have never been applied in practise. A sizeable R\&D program is required to lower the costs and investigate new technologies.

In the uncharted territory on the path to a functional Neutrino Factory, ionization cooling is allegedly the largest novelty in accelerator physics. Ionization cooling of muons at minimum-ionizing energy has never been realised in practise and has yet to be demonstrated. It makes significant contributions to both the performance (up to a factor of 10 in neutrino intensity\,\cite{bib:icool_perf}) and cost (as much as 20\,\%) of a Neutrino Factory.

The international Muon Ionization Cooling Experiment (MICE) collaboration has been created to carry out this program. It consists of accelerator physicists and experimental particle physicists from Europe, Japan and the US. The goals of the experiment are:

\begin{itemize}
\item to build a section of a cooling channel that is long enough to provide a measurable cooling effect (up to 10\,\% reduction in transverse emittance) but short enough to be affordable and sufficiently flexible to allow a variety of beam momenta, optics and absorbers to be investigated;
\item to use particle detectors to measure the cooling effect with high precision, achieving an absolute accuracy on the measurement of emittance of 0.1\,\% or better. The beam intensity will be such that a single particle will pass through the experiment every 100 ns or so.
\end{itemize}

The appeal of the Neutrino Factory physics and its concepts are presented. The ionization cooling technique and the design of the MICE cooling channel at are described in detail.

\subsection{Neutrino factory}
\subsubsection{Physics}
The potential of a Neutrino Factory is unprecedented in neutrino physics\,\cite{bib:nu_fact}. It could provide measurements of the the neutrino mixing matrix elements with an unchallenged precision. It would have a high enough resolution to unravel a potential leptonic CP violation\,\cite{bib:nu_fact_cp} and related studies of slow muon physics would open the way to a lepton collider of extremely high energy\,\cite{bib:mu_collider}. 

A muon storage ring providing neutrino beams is the ultimate tool for the measurement of the $U_{\text{PMNS}}$ elements, as it offers a well defined energy spectrum as well as a high purity neutrino beam. The flavour composition of the beam is well known and the beam is focused and intense. The production of very high energy $\nu_e$s allows the study of the $\nu_e\rightarrow\nu_\tau$ mixing channel.

The fundamental goal of a Neutrino Factory is to measure the oscillation parameters:
\begin{itemize}
\item very precise measurement of $\Delta m_{23}^2$ and $\theta_{23}$;
\item measurement of the small mixing angle $\theta_{13}$ with a precision better than half a degree;
\item determination of the ordering of neutrino masses, i.e. the sign of $\Delta m_{23}^2$, made possible by the MSW effect\footnote{The MSW (Mikheyev-Smirnov-Wolfenstein) effect is the effect of transformation of one neutrino species (flavour) into another one in a medium with varying density. Three basic elements of the effect include the refraction of neutrinos in matter, the resonance (level crossing) and the adiabaticity. The effect depends on the neutrino masses hierarchy \cite{bib:msw}.} on the neutrinos during their passage through the earth and its influence on the ratio $\mathcal{R}=N(\nu_e\rightarrow\nu_\mu)/N(\bar{\nu}_e\rightarrow\bar{\nu}_\mu)$;
\item  search for CP violation through the precise measurement of the same appearance rate asymmetry $\mathcal{R}$ as function of the energy $E$ and baseline $L$.
\end{itemize}

\begin{figure}[!htb]
  \centering
  \begin{minipage}[b]{.45\textwidth}
    \centering
    \includegraphics[width=.7\textwidth]{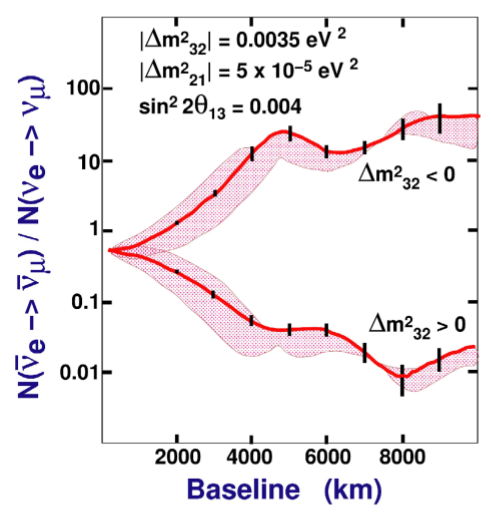}
    \caption{Neutrino--antineutrino appearance asymmetry as a function of baseline. The two possible orderings are shown together with the the CP phase uncertainty (light red bands).}
    \label{fig:nufact_hierchy}
  \end{minipage}
  \hfill
  \begin{minipage}[b]{.45\textwidth}
    \centering
    \includegraphics[width=\textwidth]{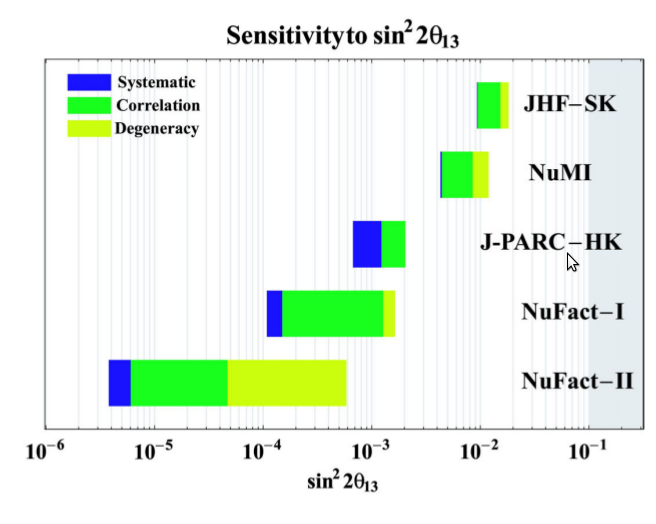}
    \caption{Sensitivity of different Neutrino Factory designs (NuFact I and II respectively) to $\sin^2\theta_{13}$ in comparison with other proposed future facilities.}
    \label{fig:nufact_theta13}    
  \end{minipage}
\end{figure}

Figure\,\ref{fig:nufact_hierchy} shows the predicted neutrino--antineutrino asymmetry ratio $\mathcal{R}=N(\nu_e\rightarrow\nu_\mu)/N(\bar{\nu}_e\rightarrow\bar{\nu}_\mu)$ as a function of the baseline \cite{bib:mice_proposal}. At very short baselines ($L\simeq0$), neither the matter effects nor the CP violation influence the appearance rate: the ratio is 0.5 which reflects the different neutrino and antineutrino cross sections. As the baseline grows more remote, the ratio $\mathcal{R}$ increases (resp. decreases) due to the MSW effect if the sign of the mass splitting $\Delta m_{23}^2$ is negative (resp. positive). At long baseline, the CP violation enters in the equation and influences the ratio slightly (indicated by the light red bands). The high precision measurement of $\mathcal{R}$ will provide both the sign of $\Delta m_{23}^2$ and the determination of the CP phase $\delta$.

The Neutrino Factory physics potential in terms of the small mixing angle $\sin^2\theta_{13}$ sensitivity has been estimated in comparison with the other major neutrino experiments considered\,\cite{bib:nu_lbl_exp}. The alternatives include JHF-SK, a combination of the Japan Proton Accelerator Research Complex 50\,GeV proton driven superbeam with the existing SuperKamiokande detector\,\cite{bib:jhf_sk} and the similar J-PARC--HK with the proposed HyperKamiokande detector\,\cite{bib:jhf_sk2}. Figure\,\ref{fig:nufact_theta13} summarises the superiority of the Neutrino Factory over any other combination of beams and detectors. The blue part of the bars indicates the statistical sensitivity limit; it is reduced if the correlations with other oscillation parameters and degeneracy errors are included. These additional sources of error can be addressed and the final achievable sensitivity is given by the leftmost edge: in both cases a fully developed Neutrino Factory overcomes the sensitivity of the other experiments by up to two orders of magnitude.

Precise measurements of deep inelastic scattering in neutrino physics will be made possible by the use of a Neutrino Factory. The electroweak sector of the Standard Model, in particular the determination of $\sin^2\theta_\text{W}$, could be tested from the measurements of both electron and muon neutrino cross sections. Non-neutrino science would also be possible; intense beams of muons with momenta of order $100$\,MeV/$c$ and a variety of time structures can be provided for slow muon physics studies. Both muon lifetime high precision measurements and magnetic muon studies will allow many parameters of the SM to be determined with unprecedented precision.

\subsubsection{Facility design}
As the neutrino beams at a Neutrino Factory\,\cite{bib:nu_fact} will be produced from the decay of muons circulating in a storage ring, the primary aim of the accelerator complex is to achieve as high a muon intensity as possible. The muon production starts with a high power proton source to create intense bunches of protons fired into a target. The pions created in the collision are subsequently captured and transported along a decay channel, where they decay to muons. The resulting muon beam has a large size and a large spread in longitudinal and transverse momentum, i.e. a large emittance, which must be reduced to avoid a large fraction of the muons being lost during acceleration and subsequent injection into the storage ring. The reduction of the momentum spread and transverse emittance takes place in two stages, called respectively rotation and cooling. The muons are then accelerated in a series of accelerators, before being injected into the storage ring.

\begin{figure}[!htb]
  \centering
  \includegraphics[width=.9\textwidth]{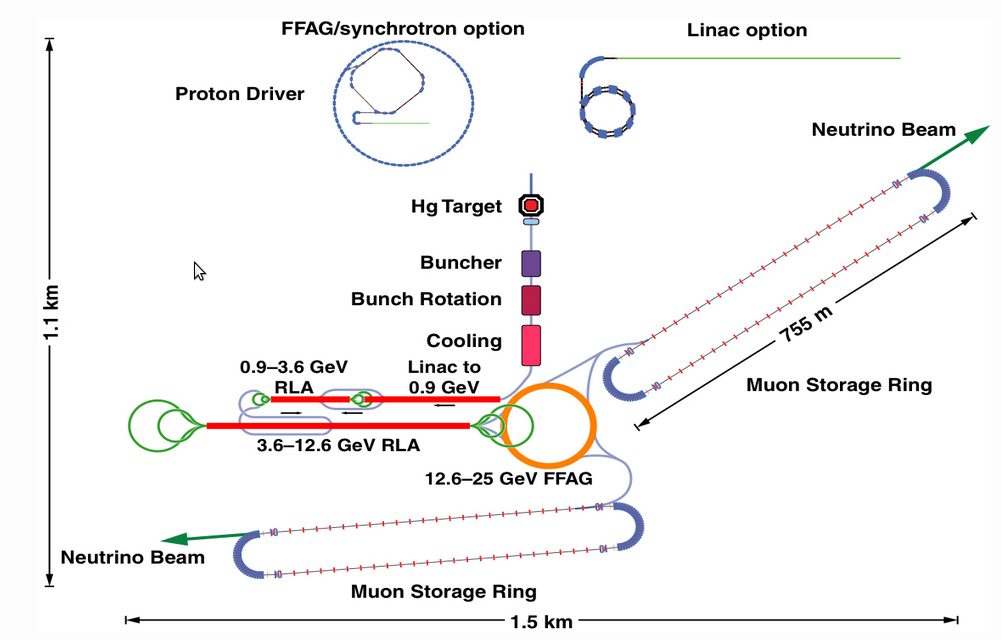}
  \caption{Schematic layout of the Neutrino Factory.}
  \label{fig:nu_fact}
\end{figure}

A number of different designs exist for the Neutrino Factory \cite{bib:nu_fact_status}. Although there are substantial differences between them, each design consists of the same basic components. Figure\,\ref{fig:nu_fact} shows how these components are laid out in the CERN design. A proton driver (synchrotron or linac) produces the necessary very high beam power (4\,MW). To minimize the longitudinal emittance of the initial muon beam, the proton bunches must be no more than a few nanoseconds long. Due to the high beam power and small size, the power density in the target far exceeds that of any comparable facility. Building a target that can withstand the mechanical and thermal stresses that such a beam will create is a major challenge and is the subject of an active R\&D program\,\cite{bib:merit}. The produced pions are then magnetically captured and focused by a powerful magnet.

The pions decay to produce muons in a decay channel that is 30--40\,m long. The large momentum spread of the decay muons will be reduced using phase rotation during which early (high energy) particles are de-accelerated and late (low energy) particles are accelerated using a system of RF cavities. The muons are then segmented into RF bunches and the transverse emittance reduced in an ionization cooling channel.

\subsection{Ionization cooling}
\label{section:icool}

Cooling the high emittance muon beam produced at the end of pion decay pipe in the Neutrino Factory is a essential step to achieve the high intensity, monochromatic and well collimated beam necessary for the completion of its physics goals. Four beam cooling techniques have been theorised and studied in the past\,\cite{bib:beam_cool}:
\begin{itemize}
\item radiation cooling takes advantage of the natural phenomenon of synchrotron radiation emitted by all relativistic charged particles accelerated or stored in a ring;

\item electron cooling consists in the injection of a well collimated electron beam in parallel of the primary beam (typically a heavy ion beam). The cooling is achieved through the multiple Coulomb scattering between the two beams.

\item stochastic cooling is based on an active feedback system that reduces the beam emittance by correcting the motion of the particles, e.g. by tuning the surrounding magnetic field;

\item ionization cooling uses the energy loss caused by the passage of charged particles through a relatively dense medium.
\end{itemize}
The first three methods are efficient in the case of $e^+e^-$ and hadron colliders but have never been applied to the cooling of muons. The mass of the muon ($\sim200$ times larger than the electron) makes cooling by radiation damping impossible while its short lifetime excludes the use of either the second or the third method. Ionization cooling, given the long interaction length of muons, is the only viable choice to cool muon beams.

\begin{figure}[!htb]
  \centering
  \includegraphics[width=.8\textwidth]{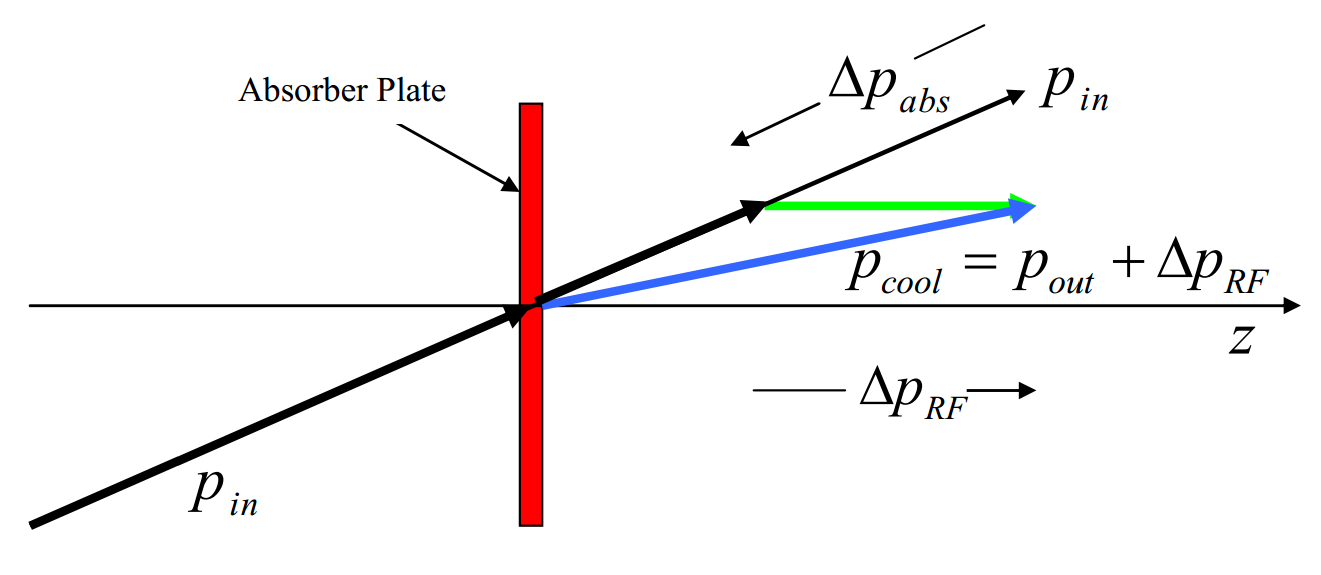}
  \caption{Conceptual representation of the principle of ionization cooling. Each particle loses total momentum by ionizing an absorber, but only the longitudinal momentum is restored by RF cavities.}
  \label{fig:icool}
\end{figure}

A conceptual representation of angular spread reduction through ionization cooling is provided in figure\,\ref{fig:icool}. The muon energy loss in the absorber is described per unit of distance by the relativistic Bethe-Bloch formula\,\cite{bib:sigmund}. The muon loses momentum both in the transverse and longitudinal frame but only the longitudinal component is restored by re-accelerating the beam in RF cavities.

The main parameter used to quantify the cooling is the geometric emittance, $\epsilon$, defined as the volume occupied by the beam in phase space. It can be expressed as $\epsilon = D^\frac16$, with $D$ the determinant of the six-dimensional covariance matrix of the beam particles in the coordinate system $(x,y,t,dx/dz,dy/dz,cdt/dz)$ with $z$ the beam direction. Taking into account the natural decrease of the beam size with the acceleration, it is convenient to define the normalized emittance, $\epsilon_n$, as a function of $(x,y,t,p/mc\cdot dx/dt,p/mc\cdot dy/dt,p/m\cdot dt/dz)$.

The 4D transverse emittance is defined as in the transverse phase space $(x,y,dx/dz,dy/dz)$. In a solenoid channel, this is equivalent to the 2D transverse emittance, $\epsilon_x$ of $(x,dx/dz)$, due the cylindrical symmetry. The covariance matrix is greatly simplified and the squared transverse emittance along the $x$ axis reads
\begin{equation}
\epsilon_x^2 = D =\det\begin{pmatrix}
<x^2> & <x\theta> \\
<\theta x> & <\theta^2> \\
\end{pmatrix} = <x^2><\theta^2>-<x\theta>^2,
\end{equation}
with $\theta=dx/dz$ the angular divergence of the particle in the $x$--$z$ plane. Following the process described thoroughly in\,\cite{bib:mu_collider_feas}, one can derive the rate of change in transverse normalised RMS emittance as
\begin{equation}
\frac{d\epsilon_{N}}{dz}\simeq-\frac1{\beta^2}\frac{\epsilon_{N}}{E_\mu}\left|\frac{dE_\mu}{dz}\right|+\frac{\beta_\perp(13.6\,\text{MeV})^2}{2\beta^3E_\mu m_\mu c^2}\frac1{X_0}
\end{equation} 
with $\epsilon_N$ the input normalised emittance, $|dE_\mu/dz|$ the rate of energy loss, $\beta_\perp$ the betatron function, $X_0$ the radiation length, $E_\mu$, $\beta$ and $m_\mu$ respectively the muon energy, velocity and mass. The first term is the energy loss cooling factor, i.e. reduces the beam emittance, while the second one is the scattering heating factor. The longitudinal emittance is defined similarly in the time-energy dimensions.

To minimize the heating term, which is proportional to $\beta_\perp$ and inversely proportional to the radiation length, the optimal choice is to use pressurized liquid hydrogen as the energy absorbing medium, with $|dE_\mu/dz| = 30$\,MeV/m and $X_0 = 8.7$\,m, and to use superconducting solenoid focusing to give a small value of $\beta_\perp \sim 10$\,cm, rather than quadrupoles. This corresponds to large beam divergence at the location of the absorbers, in order to minimize the effect of scattering in the absorber.

An additional technical requirement is high-gradient re-acceleration of the muons between absorbers to replace the lost energy, so that the ionization-cooling process can be repeated many times with negligible losses. The achievable RF gradient determines how much cooling is practical before an significant fraction of the muons have decayed or drifted out of the RF bucket.

\subsection{MICE cooling channel}
The International Muon Ionization Cooling Experiment (MICE)\,\cite{bib:mice_proposal} is an R\&D project whose main goals are to study the feasibility of a Neutrino Factory based on a muon storage ring and the experimental demonstration of the ionization cooling technique.

\begin{figure}[!htb]
  \centering
  \includegraphics[width=.75\textwidth]{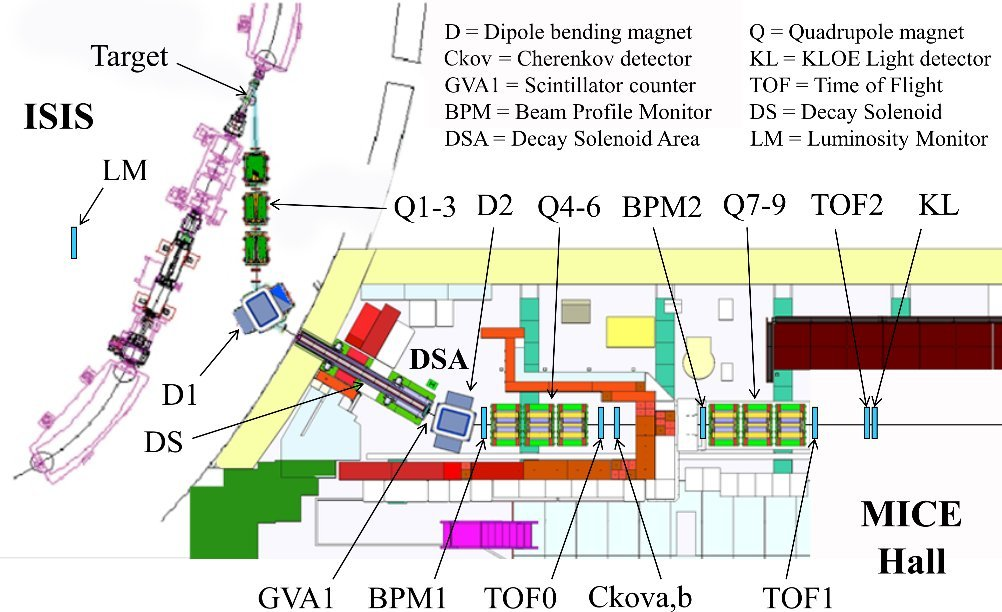}
  \caption{Layout of the MICE beam line and experiment hall. The main magnetic components needed for muon transport and the upstream diagnostics system are shown.}
  \label{fig:mice_bl}
\end{figure}

The experiment is currently under construction at the Rutherford Appleton Laboratory (RAL) in the UK. The existing ISIS synchrotron is used as an 800\,MeV proton driver to create a pion beam that further decays into the muons required in the cooling channel. A titanium target is dipped into the ISIS beam every second, producing a pion beam in the first section of the MICE beam line. The pions are captures and collimated by a triplet of quadrupoles (Q1--Q3), a dipole (D1) and are left to decay in a 5\,m decay solenoid (DS). The muons are subsequently counted by a scintillating fibre monitor (GVA1) and pass through another dipole and triplet of quadrupoles (D2+Q4--Q6) before entering the experiment. A layout of the beamline is shown in figure\,\ref{fig:mice_bl}.

\begin{figure}[!htb]
  \centering
  \begin{subfigure}[t]{.55\textwidth}
    \centering
    \includegraphics[width=\textwidth]{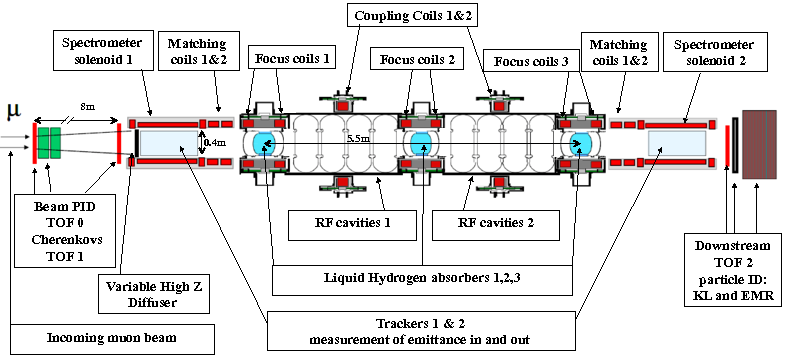}
    \caption{}
  \end{subfigure}
  \hfill
  \begin{subfigure}[t]{.4\textwidth}
    \centering
    \includegraphics[width=\textwidth]{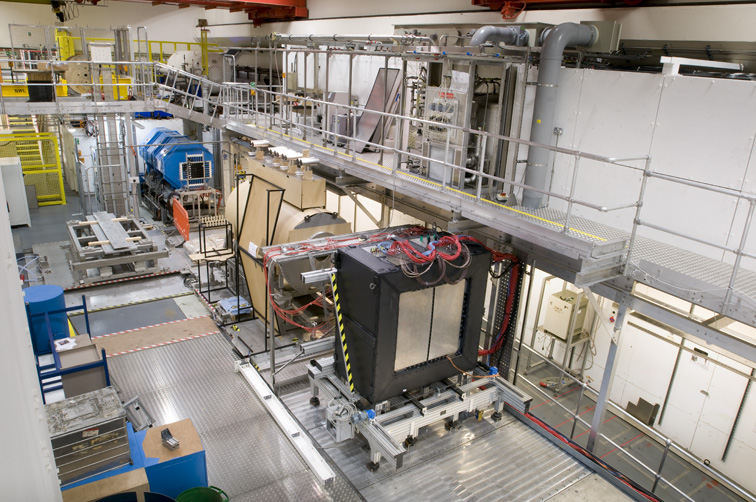}
    \caption{}
  \end{subfigure}
  \caption{(a) Schematic layout of the MICE cooling channel (beam entering from the left). (b) Photo of the MICE experimental area in October 2013. The last triplet of blue quadrupoles (Q7-Q9), TOF1 (at the end of the triplet), a dummy of SS1 and the EMR can be seen in the picture.}
  \label{fig:mice_cc}
\end{figure}

MICE typically operates with a muon beam of momenta in the range 140--240\,MeV/$c$, a $\beta_\perp=42$\,cm at the centre of the absorber and input normalized emittance\footnote{The upstream emittance can be tuned by a set of diffusers at the entrance of the cooling channel\,\cite{bib:mice_diff}.} in the range 1--10\,$\pi\cdot\text{mm}\cdot\text{rad}$. The cooling section design follows the guidelines of the US Feasibility Study-II\,\cite{bib:nu_fact_feas}. A schematic layout of the detectors and cooling section elements position is shown in figure\,\ref{fig:mice_cc} (a). The experiment in its current configuration is pictured in figure\,\ref{fig:mice_cc} (b).

The basic elements of the MICE cooling channel are three Absorber and Focusing Coil (AFCs) modules and the two RF cavity and Coupling Coil (RFCCs) stations\,\cite{bib:mice_rf}. The overall length of the channel is $\sim5.5$\,m. Each AFC module contains a liquid hydrogen absorber at a cryogenic temperature that provides the energy loss of muons, and a pair of focusing coils to reduce the betatron function, ensuring a small equilibrium emittance. Each RFCC station consists of four 201\,MHz normal-conducting RF cavities and one super-conducting solenoid.

The tracking and particle identification is accomplished by two scintillating fibre trackers, a time-of-flight (TOF) system, Cherenkov threshold counters and downstream calorimetry. The upstream PID provides pion and electron rejection while the downstream section tags muons that decayed inside the cooling channel. The overall length of the MICE experiment is $\sim11.5$\,m.

Three TOF stations \cite{bib:mice_tofs} are positioned along the cooling section to provide the time measurement in the emittance estimation. TOF0 is located at the end of the beam line, while TOF1 and TOF2 are positioned respectively at the entrance and at the exit of the cooling channel. The main task of the upstream TOFs is the pion background rejection. They also supply the trigger for the experiment in coincidence with the ISIS clock. TOF2, at the end of the channel, selects the particles passing through it for the downstream emittance measurement and the cooling efficiency estimation. The TOF stations have a common design: two planes of fast scintillator bars arranged in a X--Y configuration.

At high momentum ($>300$\,MeV/$c$), the time-of-flight difference between muons and pions becomes small over a distance of $10$\,m with respect to detector resolution. Two Cherenkov counters are used to provide a sufficiently good pion/muon separation in that regime\,\cite{bib:mice_ckov}. The active radiator is a high density silica aerogel plate that produces Cherenkov light read out by four 8 EMI 9356 KA photomultipliers. The association of the CKOV and the two first TOF station allows to achieve a beam purity of up to 99.98\%\,\cite{bib:mice_pid}.

Charged-particle tracking in MICE is provided by two scintillating fibre trackers embedded in spectrometer solenoids\,\cite{bib:mice_tracker}. They are required to measure the relative change in transverse emittance of approximately 10\,\% with a precision of \,1\%, i.e. a 0.1\,\% precision on the absolute emittance. Each spectrometer consists of a 4\,T superconducting solenoid instrumented with a 1.1\,m long tracker, composed of five planar scintillating fibre stations. One of the trackers has been tested at RAL with cosmic rays and achieved a spatial resolution of $682\pm1\,\mu$m and an efficiency of $99.82\pm0.1$\,\%.

The electron background rejection at the end of the cooling channel is based on the Electron Muon calorimeter (EMcal) station. The electrons shower in an electromagnetic preshower calorimeter, KLOE-Light (KL), while muons penetrate it. They are detected downstream in a fully active scintillating tracker-calorimeter, the Electron-Muon Ranger (EMR), extensively described in section\,\ref{section:emr}. The KL consists of $80\times80$\,cm$^2$ grooved lead layers interwoven with 1\,mm-diameter blue scintillating fibres inserted and glued in the gaps. Its thickness is $\sim4$\,cm, corresponding to about 2.5 radiation lengths. The KL relative energy resolution is $\sigma_E/E = 7\,\%/\sqrt{E(\text{GeV})}$ and its time resolution is $\sigma_t = 70$\,ps$/\sqrt{E(\text{GeV})}$\,\cite{bib:mice_kl}.

\section{Electron-Muon Ranger}
\label{section:emr}
The particle identification upstream the cooling channel in MICE is provided by the TOF dectors, the Cherenkov counters and the trackers. Simulations\,\cite{bib:mice_instrumentation} have shown that TOF2 cannot ensure alone the the rejection of electrons produced by the decay in flight of muons in the cooling channel. A detector able to accurately discriminate the electrons from the muons is required to achieve the commissioned systematic level of precision\,\cite{bib:mice_proposal}. A downstream detector system is required for particle identification at the end of the MICE channel. This detector consists of a sampling preshower calorimeter coupled to a fully active scintillator tracker-calorimeter, the Electron-Muon Ranger (EMR), which characterization is the centre of interest of this master thesis. 

In this section, the EMR and its critical role in MICE are described. The detector relies on $\sim1.5$\,tons of triangular plastic scintillating bars and electronics based on custom front-end boards, digital buffer boards and standard VME modules. Several characterization tests have been performed in the context of this thesis. Some of them were conducted during the building process and others were performed on the completed device after it was transported to RAL.

The detector is currently fully operational and located in the MICE channel. The EMR will undergo minor upgrades in the months to come as its ageing Philips PMTs\,\cite{bib:emr_philips} will be replaced by brand new Hamamatsu PMTs and many DAQ parameters will be optimized.

\subsection{Purpose}
\begin{figure}[!htb]
  \centering
  \includegraphics[width=.8\textwidth]{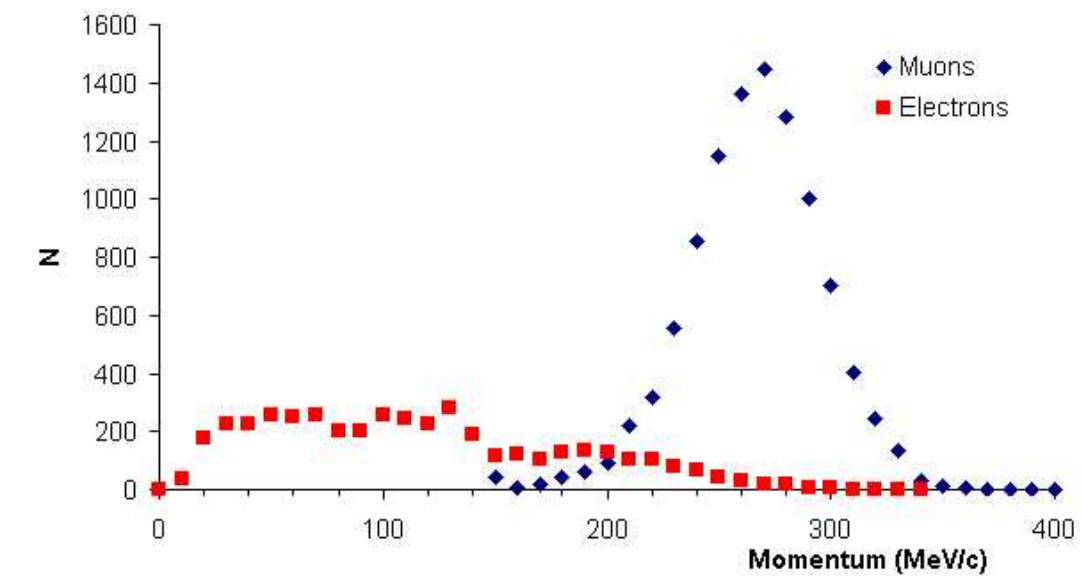}
  \caption{Momentum distribution of muons (blue lozenge) and electrons (red squares) downstream the MICE cooling channel for a 300\,MeV/$c$ central momentum beam.}
  \label{fig:e_contamination}
\end{figure}

Particle identification upstream and downstream the cooling channel is a fundamental task to achieve the required precision on the beam emittance measurement. MICE works with muon beams of tunable emittance and energy. Everything apart from muons is considered background. There are three main sources of background in MICE:

\begin{itemize}
\item pions, from which the muons are produced, that remain in the beam;

\item dark current originating from the RF cavities operating in high electric and magnetic fields. Electrons are ripped off the surface of the cavities and accelerated along the cooling channel, causing bremsstrahlung photon emission and background noise in the trackers;

\item muons decaying inside the cooling section or in one of the spectrometers. The momentum distribution of muons and electrons arriving at the end of the cooling channel for a 300\,MeV/$c$ central momentum beam is given in figure\,\ref{fig:e_contamination}.
\end{itemize}

The separation of pions from muons and the RF electron background rejection at the beginning of the cooling section are provided by the upstream TOFs, the CKOV stations and the spectrometer trackers\,\cite{bib:mice_status}. The main concern for the emittance high precision measurement is thus represented by the downstream particle identification. Kinematics cuts can reject about 80\,\% of decay electrons, but this is not enough to avoid non-negligible systematic errors on the beam emittance measurement. Dedicated detectors are necessary to separate electrons from muons. 

\begin{figure}[!htb]
  \begin{minipage}[b]{.45\textwidth}
    \centering
    \includegraphics[width=\textwidth]{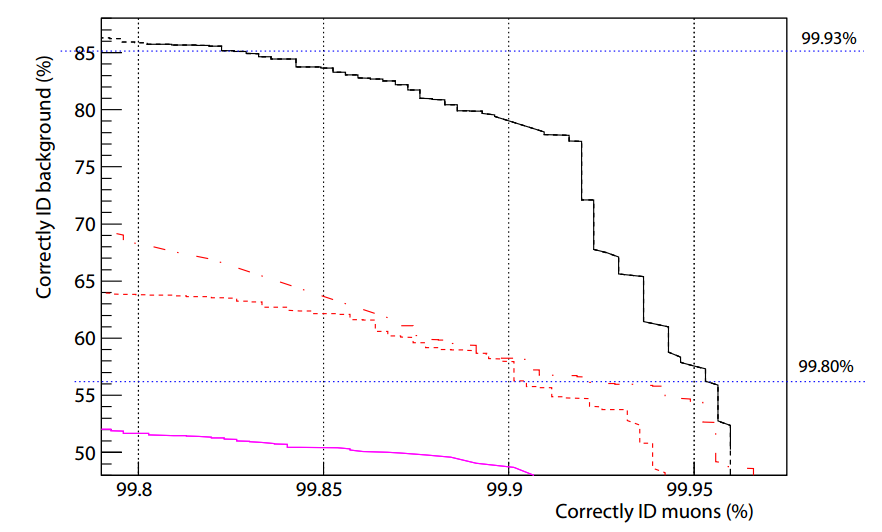}
    \caption{Background identification efficiency as a function muon acceptance for a $140\pm14$\,MeV/$c$ beam. The solid black line corresponds to SW+TOF, the dash-dotted red line to KL, the dashed red and black lines to KL and SW without TOF repectively and the purple solid line to TOF only.}
    \label{fig:e_mu_efficiency}    
  \end{minipage}
  \hfill
  \begin{minipage}[b]{.45\textwidth}
    \centering
    \includegraphics[width=\textwidth]{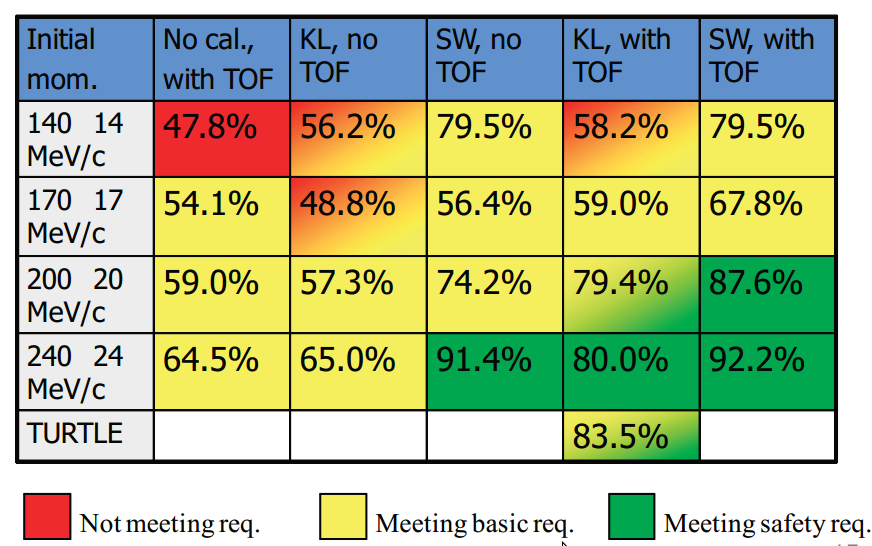}
    \caption{Background rejection efficiency at 99.9\% muon acceptance for four different beam central momenta and five designs of downstream apparatus.\\ \\ \\}
    \label{fig:e_mu_tab}
  \end{minipage}
\end{figure}

Several solutions based on a calorimeter system were proposed and their performances in terms of electron/muon separation efficiency were studied with G4MICE simulations\,\cite{bib:mice_instrumentation}. Figure\,\ref{fig:e_mu_efficiency} shows the background identification efficiency as a function of muon acceptance for three alternative designs and a $140\pm14$\,MeV/$c$ muon beam. The red line indicates the configuration in which four KL-like layers are present, the black line shows a design that foresees the use of a single KL layer followed by a fully active plastic scintillator detector (KL+SW\footnote{The original proposed design of the EMR, called SandWich (SW), was composed of 10 modules of plastic scintillator with different thicknesses. The detector design was changed mainly due to cost reduction and simplification of the manufacturing.}) and the purple line represents a solution with only TOF2. The optimal choice is the second one. Figure\,\ref{fig:e_mu_tab} shows background identification efficiency for different nominal momenta and configurations. The configuration with the SW and TOF2 yields the best possible performance and is the only one that yields acceptable background rejection efficiencies for the four beam central momenta of interest.

The downstream background rejection approach is to distinguish electrons from muons using the longitudinal profile of the electromagnetic shower at the end of the cooling section. A high Z material (e.g. a lead preshower) combined with a low Z one (e.g. scintillating plastic) is the ideal choice. Muon events are distinguished from the background events due to their significantly different topology. The electrons lose most of their energy in the preshower generating an electromagnetic cascade in the following layer while the muons penetrate the high Z material without interacting.

\subsection{Design}
\begin{figure}[!htb]
  \centering
  \begin{minipage}[b]{.45\textwidth}
    \centering
    \includegraphics[width=\textwidth]{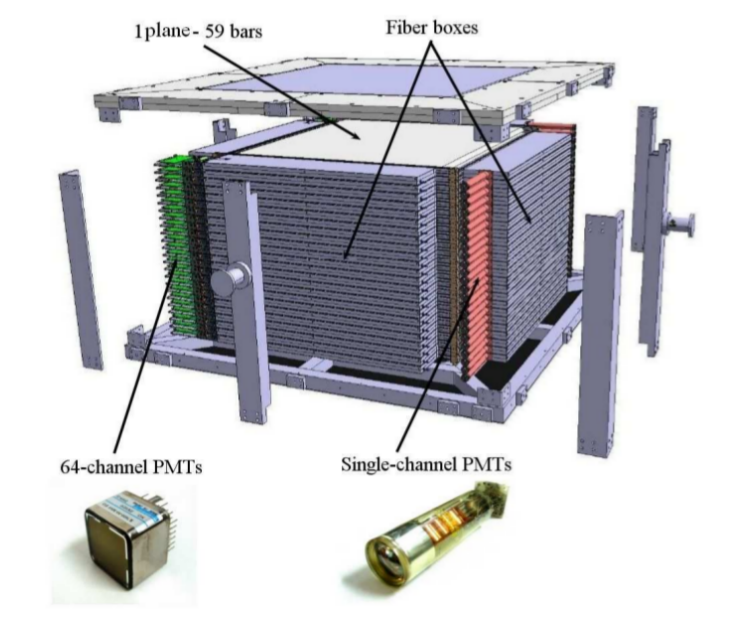}
    \caption{3D engineering view of the EMR detector and its mechanical support. The placement of the SAPMTs, MAPMTs, FEBs and fibre boxes are represented.}
    \label{fig:EMR}
  \end{minipage}
  \hfill
  \begin{minipage}[b]{.45\textwidth}
    \centering
    \includegraphics[width=.9\textwidth]{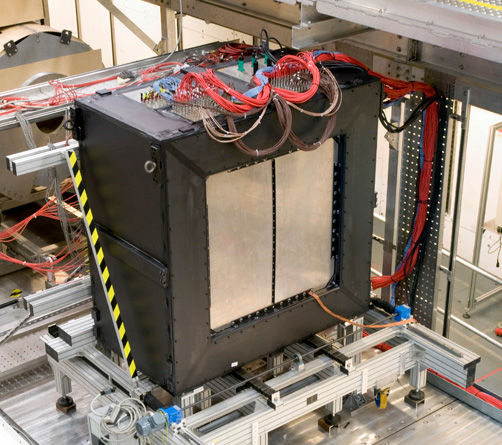}
    \caption{Picture of the back of the EMR box placed in the beam line of MICE at Step\,I. The bundled red cables supply the high voltage to the PMTs.}
    \label{fig:EMR_picture}    
  \end{minipage}
\end{figure}

Figure\,\ref{fig:EMR} depicts a 3D engineering rendering of the Electron-Muon Ranger (EMR) and figure\,\ref{fig:EMR_picture} shows a picture of the detector in the MICE hall. The EMR consists of 48 layers, organized in an X--Y geometry (24 X--Y modules), of extruded scintillator bars\,\cite{bib:emr_scintillator} made of blue-emitting DOW Styron 663\,W polystyrene +\,1\,\% PPO +\,0.03\,\% POPOP dopants\footnote{PPO (2,5-Diphenyloxazole) and POPOP (1,4-bis(5-phenyloxazol-2-yl) benzene) are scintillators. POPOP is used as a wavelength shifter; its emission spectrum peaks at 410 nm.}. The layers are positioned one after the other and are supported by a metallic frame. The whole detector is enclosed in a black aluminium box (EMR Outer Box, EOB) to shield it from ambient light.

Each layer consists of 59 1.1\,m-long bars of a triangular sections with a 3.3\,cm base and 1.7\,cm height, for a total of 2832 bars. A complete EMR module covers an active region of $\sim1$\,m$^2$. A layer weighs $\sim28$ kg and the whole scintillating volume approaches the one and a half tons mark.

\begin{figure}[!htb]
  \centering
  \includegraphics[width=.8\textwidth]{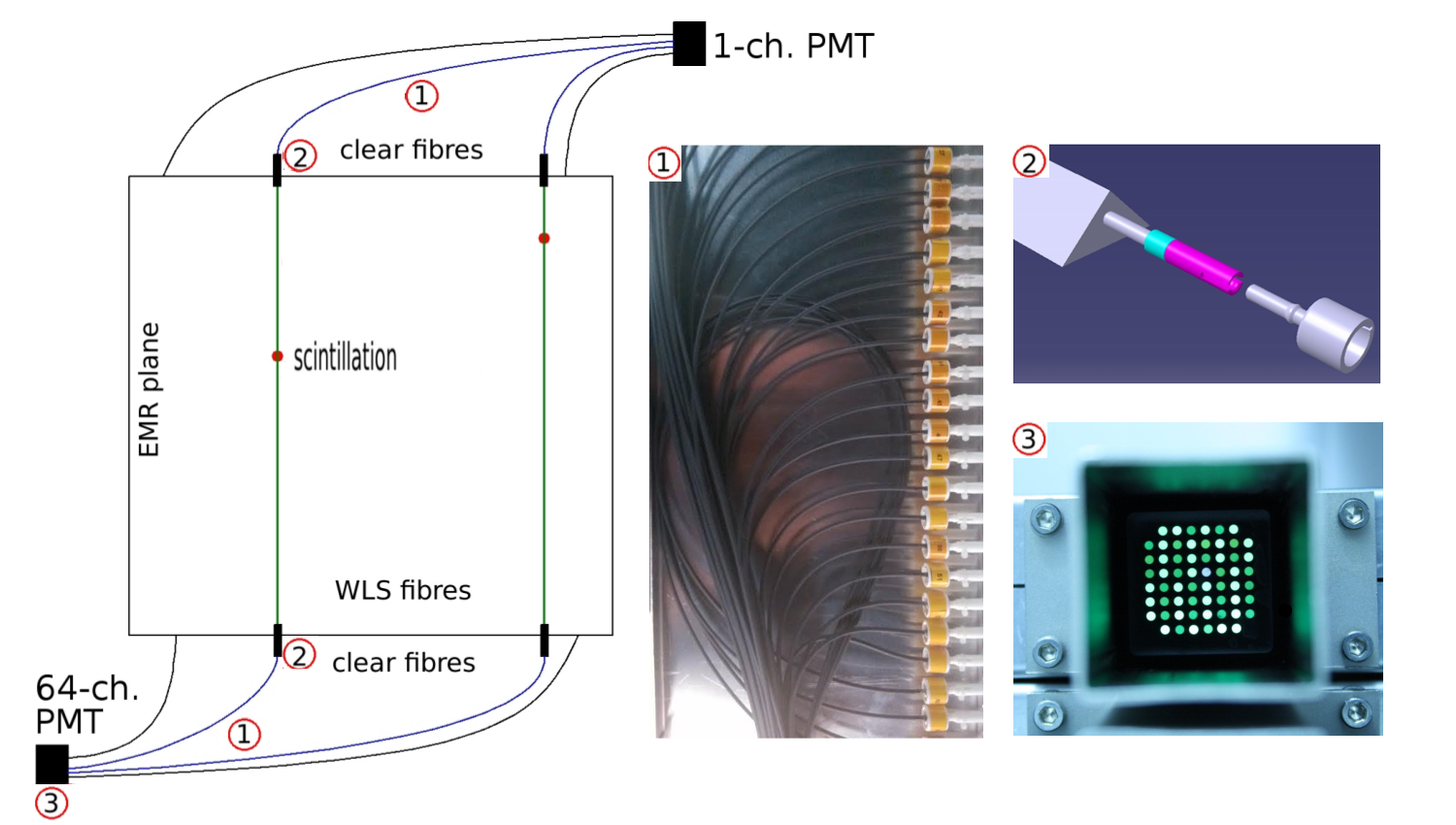}
  \caption{Scheme of the light transport in a plane of the EMR. At scintillation, light is transported by the WLS fibre to the connectors (2). It is transmitted through the clear fibres (1) to the fibre mask (3) that is fitted against the PMT. The mask represented in this figure is the MAPMT fibre mask.}
  \label{fig:light_transport}
\end{figure}

The light produced by a scintillator is conducted by one 1.2\,mm BFC-91A Wave Length Shifter (WLS) fibre\,\cite{bib:emr_wlsfibres}. Each fibre goes through the whole bar and has a polished end on either side. A connector is screwed onto the end sections, to which a clear 1.5\,mm polystyrene fibre\,\cite{bib:emr_clfibres} is be attached. The clear fibre connectors are clipped against the polished end of the WLS fibres and have a variable length depending on the distance between their end point and the photomultiplier\,\cite{bib:emr_design_change}. To protect the fibres and support the photomultipliers and their electronics, each layer is equipped with an aluminium box on each side that also provides light tightness. Figure\,\ref{fig:light_transport} summarises the light transport process from production to detection.

The clear fibres are connected on both sides to two different photomultiplier systems. On one side, the fibres exiting from each bar are grouped together and are connected, through a dedicated mask, to the single-anode Philips XP 2972 PMT (SAPMT)\,\cite{bib:emr_philips}, which purpose is to measure the total charge deposited in one plane, i.e. the energy loss. The front-end electronics of the SAPMT have been adapted to put it directly inside the metallic support that provides the against the magnetic field (mu metal). Each SAPMT is powered by a 1800\,V high voltage power supply and connected analogically to a wave form digitizer. On the other side, each clear fibre coupled, through a dedicated square mask, to one specific channel of a multi-anode green enhanced PMT (MAPMT) R7600-00-M64 (H7546B assembly, Hamamatsu\,\cite{bib:hamamatsu_pmt}). The rear of the MAPMT is soldered to a 4 layer rigid-flex (kapton) circuit that allows for the required mechanical flexibility. Each MAPMT is powered by a 700\,V high voltage power supply and connected to front-end electronics for further processing.

The EMR electronics chain has to cope with the MICE experimental duty cycle which consists in a 1\,ms spill every second. Within this spill, up to one good event every 5\,$\mu$s has to be recorded. In this time scale, the EMR electronics chain has to sample and discriminate the signal of each MAPMT channel, assign a time stamp to every bar over threshold, store data in a digital form and make them available for the readout at the end of the spill.

The EMR electronics chain is divided in three main elements:

\begin{enumerate}
\item the Front-End Boards (FEBs), located near the fibre masks, route the MAPMT signals to an ASIC that conditions the signals and send them to the second element in the chain. The signal from the MAPMT is shaped, the time-over-hreshold is measured and digitized. The choice to use only the digital information is due to time constraints;

\item the Digital Buffer Boards (DBBs) are data storage modules coupled to each FEB. They sample the digital outputs of the ASIC with a 400\,MHz clock and, in presence of a spill gate, store the channel information with a timestamp and a time-over-threshold measurement to send them to the DAQ system in the interspill period;

\item the Data AcQuisition system (DAQ) consists of a VME crate hosting the Configuration Boards (VCBs), the Readout Boards (VRBs) and the flash Analogue-to-Digital Converters (fADCs). The VCB main task is to flash the FEBs' firmware, i.e. set the ASIC mask (e.g. pre-amplifier gain, shaper parameters or discriminators threshold). The configuration of the DBBs (e.g. clock rate or data format) is also managed by this board. The boards are configured at the beginning of each run. The MICE particle trigger is sent to the detector and the readout boards straight from the control room through a NIM shaper. The clock synchronization between the boards is performed using the trigger signal.
\end{enumerate}

\begin{figure}[!htb]
  \centering
  \includegraphics[width=.7\textwidth]{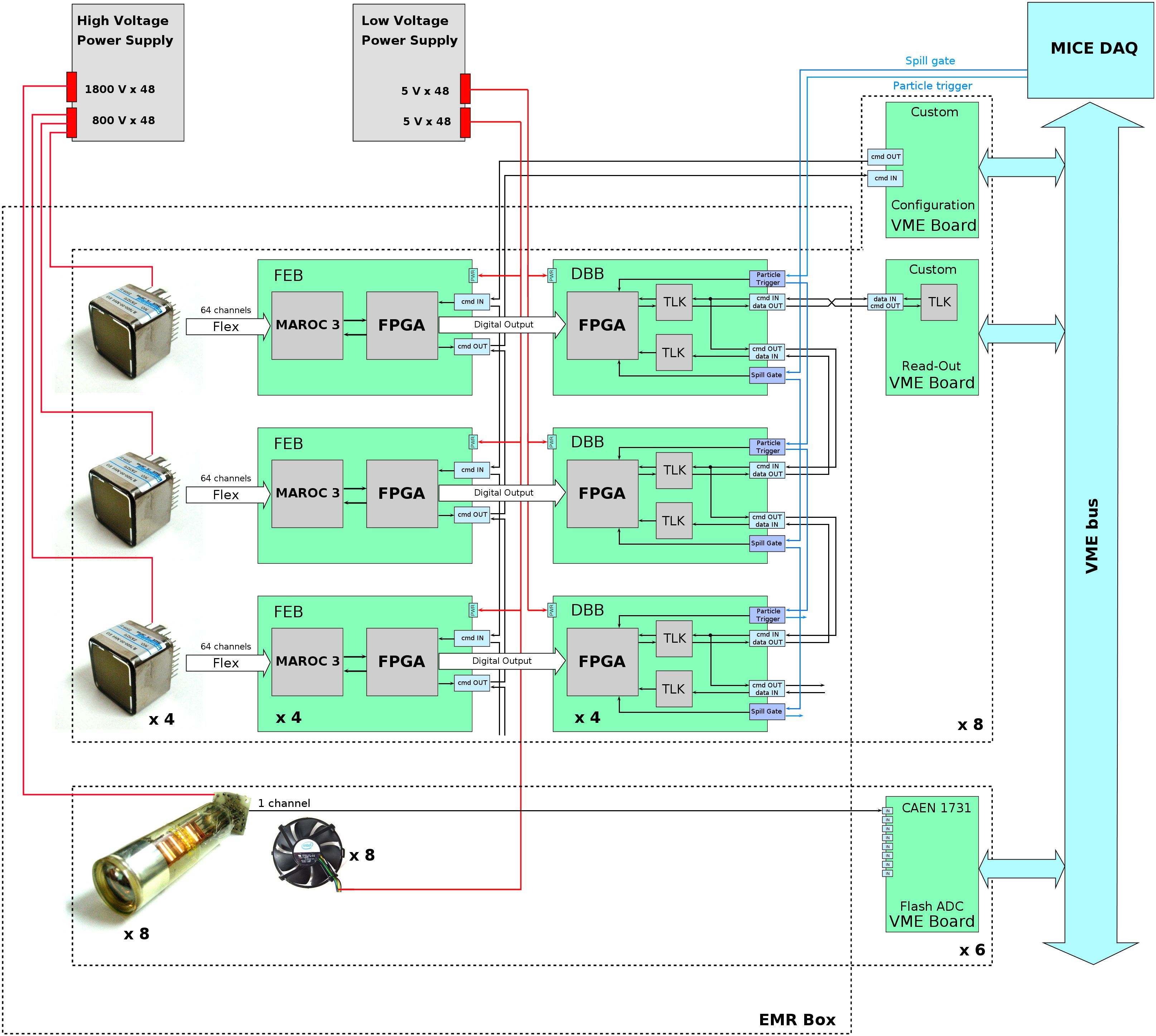}
  \caption{Scheme of the electronics chain for the readout and setting of the EMR detector. The SAPMTs are connected to fADCs housed in the VME crate. The MAPMTs signals are sampled and digitized by the Front-End Boards (FEBs) and the Digital Buffer Board (DBBs).}
  \label{fig:electronics_chain}
\end{figure}

A schematic representation of the complete electronics chain is shown in figure\,\ref{fig:electronics_chain}. The analogue signal of each single channel PMT is sampled and digitized by a V1731 Wave Form Digitizer (WFD, CAEN\,\cite{bib:emr_fADC}) housed in the VME crate.

\subsection{Reconstruction software}
\label{section:recon}
\begin{figure}[!htb]
  \centering
  \includegraphics[width=.8\textwidth]{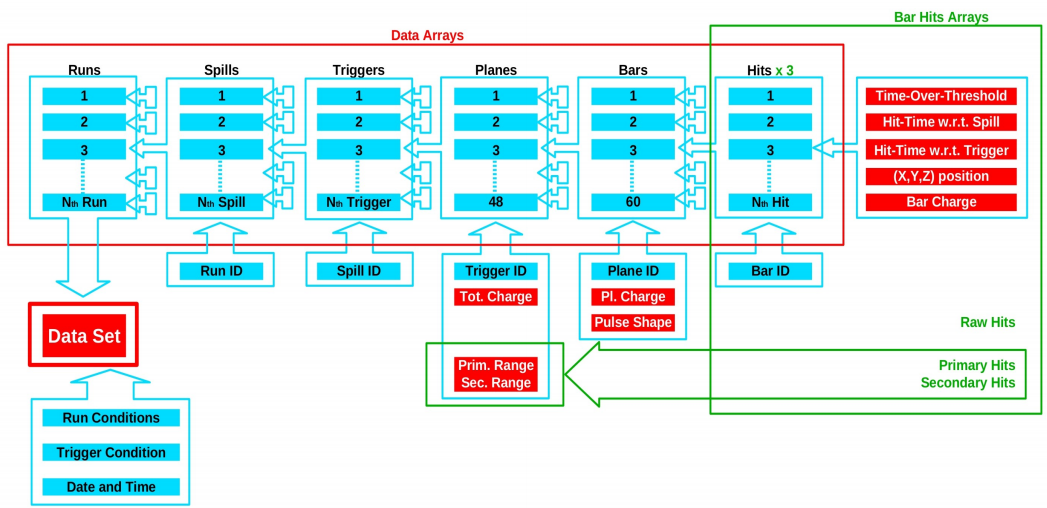}
  \caption{Data structure of the EMR. Each subsection contains its own set of variables. Some of them are defined during data acquisition such as the total plane charge or the bar time-over-threshold measurement but most of them are filled during reconstruction, e.g. the particle track space points, the range or bar charge.}
  \label{fig:data_structure}
\end{figure}

The EMR raw data are structured as shown in figure\,\ref{fig:data_structure} and consists in an array of subcategories. Each run is divided in a collection of raw ROOT files containing around 50 MICE spills. A spill contain the data arrays for all the active detectors in the cooling channel at the time of data acquisition. 

For a given particle trigger ($\sim50$ per spill with no DS during Step I), the array of planes that have been hit is stored. Each plane hit contains the information on the total integrated charge for this trigger and the number of bars hit in the plane. Each bar within the plane holds the time-over-threshold measurements and timestamps of the hits that were recorded in it. 

The EMR reconstruction will be useful in the signal acquisition efficiency analysis section. Its process is still under refinement but is already fully functional and divided in the following steps:

\begin{enumerate}
\item the bar hits are sorted according to the delay between their time stamp $t_\text{hit}$ and the trigger time $t_\text{trig}$, i.e. $\Delta t=t_\text{hit}-t_\text{trig}$. If the hit is close to the trigger, the hit corresponds to the primary particle trail and is stored in the first hit array. If the hit happens shortly after the primary time interval, it is associated with electronic noise and is stored in a second array. If the time does not correspond to a trigger, it is moved to a third array which contains the decay candidates as well as random noise;

\item for each hit in a given plane , the missing coordinate is reconstructed as the average of the two adjacent planes. In the case of an X plane, the $y$ coordinate is the average of the position of the hits in the adjacent Y planes;

\item the hits in the third array are associated in bunches according to their time stamps. The primary and secondary tracks are fitted with a straight line in both projections and their end points are reconstructed. The software tries to associate piecewise each decay candidate to one of the primary track within the spill by geometrically matching the end point of the primary track with either end of the secondary one.
\end{enumerate}

\section{MAPMT readouts quality test}
\subsection{Experimental set-up}
To check the readout electronics before assembling the EMR, a test bench was developed using an LED driver. This set-up provides a steady light pulser which intensity is constant from one readout chain to another.  A schematics of the EMR readout electronics is represented in figure\,\ref{fig:electronics_chain}. The light coming from the LED pulser shines on all the channels of the MAPMT. These primary signals are shaped and digitized by the FEB and DBB before being sent to a VME readout board (VRB). A VME configuration board (VCB) is used to flash the firmware or modify the readout of the parameters of the tested FEB.

The time-over-threshold (ToT) is recorded in each MAPMT channel for $10^6$ LED pulses. This measurement is obtained by shaping the MAMPT primary signal, setting an arbitrary threshold and recording how long the signal stays over this given value. It is output in ADC counts; 1\,ADC count corresponds to 2.5\,ns with a 400\,MHz sampling clock.

\subsection{Results}
\begin{figure}[!htb]
	\begin{minipage}[b]{.45\textwidth}
		\centering
		\includegraphics[width=\textwidth]{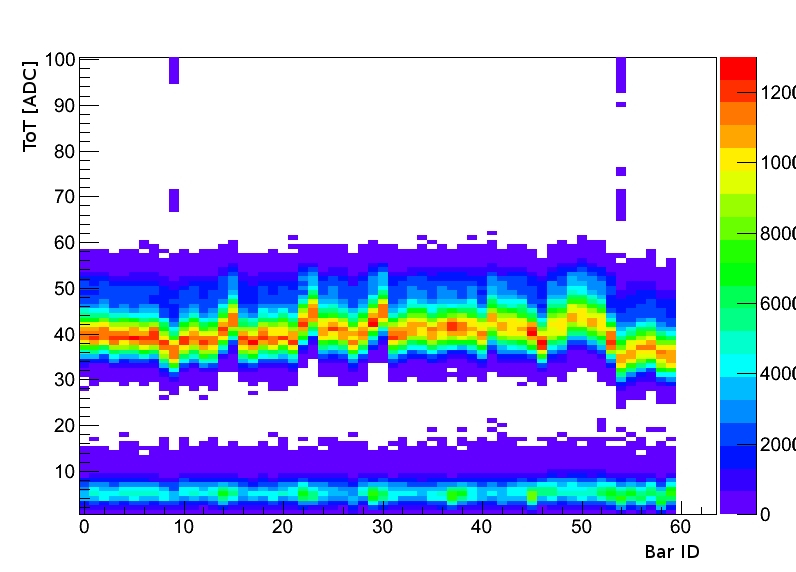}
		\caption{Time-over-threshold distribution of a functional readout chain. Each channel responds normally and gives a ToT distribution similar to the others.}
		\label{fig:readout_hist}
	\end{minipage}
	\hfill
	\begin{minipage}[b]{.45\textwidth} 
		\centering
		\includegraphics[width=\textwidth]{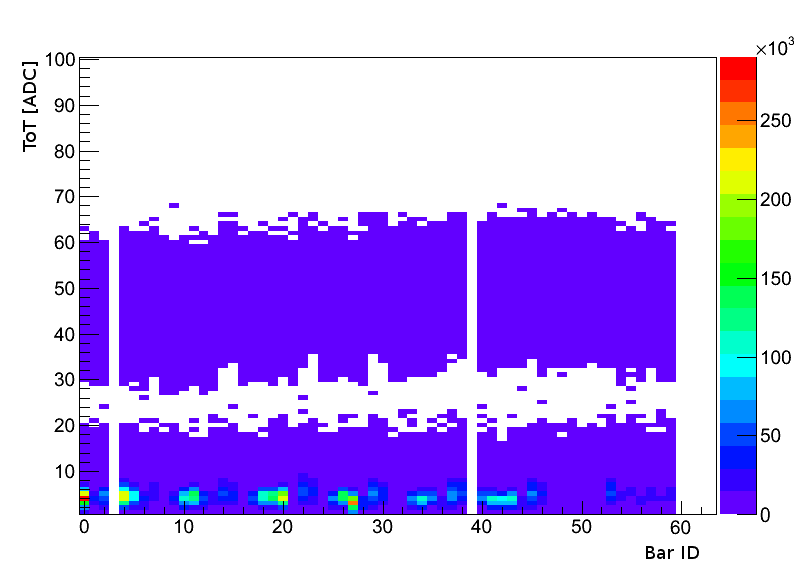}
		\caption{Time-over-threshold distributions of a malfunctioning readout chain. Two channels are dead and the others record unacceptable levels of low ToT hits, i.e. noise.}
		\label{fig:readout_bad}
	\end{minipage}  
\end{figure}

The distribution of time-over-threshold is displayed for each channel of an operational readout chain in figure\,\ref{fig:readout_hist}. In this case, none of the channels are flawed, they all record similar signals and have comparable distributions. The most numerous are the very high energy hits corresponding to the primary LED signal while the low ToT band corresponds to the trailing noise and crosstalk of each signal. The arched behaviour that is observed on the right side of the histogram every seven or eight bars is explained by the MAPMT mask structure. Each bunch corresponds to a distinct row of the MAPMT mask; the channels at the edges of these arches are on the outside of the mask and receive a little less light and are less cross-talked to, as explained in section \ref{section:xt}.

An example of a malfunctioning readout chain is shown in figure\,\ref{fig:readout_bad}. Two channels are dead as they do not record a single hit. The others do not exhibit an acceptable behaviour as the distribution does not peak where it is expected. The large amount of low energy hits suggests an unacceptable level of noise. FEBs presenting this type of behaviour are rejected from the pool and another board is used. At the end of testing, all the FEBs integrated in the EMR are functioning properly. An exhaustive set of time-over-threshold distributions is given for each plane in\,\cite{bib:electronics_qt}.

\section{Clear fibre luminosity}
\subsection{Experimental set-up}
To measure the luminosity of each fibre with respect to the others, a Canon\textregistered\ EOS 1000D camera is used as primary measuring tool. Each time a plane is assembled, the camera is placed right in front of the fibre bundle mask that is to be coupled with an MAPMT. The camera is placed at a constant distance from the bundle and the following settings are kept unchanged:

\begin{itemize}
\item exposure of 0.05\,s;
\item aperture of 5.00\,EV (f/5.7);
\item no flash.
\end{itemize}

\begin{figure}[!htb]
  \centering
  \includegraphics[width=.75\textwidth]{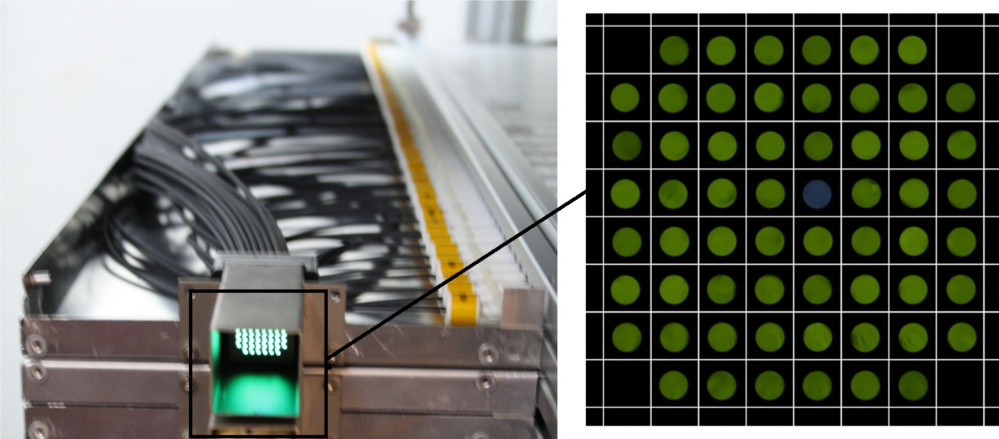}
  \caption{Close up picture of a fibre bundle mask layered with a grid allowing for the measurement of the luminosity of each fibre.}
  \label{fig:lumi_meas}
\end{figure}

A black light-proof heavy duty fabric cover is lowered over the whole detector so that the measurement is not influenced by the ambient luminosity. LED light sources are placed on the top part of the cover and shine directly on the 59 topmost bars of the EMR. When everything is in place, a picture of the fibre bundle mask is taken. The picture is overlaid with a grid as depicted in figure\,\ref{fig:lumi_meas} and the luminosity in each compartment is measured from the amount of photons recorded in that region of the CMOS sensor.

\subsection{Results}
\begin{figure}[!htb]
 \begin{minipage}[b]{.45\textwidth}
  \centering
  \includegraphics[width=\textwidth]{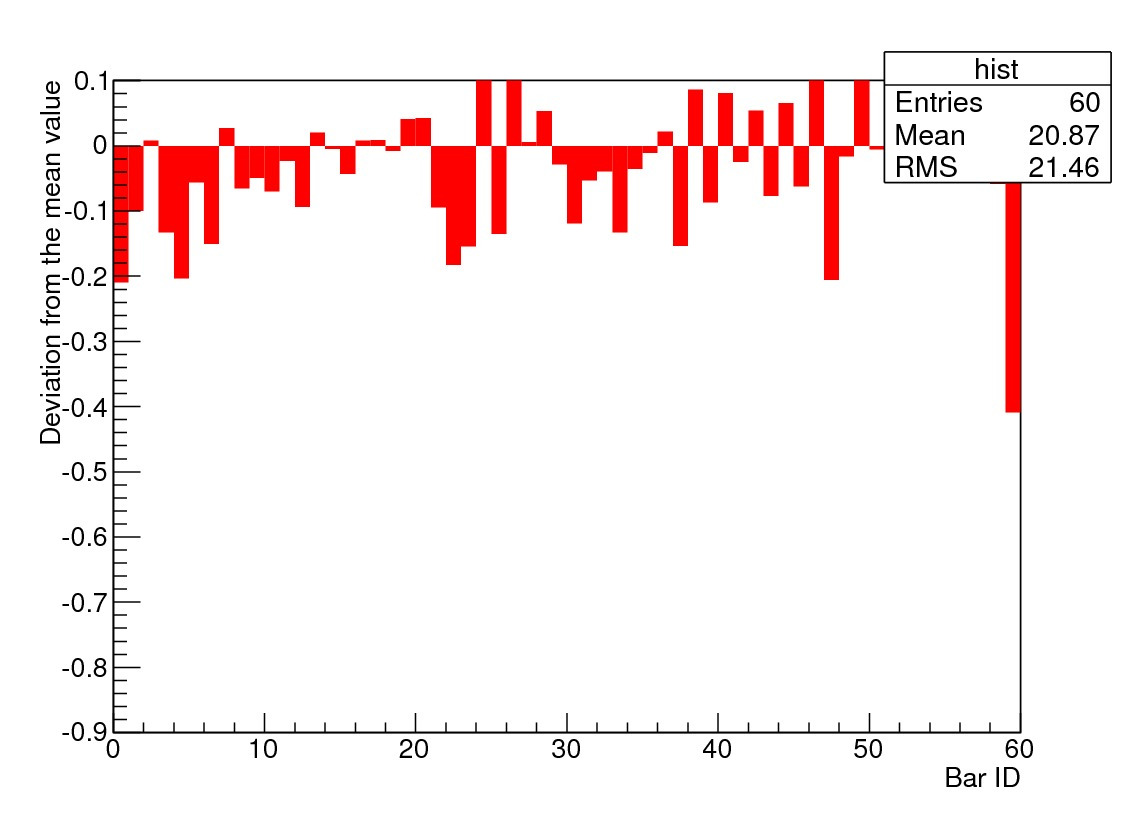}
  \caption{Deviation from the average luminosity $\Delta L_i/\overline{L}$ in a functional chain. Most values are contained within 20\,\% of the average. The last bin corresponds to the test channel and was not lit at the time of data acquisition.}
  \label{fig:lumi_hist}
 \end{minipage}
 \hfill
 \begin{minipage}[b]{.45\textwidth}
  \centering
  \includegraphics[width=\textwidth]{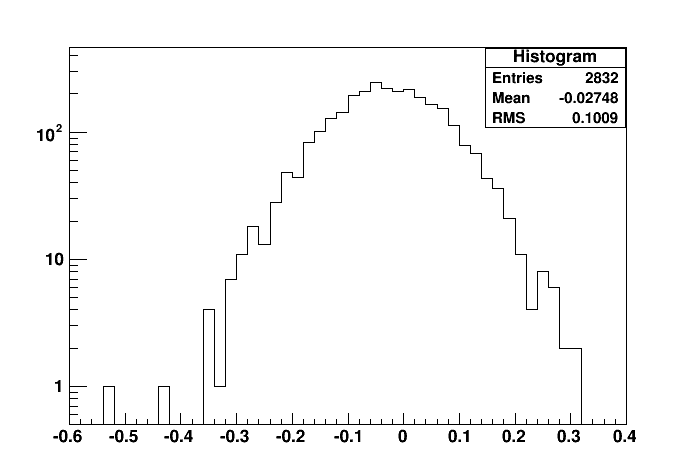}
  \caption{Distribution of the deviation from the average luminosity $\Delta L_i/\overline{L}$ in the 2832 channels. Two fibres yield a luminosity more than 40\,\% under the average signal and could be damaged.}
  \label{fig:lumi_dist} 	
 \end{minipage} 
\end{figure}

The measured luminosity is averaged for each plane and a the relative deviation from this average is computed for each bar. The relative deviation of channel $i$ reads
\begin{equation}
\frac{\Delta L_i}{\overline{L}} = \frac1{\overline{L}}(L_i -  \overline{L}).
\end{equation}
with $\overline{L}=\frac1{59}\sum_{i=1}^{59}L_i$, the average luminosity in a given plane. The deviation is represented as a function of the channel ID for a functional chain in figure\,\ref{fig:lumi_hist}. The rightmost entry corresponds to channel 0 and is to be ignored as it corresponds to the test channel that was not lit when the picture was taken.

The distribution of deviations for all the planes and their 2832 bars is given in figure\,\ref{fig:lumi_dist} and categorised in four intervals in table\,\ref{tab:lumi_dist}. Only two fibres are more than 40\,\% under the average luminosity in their respective plane and could be damaged. The rest of them have a satisfactory transmission. Calibration is required but this does not influence the range reconstruction in the EMR as hits are not lost due to this deviation. An exhaustive set of luminosity histograms is provided for each plane in \cite{bib:electronics_qt}.

\begin{table}[!htb]
  \centering
    \begin{tabular}{c|c}
      Deviation $\Delta L/\overline{L}$ & Counts \\
      \hline
      $<-0.4$ & \color{red}{2} \\
      $\left[-0.4,-0.2\right]$ & 130 \\
      $\left[-0.2,0.2\right]$ & 2667 \\
      $>0.2$ & 33 \\
    \end{tabular}
  \caption{Categorisation of the channel relative luminosity deviation.}
  \label{tab:lumi_dist}   
\end{table}

\section{Channel mismatch analysis}
\subsection{Potential causes}
\begin{figure}[!htb]
  \centering
  \begin{subfigure}[]{\textwidth}
    \centering
    \includegraphics[width=.8\textwidth]{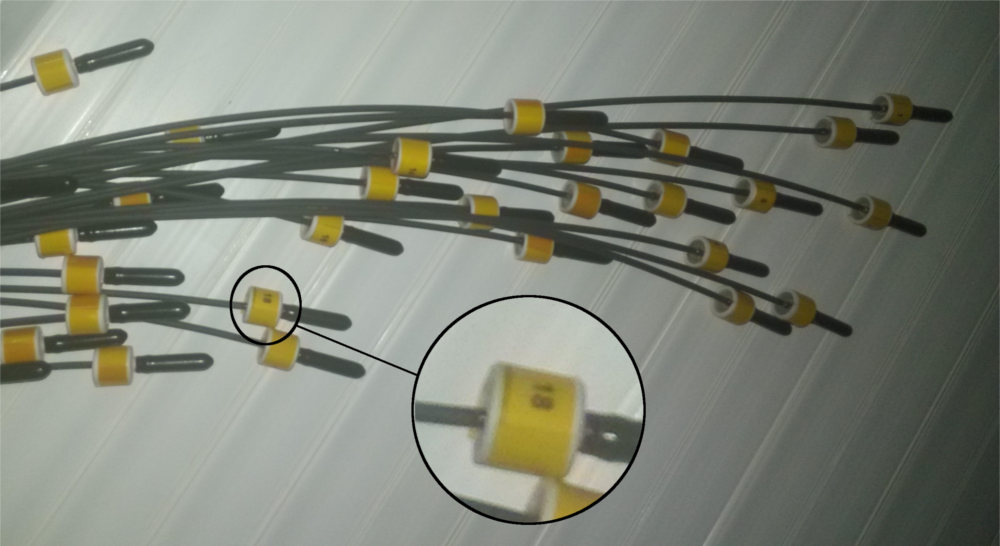}
    \caption{}
  \end{subfigure}
  \vspace{5 mm}
  \linebreak
  \begin{subfigure}[t]{.55\textwidth}
    \flushright
    \includegraphics[width=\textwidth]{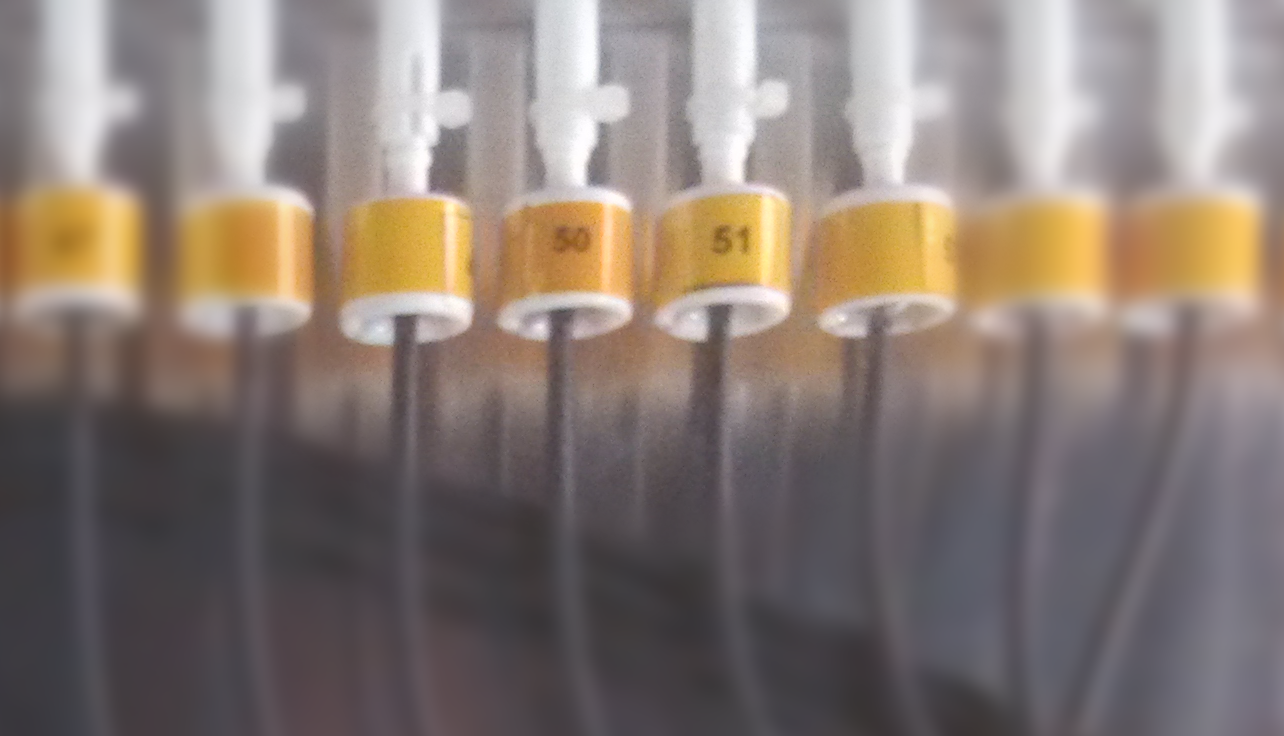}
    \caption{}
  \end{subfigure}
  \hfill
  \begin{subfigure}[t]{.32\textwidth} 
    \flushleft
    \includegraphics[width=\textwidth]{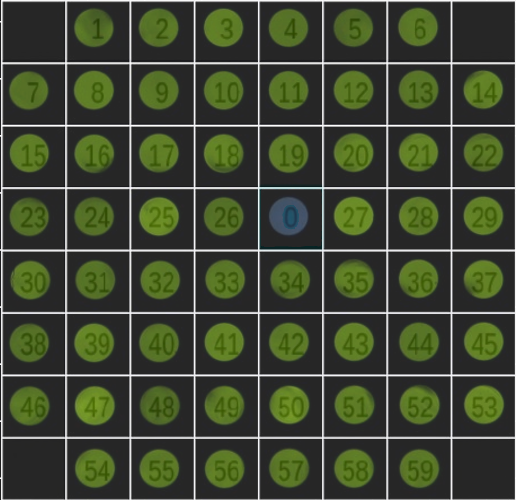}
    \caption{}
  \end{subfigure}
  \caption{Channel mismatch hot spots: (a) clear fibre bundle magnified on one particular label that may be incorrect; (b) connectors between the scintillating bars and the clear fibres, the connectors of two bars could be swapped; (c) clear fibre bundle mask, the fibres might be fitted in the wrong hole.}
  \label{fig:ch_mismatch}
\end{figure}

After the changes applied to the EMR design\,\cite{bib:emr_design_change}, a connector was added between the WLS fibres glued inside the scintillating bars and the clear fibres going to the MAPMTs. A mismatch occurs if, during the building process, the person in charge of connecting the fibres to the bars inadvertently connected a fibre to the wrong bar. Alternatively, in the manufacture of the fibre bundle itself, fibres might have been fitted in the wrong spot in the mask. Each fibre was manually tagged with a number ranging from 1 to 59 corresponding to each bar in the plane and is vulnerable to mislabelling. Finally, a short circuit or other manufacturing problem in the FEB, DBB or VRB could result in an electronic mismatch and would cause artefacts in the detector as well. These scenarios are depicted in figure\,\ref{fig:ch_mismatch}.

\subsection{Data acquisition}
To identify mismatched channels, it is necessary to cover the whole detector with a large amount of particle tracks to reach high statistical significance. A mismatch channel will be displaced from a true track significantly more often than a correctly mapped channel. Two options are available: beam data recorded in October 2013 for MICE Step I or cosmic muon data. The former does not cover the whole detector as some of the muons and pions stop in it and the greater part of the beam is located at the centre of the EMR. Cosmic muons are perfectly suited for this procedure as they always go through the detector without stopping and have no preferential location.

When the data sample used in this paper was recorded, the EMR was completed and located in the MICE hall at RAL. The detector was positioned up right, planes vertical, perpendicular to the ISIS beam. Two pairs of planes (15--16 or 31--32) were used as the particle triggers in coincidence with a spill gate of 3\,ms generated by the DAQ program at a frequency as high as the DAQ process allows it. In logic terms, $\text{trigger}=((p15\cap p16)\cup(p31\cap p32))\cap \text{spillGate}$.

\begin{figure}[!htb]
  \centering
  \begin{subfigure}[t]{\textwidth}
    \centering
    \includegraphics[width=.8\textwidth]{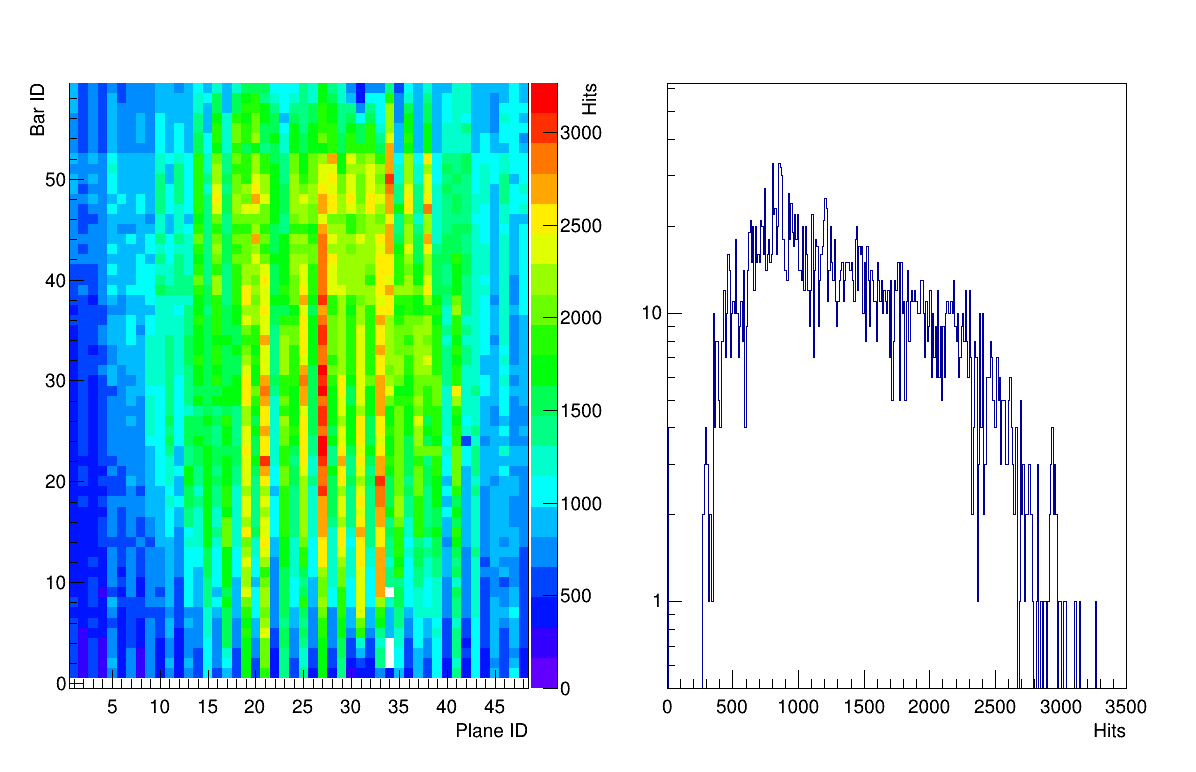}
    \caption{}
  \end{subfigure}
  \vspace{5 mm}
  \linebreak
  \begin{subfigure}[t]{\textwidth}
    \centering
    \includegraphics[width=.8\textwidth]{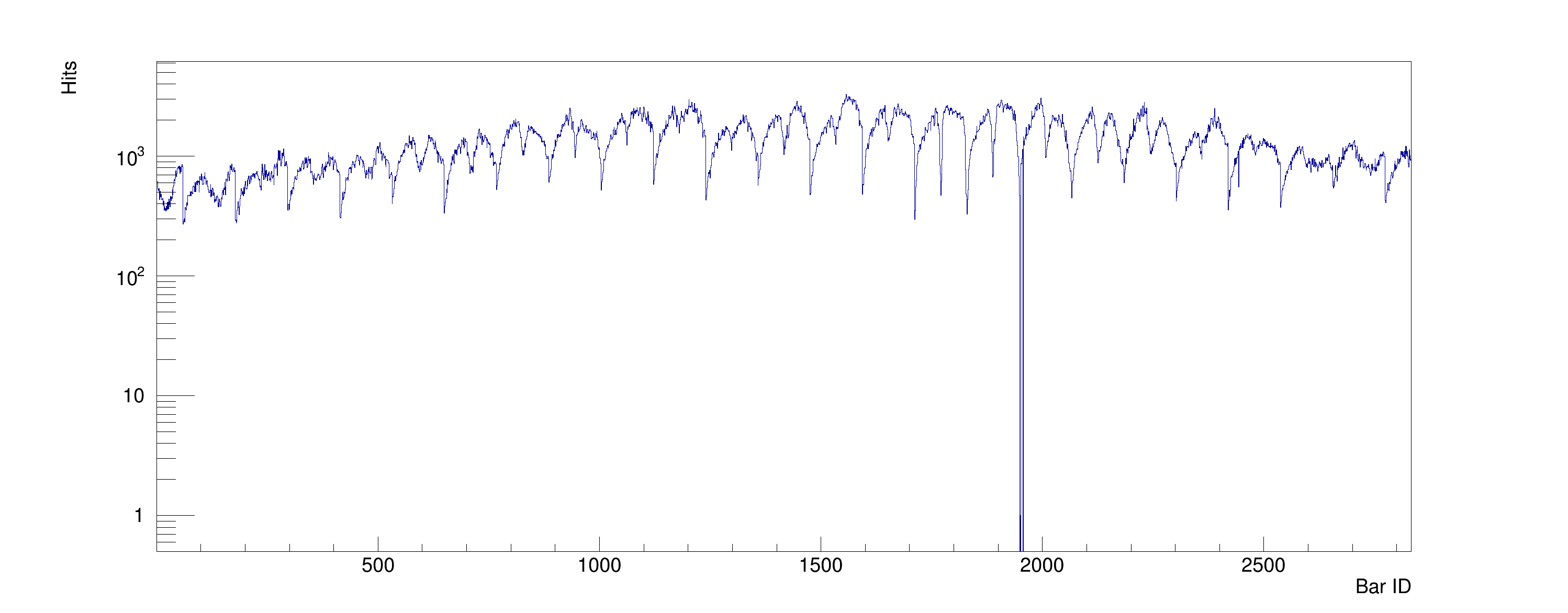}
    \caption{}
  \end{subfigure}
  \caption{Integrated amount of hits recorded in order to probe potential mismatched channels in the EMR. (a) Values given in each channel as a function of the plane ID and bar ID (left) and distribution of the amount of hits (right). (b) Values given as a function of the global channel ID.}
  \label{fig:hits}
\end{figure}

Data were recorded for 60 hours and yielded over $2.2\times10^5$ particle triggers. This corresponds to a trigger frequency of $\sim$1\,Hz. The time-over-threshold (ToT) measurements were recorded along with the timestamps of the 2832 EMR channels for each trigger and are the only measurement used in this analysis. The integrated amount of hits recorded in each bar is represented in figure\,\ref{fig:hits}. No hits were recorded in a total of four bars identified as dead channels: 2, 3, 4 and 9 of plane 34. In each channel, a number $N_i$ of hits was recorded of order $\sim10^3$. The statistical significance of a mismatch goes as $1/\sqrt{N_i}$, so that an accuracy of order 1\,\% should be reached.

\subsection{Hit pre-selection}
Channel mismatch is not the only potentially unwanted phenomenon occurring in the EMR detector. One of the other subjects of this study, the crosstalk, is another cause for artificial hits. Electronic noise appears as a signal and is recorded along with physical hits. Two cuts are applied to the cosmic muon sample in order to clean up reject unwanted hits.

The first cut concerns the time elapsed between the trigger time and the hit time, i.e. $\Delta t=t_\text{hit}-t_\text{trig}$. There is a delay before a bunch of hits generates a trigger and this delay is more or less constant. Restricting $\Delta t$ to a small interval gets rid of most of the noise that is distributed randomly or trailing later after a trigger. The interval chosen is $-94$\,ADC counts $<\Delta t<$ $-80$\,ADC counts.

The second cut is on the the time-over-threshold. In the presence of crosstalk, not all the energy is transferred to the neighbouring channel and most of the signals generated have lower intensity then the true hit. A cut is applied on the energy as well to focus on the primary hits generated by minimum ionizing (MIP) cosmic muons passing through the scintillating volume. The energy lower limit chosen in this analysis is 6 ADC counts.

\begin{figure}[!htb]
  \centering
  \includegraphics[width=.75\textwidth]{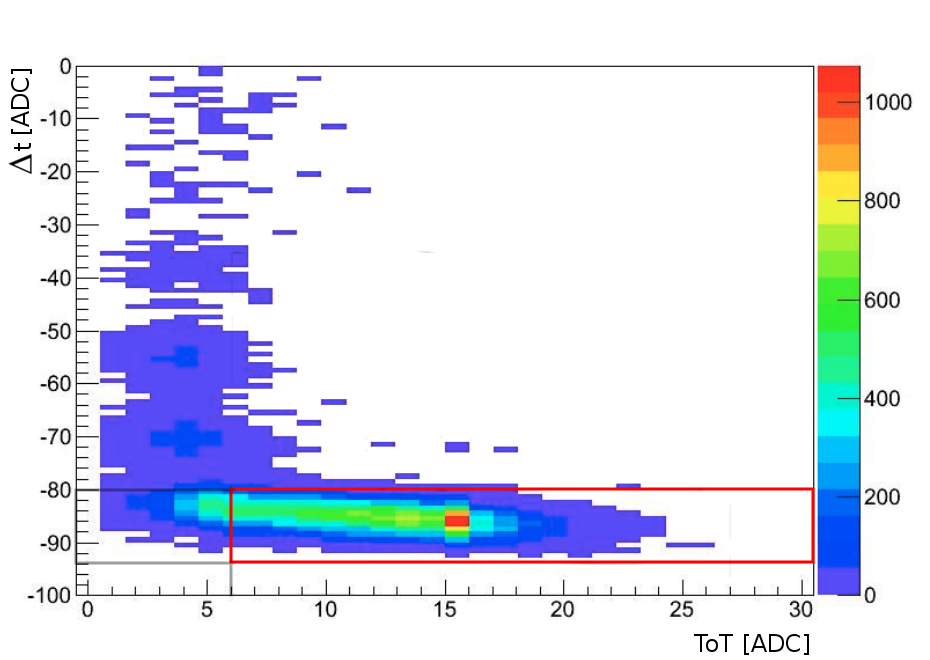}
  \caption{2D histogram of the delay, $\Delta t$, as a function of the time-over-threshold, ToT, distribution of the data sample used for the mismatch analysis. The red lines delimit the area of the eligible bar hits. The unit of both axis is ADC counts (2.5 ns).}
  \label{fig:deltatvstot}
\end{figure}

In figure\,\ref{fig:deltatvstot}, the energy and time distribution has been combined in one 2D histogram. The largest density lies at the typical MIP energy deposition ($ToT\simeq15$\,ADC counts) and typical $\Delta t\simeq-85$\,ADC counts. The energy--time area used to reconstruct the tracks is limited by the red lines.

\begin{figure}[!htb]
  \centering
  \begin{subfigure}[]{\textwidth}
    \centering
    \includegraphics[width=.9\textwidth]{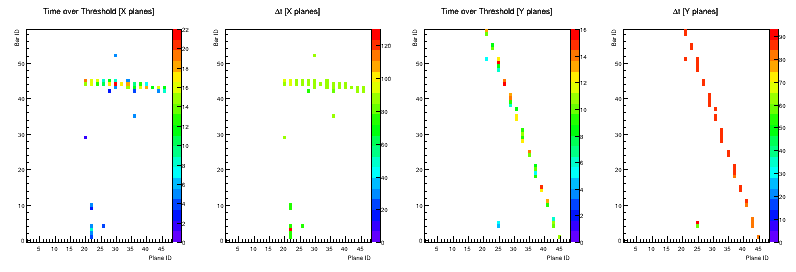}
    \caption{}
  \end{subfigure}
  \vspace{5 mm}
  \linebreak
  \begin{subfigure}[t]{\textwidth}
    \centering
    \includegraphics[width=.9\textwidth]{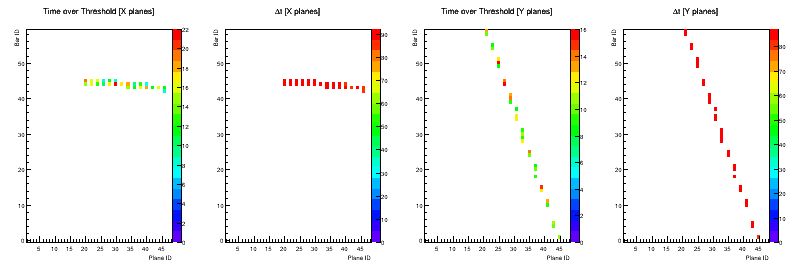}
    \caption{}
  \end{subfigure}
  \caption{Single event display of a cosmic muon going through the EMR, (a) before applying any cuts to the data, (b) after applying a selection on the timing and the time-over-threshold.}
  \label{fig:cuts}
\end{figure}

This process is very important for the success of the fitting procedure described in the next subsection. The selection allows for the linear fits to be generally more successful and the tracks to be cleaner. The way this selection affects the tracks is shown in a single event displays in figure\,\ref{fig:cuts}. Each row represents the same event but the top one has not undergone any cuts while the bottom one has. For each event, the energy deposition and the timing are represented in the X planes ($z$--$x$ projection) on the left and Y planes ($z$--$y$ projection) on the right.

\subsection{Track reconstruction}
To reconstruct cosmic tracks and calculate the distance of each bar from the particle trail, it is necessary to fit each array of hits with straight lines in each projection. Positions and uncertainties have to be attributed to each bar hit in the EMR. Due to the triangular geometry of the sections of the scintillators, the middle of a given bin is not the optimal estimate of the central coordinates of the corresponding bar. The true average location of the track in a bar is at the centre of mass (COM) of the triangle and its uncertainties extend rectilinearly towards its boundaries.

\begin{figure}[!htb]
  \centering
  \includegraphics[scale=0.30]{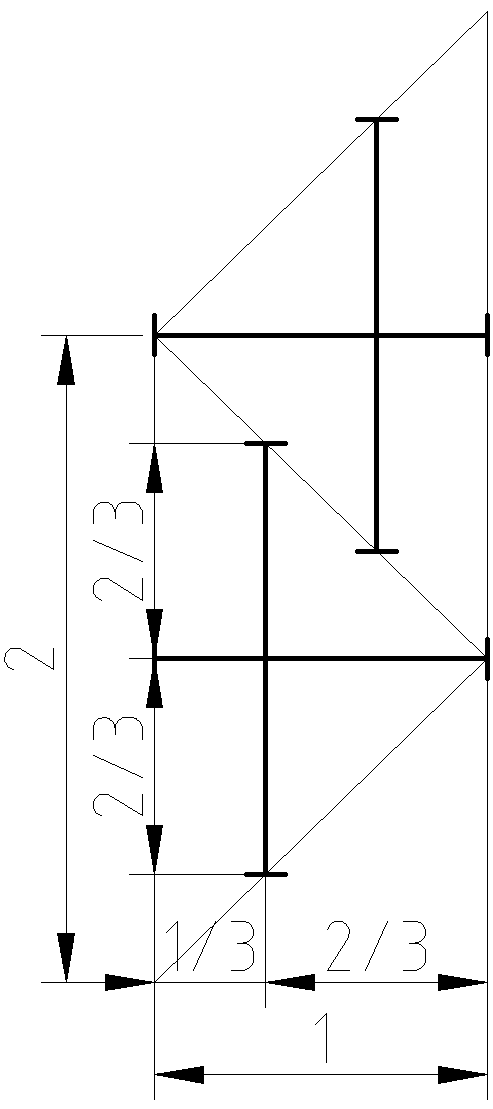}
  \caption{Asymmetric errors in two adjacent scintillating bars. The intersection point of a cross constitutes the centre of mass of a scintillating bar and the bold lines represent the uncertainty on that point. EMR bars have triangular sections, with a width of 2\,b.u. and a height of 1\,b.u.}
  \label{fig:error}
\end{figure}

The section of an EMR bar is almost an isosceles triangle. Its base measures 33\,mm and its height 17\,mm. In this analysis, the distances are computed in terms of bin units. A bin unit, b.u., is equivalent to 17 mm, i.e. the thickness of a plane. In this metric, the average distance of a mismatched channel from a track should always average to a integer amount. The width of a bar within a plane is taken to be exactly $2$\,b.u. The space points are centred in the triangle COM located $1/3$\,b.u. from its base and the error bars are prolonged from that point outwards to the edges of the triangle. Figure\,\ref{fig:error} represents the error bars and the measurements in b.u. of two adjacent bars. 

Asymmetrical errors are associated differently following the orientation of the bars. The tracks are still represented on event displays with square bins but the calculation behind the distance measurement involves triangular ones. In this binning choice, adjacent $y$ error bars overlap, which is expected for bars that do so in the detector.

\begin{figure}[!htb]
  \centering
  \includegraphics[width=.9\textwidth]{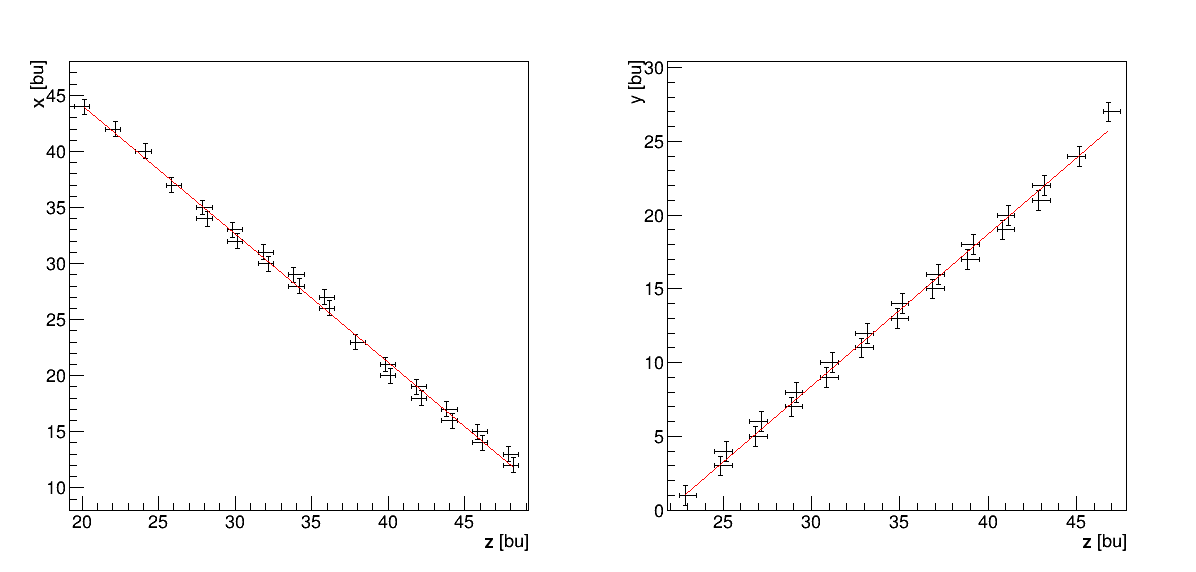}
  \caption{Successful fits of a cosmic muon track in the $z$--$x$ (left) and $z$--$y$ (right) projections; the linear regression is represented in red. Most of the points are closer to track due to the triangular bin geometry adopted for this analysis.}
  \label{fig:goodfits}
\end{figure}

After assigning positions and uncertainties to all the bars hit, the track is fitted with a first order linear function of the form $f(z)=az+b$. The X planes and Y planes are fitted separately as they represent the $z$--$x$ and the $z$--$y$ projections of the three dimensional trail of a cosmic muon, respectively. It is not relevant to combine the two fits into one as a mismatch only happens within the same plane. The fitting algorithm used is the least squares method involving the minimization of $\chi^2$ defined as
\begin{equation}
\chi^2=\sum_i\frac{(q_i-(az_i+b))^2}{\sigma^2},
\end{equation} 
with $q_i=x_i,y_i$ and $z_i$ the coordinates of a given bin centre and $\sigma=2/3$\,b.u. the uncertainty on the $q$ coordinate. In figure\,\ref{fig:goodfits}, two successful fits of the same track are displayed in its $z$--$x$ and $z$--$y$ projections. The choice of triangular bars significantly influences the calculation of $\chi^2$ as most points are closer to the track than they would be, had the location been taken as the centre of their respective square bins.

\begin{figure}[!htb]
  \centering
  \includegraphics[width=.9\textwidth]{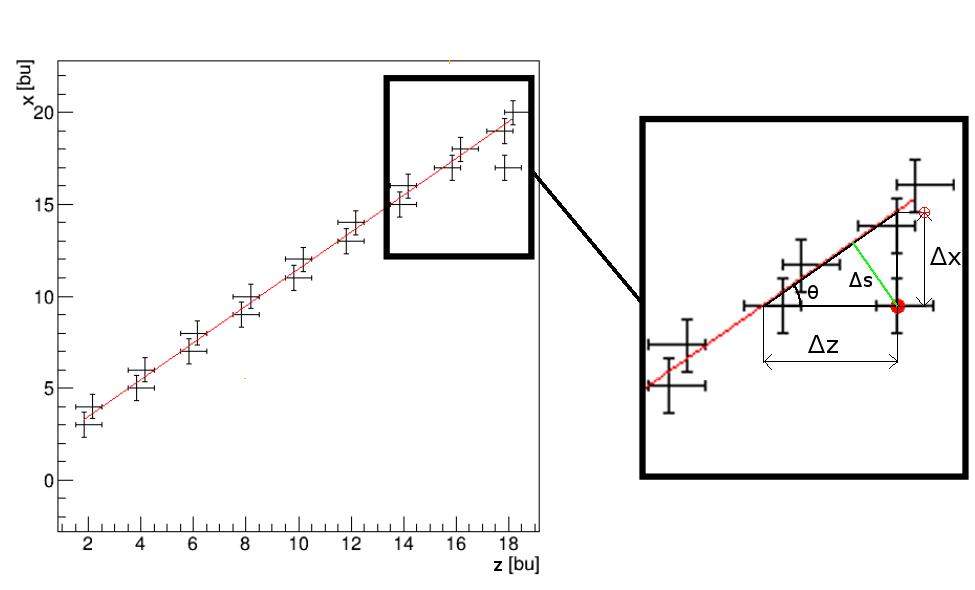}
  \caption{Illustration of the distance measured in the fitting process.}
  \label{fig:distance}
\end{figure}

An additional precaution is taken to make the fit perfectly accurate. Despite the pre-selection, a hit may occur randomly far away from the track. It might be caused by high intensity crosstalk, synchronized noise or an actual mismatch. These hits influence the location of the linear regression quite significantly and need to be excluded from the fitting procedure. In order to distinguish these hits from the main track, their distances from the primary fit is calculated for each one of them. This measurement, $\Delta s$, is represented by the green line in figure\,\ref{fig:distance} and reads
\begin{equation}
\Delta s = \Delta q|\sin(\theta)| = \Delta q|\sin(\arctan(a))|,
\end{equation}
with $\Delta q=|q-(az+b)|$, $q=x,y$ and $a$ and $b$ the slope and $y$-intercept of the fitted line.

The hits for which the distance calculated exceeds 2\,b.u. are rejected from the fitting sample but kept to be processed in the mismatch analysis. The graph is then fitted one more time, including only the hits located close to the track to achieve maximum accuracy.

\subsection{Hit post-selection}
An EMR hit that has passed preliminary cuts needs to undergo a final selection process before it is included in the analysis. Three additional criteria have to be fulfilled:

\begin{enumerate}
\item the track initial fit $\chi^2$ value cannot exceed 250. This prevents very messy events such as electromagnetic showers from being included into an analysis requiring clean tracks;

\item the muon has to hit at least 10 planes in a given projection. This gives an extra criterion bounding the tracks to be more or less horizontal and not to be random noise in the EMR, which would make the position of the hits with respect to a fit irrelevant;

\item a hit has to be part of a plane with a maximum of 2 hits. A hit recorded in a vertical event can have a very big $\Delta q$ even though it is very close from the track. As this is the measure used to determine a mismatch, it is crucial to avoid this situation.
\end{enumerate}

\begin{figure}[!htb]
  \centering
  \begin{subfigure}[t]{.3\textwidth}
    \includegraphics[width=\textwidth]{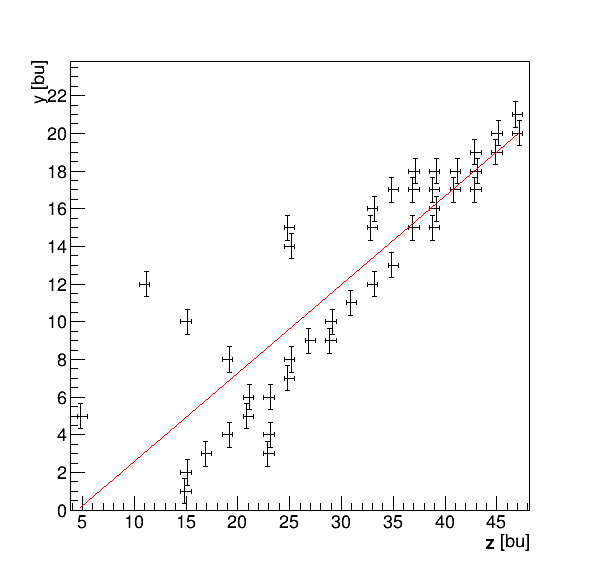}
    \caption{}
  \end{subfigure}
  \hfill
  \begin{subfigure}[t]{.3\textwidth} 
    \includegraphics[width=\textwidth]{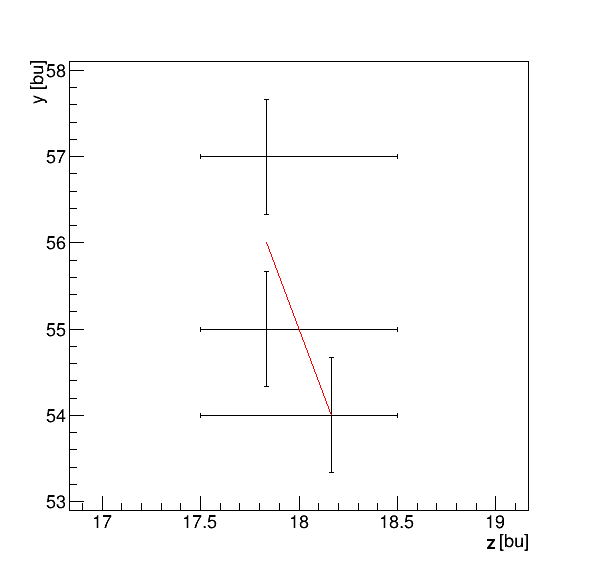}
    \caption{}
  \end{subfigure}
  \hfill
  \begin{subfigure}[t]{.3\textwidth} 
    \includegraphics[width=\textwidth]{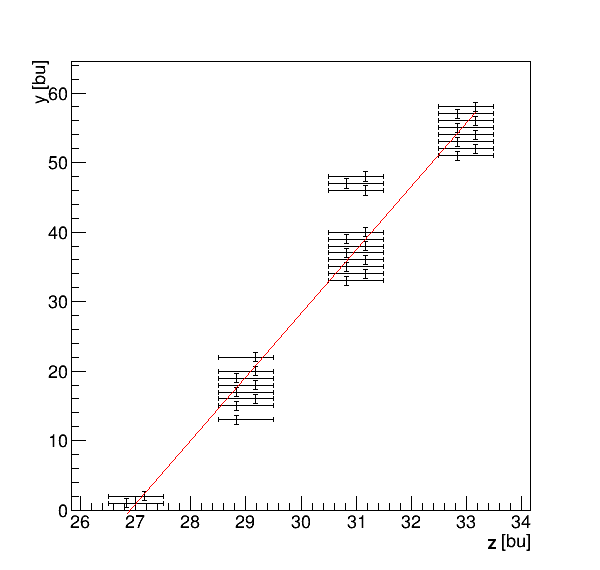}
    \caption{}
  \end{subfigure}
  \caption{Single projection of event displays of tracks and hits excluded from the analysis: (a) the event is too messy (e.g. electromagnetic shower); (b) not enough planes are hit to form an appropriate track; (c) too many bars are hit within the same plane (vertical track) to appreciate the mismatch.}
  \label{fig:badhits}
\end{figure}

For each criterion, an example of a track that does not match it is represented in figure\,\ref{fig:badhits}. These types of events or hits are rejected from the analysis.

\subsection{Mismatch figures-of-merit}
The identification procedure consists in using the average $\Delta q,\,q=x,y$, as the primary variable. It represents the absolute distance between the COM of a bar and the reconstructed track within a plane, i.e. $\Delta q=|q_m-(az_m+b)|$ for a given COM of coordinates $(z_m,q_m)$.

The essential figures-of-merit for each channel are the mismatch ratios. For each given integer distance $i$\,b.u., calculate a ratio, $R_i$, that corresponds to the probability that a hit is within acceptable range of a distance $i$\,b.u. off track. For $f(\Delta q)$ the distance distribution, that ratio is defined as
\begin{equation}
R_i=\frac{\int_{i-2/3}^{i+2/3}f(\Delta q)\text{d}(\Delta q)}{\int_{0}^{2/3}f(\Delta q)\text{d}(\Delta q)+\int_{i-2/3}^{i+2/3}f(\Delta q)\text{d}(\Delta q)}.
\end{equation}
In the results section, only two of these ratios are presented. For the first one, $R_1$, it was chosen to change the lower boundary of the first bin to $(1-1/3)$\,b.u. to cancel out the overlap and simplify the computation of the ratio estimation. The ratio subsequently reads
\begin{equation}
R_1=\frac{\int_{2/3}^{5/3}f(\Delta q)\text{d}(\Delta q)}{\int_{0}^{5/3}f(\Delta q)\text{d}(\Delta q)}\,.
\end{equation}
The other ratio covers the rest of the distribution, that is
\begin{equation}
R_{\geq2}=\frac{\int_{4/3}^{+\infty}f(\Delta q)\text{d}(\Delta q)}{\int_{0}^{2/3}f(\Delta q)\text{d}(\Delta q)+\int_{4/3}^{+\infty}f(\Delta q)\text{d}(\Delta q)}\,.
\end{equation}
If a bar is mismatched by more than one b.u., it appears clearly in the last ratio, as all the hits are located around a given distance $i$\,b.u. from the track and it converges to one. Given an average distance $\overline{\Delta q}$, the distribution of hits can be degenerated to a Dirac delta to get
\begin{equation}
R_{\geq2}=\frac{\int_{4/3}^{+\infty}\delta(\Delta q-\overline{\Delta q})\text{d}(\Delta q)}{\int_{0}^{2/3}\delta(\Delta q-\overline{\Delta q})\text{d}(\Delta q)+\int_{4/3}^{+\infty}\delta(\Delta q-\overline{\Delta q})\text{d}(\Delta q)}=1,\,\,\forall\overline{\Delta q}>4/3\,.
\label{eq:rgeq2}
\end{equation}
It is trivial that this computation holds for broader distributions, as long as it does not tail out much more than 1\,b.u. away from its centre point. The adjacent ratio $R_1$ is more complicated to estimate and is developed in section\,\ref{section:r1estimate}.

\subsection{$R_1$ mismatch ratio estimation}
\label{section:r1estimate}
The use of triangular bars makes the identification of a mismatch between adjacent bars more challenging. This geometry bounds a track to hit two adjacent bars per plane or more and the post-selection process constrains it to exactly two. Considering the error bars chosen in a previous subsection represented in figure\,\ref{fig:error}, a track could go through both bars and not be within $2/3$\,b.u. of the COM (black dot) of either one of them. An estimation of the expected mismatch rate in normal and mismatched channels is necessary to be able to distinguish them.

Determining the ratio $R_1$ in the EMR requires the computation of three functions:
\begin{enumerate}
	\item the angular distribution of the area accessible to a particle that hits exactly one adjacent pair of bars, $f_A(\theta)$, as a bigger area is equivalent to more events, i.e. a higher probability density;
	\item the angular mismatch probability density function (PDF), $f_M(\theta)$;
	\item the weighted average of the area distribution, $w(\theta)=f_A(\theta)\times f_C(\theta)$, with $f_C(\theta)$ the cosmic muon angular distribution.
\end{enumerate} 
In all of these calculations, the zenith angle $\theta$ and the standard bar units described in the previous subsection are used. The total area covered by the section of two bars is 2\,b.u.$^2$.

\subsubsection{Normal channels}
\begin{figure}[!htb]
  \centering
  \includegraphics[width=.75\textwidth]{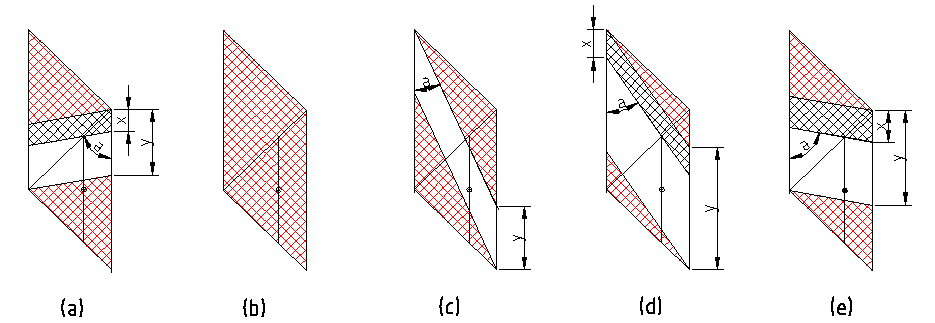}
  \caption{Visual representation of the area and mismatch distributions in five angular regimes. (red mesh) Area not accessible to the particles; (plain white) area accessible that is within $2/3$\,b.u. of the bottom bar centre; (black mesh) accessible area that is not within $2/3$\,b.u. of the bottom bar centre.}
  \label{fig:areas}
\end{figure}

In figure\,\ref{fig:areas}, each pair of bars represent one of five range of angles corresponding to as many regimes for which the distributions are summarised in table\,\ref{tab:areas}. In each case, the red meshed zone represents area that is not accessible to the particle in that regime, as it causes more or less than exactly two hits. The white and black meshed spaces are accessible but the black part contributes to the mismatch ratio as it is farther away than $2/3$\,b.u. from the centre of the bottom bar. $y$ and $x$ are the thickness the accessible zone and of the mismatched zone, respectively. The area accessible is a function of $y$ and the mismatch PDF is exactly $x/y$.

\renewcommand{\arraystretch}{1.5}
\begin{table}
	\centering
	\begin{tabular}{c|c|c|c}
		Tag & Angular range & Area distribution $f_A(\theta)$ & Mismatch PDF $f_M(\theta)$\\
		\hline
		\textbf{(a)} & $\theta\in\left[-\frac{\pi}{2},-\frac{\pi}{4}\right]$ & $1-\tan\left(\frac{\pi}{2}-\theta\right)$ & $\frac13$ \\
		\hline
		\textbf{(b)} & $\theta\in\left[-\frac{\pi}{4},\arctan\frac{1}{3}\right]$ & 0 & 0 \\
		\hline
		\textbf{(c)} & $\theta\in\left[\arctan\frac13,\arctan\frac{1}{2}\right]$ & $3-\tan\left(\frac{\pi}{2}-\theta\right)$ & 0 \\
		\hline
		\textbf{(d)} & $\theta\in\left[\arctan\frac12,\frac{\pi}{4}\right]$ & $3-\tan\left(\frac{\pi}{2}-\theta\right)$ & $\frac23\left(\frac{2-\tan(\pi/2-\theta)}{3-\tan(\pi/2-\theta)}\right)$ \\
		\hline
		\textbf{(e)} & $\theta\in\left[\frac{\pi}{4},\frac{\pi}{2}\right]$ & $1+\tan\left(\frac{\pi}{2}-\theta\right)$ & $\frac13$
	\end{tabular}
	\caption{Expressions of the area distribution and mismatch PDF in the five angular regimes represented in figure\,\ref{fig:areas}.}
	\label{tab:areas}
\end{table}
\renewcommand{\arraystretch}{1}

\begin{figure}[!htb]
  \centering
  \includegraphics[width=.6\textwidth]{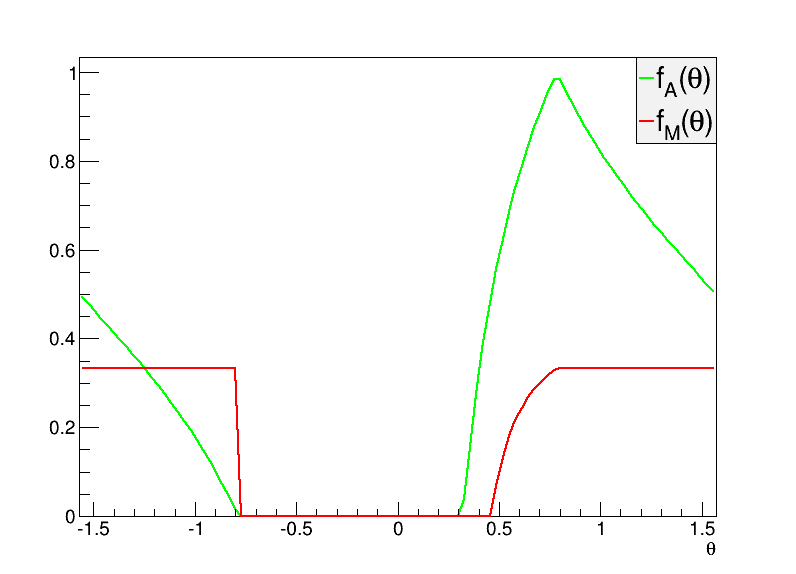}
  \caption{Area distribution, $f_A$, and mismatch PDF, $f_M$, of a cosmic muon going through the EMR at an angle $\theta$ with the zenith. At angles close to zero, there cannot be exactly two hits and both functions cancel. At an angle of $\pi/4$, the entire area is accessible. The mismatch PDF does no go over one third of the total area available.}
  \label{fig:fareapdf}
\end{figure}

\begin{figure}[!htb]
  \centering
  \includegraphics[width=.6\textwidth]{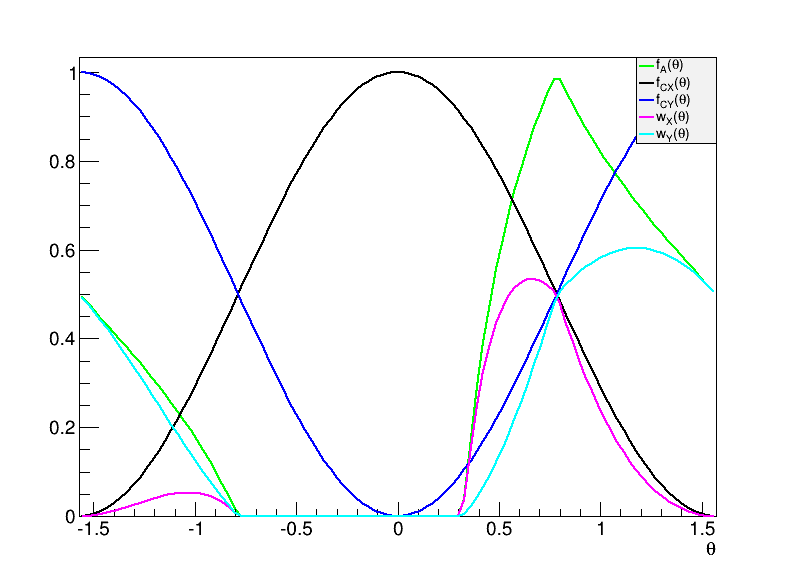}
  \caption{Weight functions of the adjacent mismatch ratios. Due to the rotation between the two plane orientations, the cosmic muon angular distributions are in phase opposition. This gives different estimation of the predicted mismatch ratio.}
  \label{fig:fweights}
\end{figure}

The area angular distribution has been normalized to a maximum area of 1 to represent a PDF and is drawn along with the mismatch PDF in figure\,\ref{fig:fareapdf}. They are the same for any given pair of bars in the EMR and as such these functions are representative of the whole detector. The area PDF must be combined to the cosmic angular distribution, $f_C$, in order to calculate the average mismatch ratio. The cosmic muon distribution is not the same for the X and Y planes. Both orientations are perpendicular to the ground but the bars are horizontal in the X planes and vertical in other. The angular distribution of the muons must be shifted by $\pi/2$ radians:  $f_{CX}(\theta)\propto\cos^2(\theta)$ \cite{bib:PDG} in $x$ and $f_{CY}(\theta)\propto\sin^2(\theta)$ in $y$. The weight functions $w_X(\theta)$ and $w_Y(\theta)$ (resp. for the X and Y planes) are drawn in figure\,\ref{fig:fweights}. They are significantly affected by the orientation of the bars.

Given the piecewise expressions of the required functions, the weighted average of the mismatch PDFs correspond to the predicted adjacent mismatch ratio in each projection, i.e.
\begin{equation}
R_1^X=\frac{\int_{-\pi/2}^{\pi/2}f_M(\theta)\times w_X(\theta)\text{d}\theta}{\int_{-\pi/2}^{\pi/2}w_X(\theta)\text{d}\theta}\simeq25.3\%,
\label{eq:r1x}
\end{equation}
\begin{equation}
R_1^Y=\frac{\int_{-\pi/2}^{\pi/2}f_M(\theta)\times w_Y(\theta)\text{d}\theta}{\int_{-\pi/2}^{\pi/2}w_Y(\theta)\text{d}\theta}\simeq32.2\%.
\label{eq:r1y}
\end{equation}

\subsubsection{Mismatched channel}
A mismatched channel cannot be orphaned as all of the 2832 channels of the EMR are read out. As a mismatch comes from a swap between two fibres, mismatched channels must in the minimal case come in pairs. The case in which two adjacent bars are swapped is treated separately from the other types. A particle going through one of the EMR plane produces hits in two adjacent bars.

For a pair of mismatched channels, there are two possible occurrences. The particle goes either through both of them or through one of them and a third channel. In the first case, the mismatch PDF obtained in the normal channel computation remains unchanged. In the second, the bar that has been swapped is systematically further away than $2/3$\,b.u. from the track, which takes the PDF to a flat 100\,\%. As these two cases are equiprobable, the average of the two yield the estimated mismatch ratio, $R_1$, for two swapped adjacent bars:
\begin{equation}
R_1^{X,m} = 62.6\,\%,
\label{eq:r1xm}
\end{equation}
\begin{equation}
R_1^{Y,m} = 66.1\,\%.
\end{equation}

\subsection{Results}
\subsubsection{Adjacent bars mismatch ratio $R_1$}
\begin{figure}[!htb]
  \centering
  \includegraphics[width=.7\textwidth]{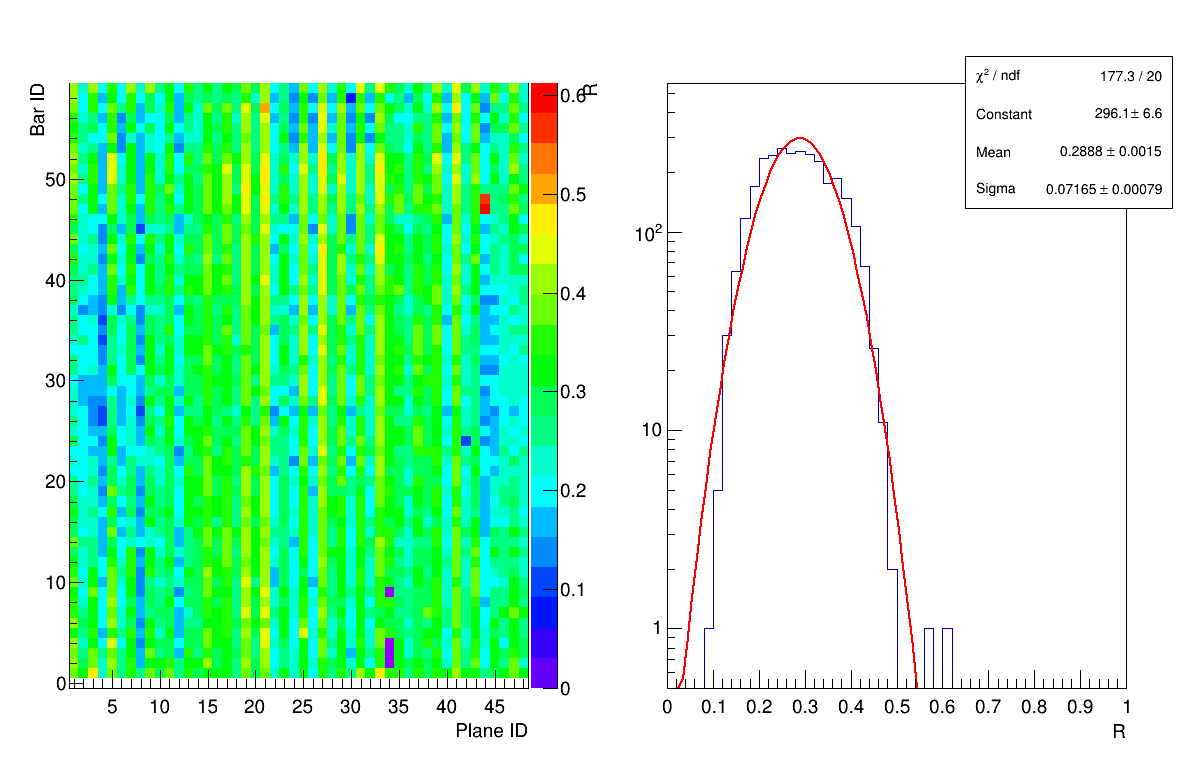}
  \caption{(left) Mismatch ratio of adjacent bars, $R_1$, as a function of plane and bar ID; (right) complete distribution of adjacent mismatch ratios.}
  \label{fig:ratio1}
\end{figure}

The ratio $R_1$ is represented in a two dimensional histogram for each plane and bar ID in figure\,\ref{fig:ratio1}. On the right is the distribution of ratios. Two red channels stand out in the left hand distribution and correspond to the outliers in the right hand histogram.

\begin{figure}[!htb]
  \centering
  \includegraphics[width=.7\textwidth]{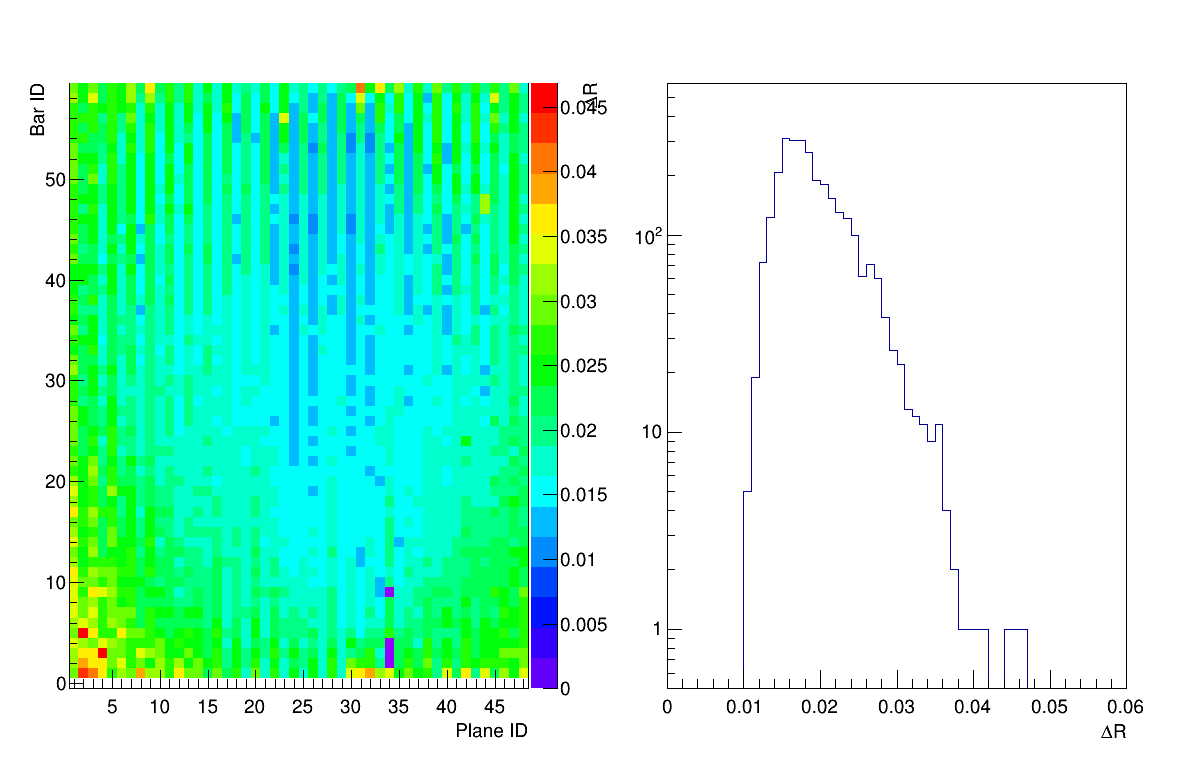}
  \caption{(left) Uncertainty on the ratio of adjacent bars, $\Delta R_1$, as a function of plane and bar ID; (right) complete distribution of adjacent mismatch ratio uncertainties.}
  \label{fig:ratio1err}
\end{figure}

The broad Gaussian distribution is centred in $28.9\pm0.15$\,\%, which is compatible with the prediction made earlier. The estimated values of the ratio $R_1$ for the two plane orientations in equations \ref{eq:r1x} and \ref{eq:r1y} give an average of 28.7\,\%. The width of the distribution is due to the fact that not every particle going through a bar produces a hit (see section\,\ref{section:eff}). A particle sometimes passes through more than two bars but only leaves two hits, complicating the geometry of the estimation. Crosstalk also influences the ratio as it produces artificial hits away from the track (see section\,\ref{section:xt}). The uncertainty on the mismatch ratio for each bar and its error distribution are compiled in figure\,\ref{fig:ratio1err}. Despite the width and uncertainty, two channels clearly stand out and their ratio are summarised in table\,\ref{tab:mismatch}.

\begin{table}[!htb]
\centering
\begin{tabular}{c|c|c}

Plane ID & Bar ID & $R_1$ \\
\hline
44 & 47 & 62.5$\pm$3.5\% \\

44 & 48 & 57.2$\pm$3.2\% \\

\end{tabular}
\caption{Mismatch channel IDs and corresponding mismatch ratios}
\label{tab:mismatch}
\end{table}

This anomaly is significant for several reasons. With the recorded levels of uncertainty, these measurements correspond to two individual 4\,$\sigma$ deviations from the centre of the distribution. The probability of such a deviation occurring in two adjacent channels is prohibitively low unless the two channels are mismatched. The other argument is that the measured ratio are compatible with the predicted ratio in an X plane given in equation\,\ref{eq:r1xm}.
 
\subsubsection{Other mismatches ratio $R_{\geq2}$}
\begin{figure}[!htb]
  \centering
  \includegraphics[width=.7\textwidth]{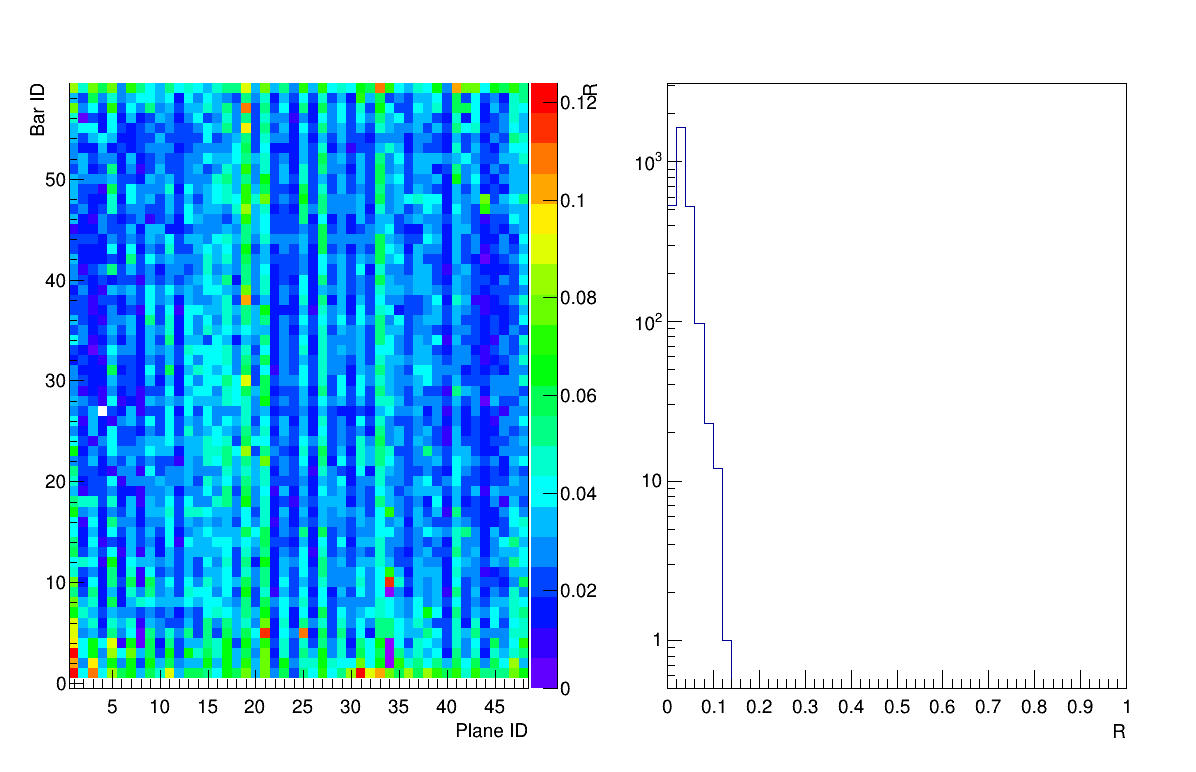}
  \caption{(left) Mismatch ratio of bar displaced by at least 2\,b.u., $R_{\geq2}$, as a function of plane and bar ID; (right) complete distribution of the corresponding mismatch ratios.}
  \label{fig:ratio2}
\end{figure}

The same mismatch ratio analysis was conducted for any other possible displacements, $R_{\geq2}$, and compiled in two histograms in figure\,\ref{fig:ratio2}. It appears that all levels of mismatch are low and not compatible with the 100\,\% that is expected from this type of mismatch as computed in equation\,\ref{eq:rgeq2}. No mismatch was registered for bars that are further than adjacent.

\section{Crosstalk analysis}
\label{section:xt}
\subsection{Causes}
Crosstalk in the EMR happens in one critical location: the MAPMT. All the fibres coming from 59 bars come together in a single compact location and signals can potentially interfere with each other.

\begin{figure}[!htb]
  \centering
  \begin{subfigure}[t]{.45\textwidth} 
    \includegraphics[width=\textwidth]{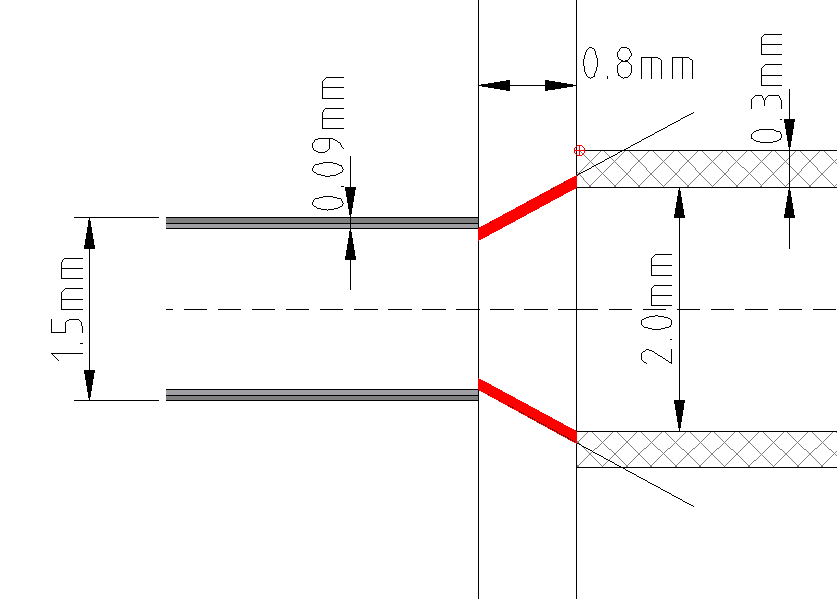}
    \caption{}
  \end{subfigure}
  \hfill
  \begin{subfigure}[t]{.40\textwidth} 
    \includegraphics[width=\textwidth]{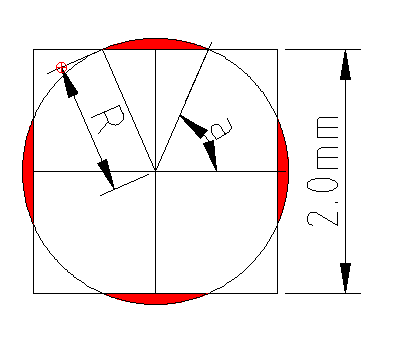}
    \caption{}
  \end{subfigure}
  \caption{Illustration of OXT in an MAPMT: (a) side view of a clear multi cladding fibre shining on one channel of the MAPMT; (b) Circle of light coming from the clear fibre and shining on the MAPMT channel. The red filled area represent light leaking to adjacent channels.}
  \label{fig:oxt}
\end{figure}

The first type of interference encountered is optical crosstalk (OXT). A single fibre of the bundle shines on more than one channel of the MAPMT mask. This phenomenon is depicted in figure\,\ref{fig:oxt}. The clear fibres are multi-cladding fibres, 1.5\,mm in diameter with a 0.72 numerical aperture\,\cite{bib:emr_clfibres}. The thickness of the layer of glass on top of the MAPMT channels reaches 0.8\,mm\,\cite{bib:hamamatsu_pmt}. This geometry allows part of the light in the core of the fibre to leak onto other surrounding channels, even in the case of fibres perfectly aligned and in contact with the photocathode. In the assumption that the fibre is perfectly against the mask and the light is uniformly distributed in the fibre, the fraction of light leaked is estimated as represented in figure\,\ref{fig:oxt} to be
\begin{equation}
R_L=\frac{A_e}{A_c}=\frac{4(\pi R^2(\pi/2-a)/\pi-R\sin(\pi/2-a))}{\pi R^2}\simeq6\%
\end{equation}
with $A_e$ the red area outside of the square MAPMT channel, $A_C$ the total area of the circle of light, and $R=1.09$\,mm the radius of the circle of light.

\begin{figure}[!htb]
  \centering
  \begin{subfigure}[t]{.45\textwidth} 
    \includegraphics[width=\textwidth]{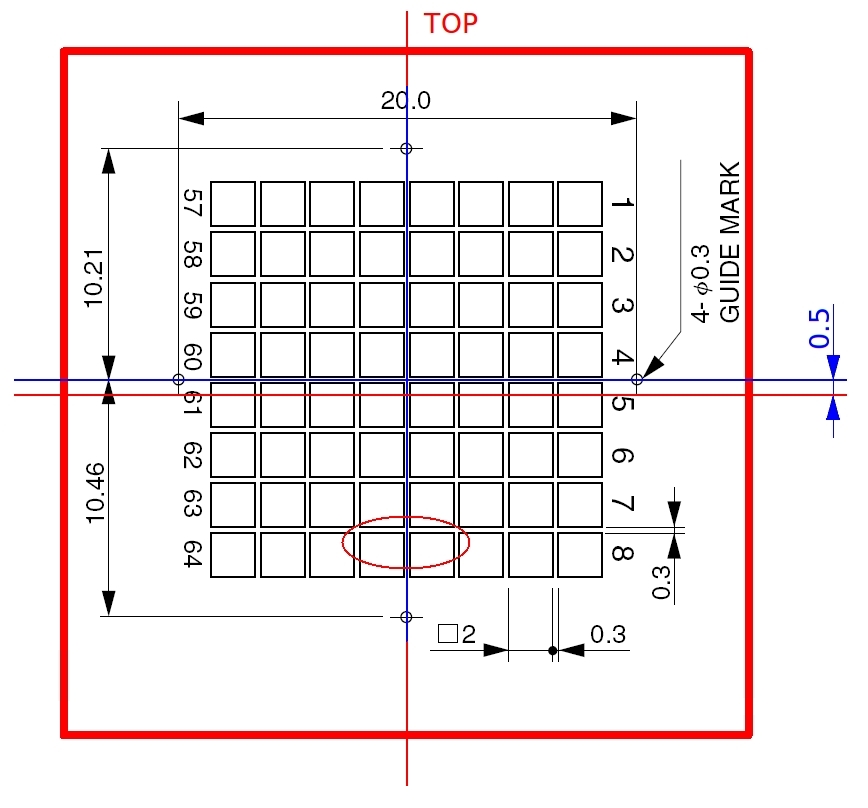}
    \caption{}
  \end{subfigure}
  \hfill
  \begin{subfigure}[t]{.4\textwidth} 
    \centering
    \includegraphics[width=\textwidth]{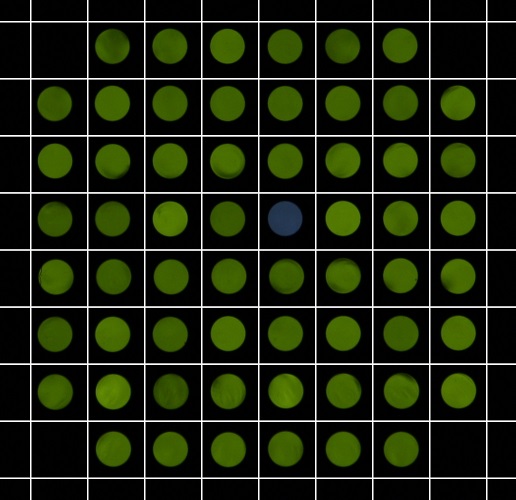}
    \caption{}
  \end{subfigure}
  \caption{Causes for crosstalk asymmetry: (a) technical drawing of the MAPMT mask, the red line represents a shift of the fibre bundle of 0.5 mm with respect to the mask; (b) clear fibre bundle, the fibres are not all glued in the same position in their grid compartment.}
  \label{fig:irreg}
\end{figure}

The level of light leak varies for different reasons. The whole PMT mask might be shifted with respect to the centre of the fibre bundle, as shown in figure\,\ref{fig:irreg} (a). In that case, the level of crosstalk measured in the surrounding channels of a given fibre is not isotropic and the level of asymmetry is used to determine the misalignment. In addition, a fibre is not necessarily glued perfectly in the centre of its slot in the bundle. This effect is seen on the picture of the fibre mask in figure\,\ref{fig:irreg} (b), where the fibres are not always exactly in the middle of the grid compartment. It does not affect the general crosstalk levels but shifts the values slightly from one fibre to another.

\begin{figure}[!htb]
  \centering
  \begin{subfigure}[t]{.3\textwidth} 
    \includegraphics[width=\textwidth]{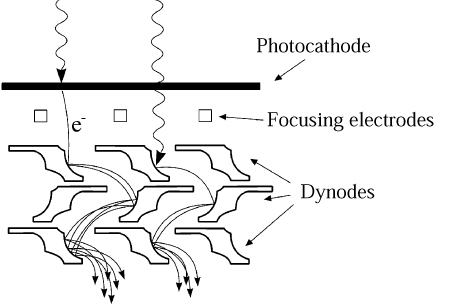}
    \caption{}
  \end{subfigure}
  \hfill
  \begin{subfigure}[t]{.3\textwidth} 
    \includegraphics[width=\textwidth]{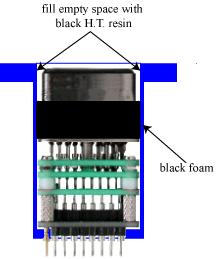}
    \caption{}
  \end{subfigure}
  \hfill
  \begin{subfigure}[t]{.3\textwidth} 
    \includegraphics[width=\textwidth]{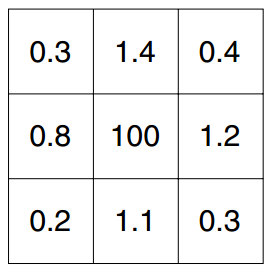}
    \caption{}
  \end{subfigure}
  \caption{Illustration of AXT in an MAPMT: (a) schematic drawing of electrons leaking from one accelerating chain to another; (b) longitudinal section of an MAPMT; (c) AXT levels measured in house in the H7546A Hamamatsu MAPMT.}
  \label{fig:anode_ct}
\end{figure}

The other source of interference is anode crosstalk (AXT). A photoelectron may leak from a dynode to an adjacent accelerating structure and generate a signal in another channel as represented in figure\,\ref{fig:anode_ct} (a) and (b). It is a well understood phenomenon of which the levels have been measured in-house by Hamamatsu\,\cite{bib:hamamatsu_handbook}. They represent of order 1\,\% of the primary signal in the adjacent channels for a 1\,mm clear fibre lit by an incandescent light (figure\,\ref{fig:anode_ct} (c)). This value depends mostly on the luminosity of the light and on the surrounding magnetic field. In the current set up, no magnetic field but the Earth's is present and it does not affect the analysis. This crosstalk constitutes a background for the crosstalk analysis and is not sensitive to a misalignment of the fibres.

\subsection{Data acquisition}
\begin{figure}[!htb]
  \centering
  \begin{minipage}[b]{.45\textwidth} 
    \centering
    \includegraphics[width=.8\textwidth]{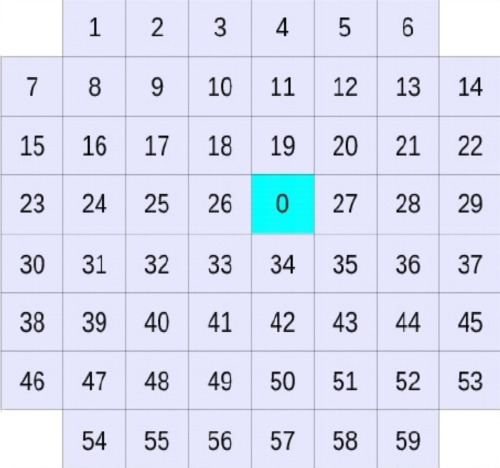}
    \caption{60 channels MAPMT mask used in the EMR; adjacent bars in the detector mostly produce light in adjacent channels.}
    \label{fig:mask_num}    
  \end{minipage}
  \hfill
  \begin{minipage}[b]{.45\textwidth} 
    \centering
    \includegraphics[width=.8\textwidth]{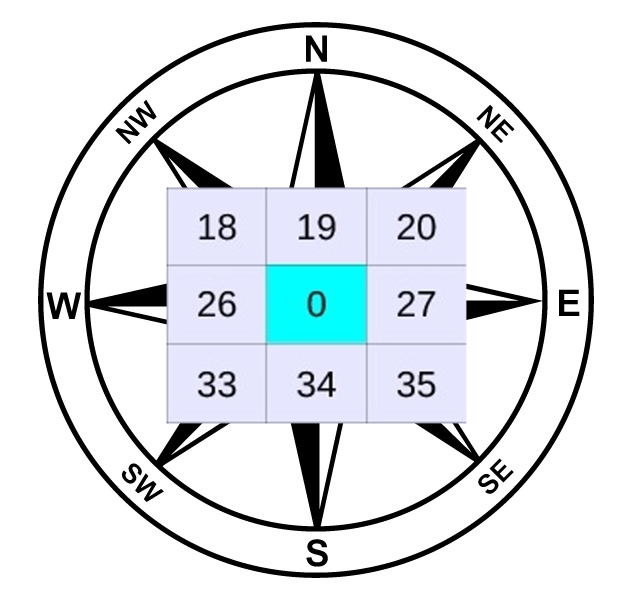}
    \caption{Representation of channel 0 and the its surrounding channels; cardinal points are associated to each one of them.}
    \label{fig:mask_coord}        
  \end{minipage}
\end{figure}

To assess the level of crosstalk, cosmic and beam data are poorly suited in the EMR. A real particle track always hits at least two bars within the same plane. Due to the channel ordering on an MAPMT numbered mask, shown in figure\,\ref{fig:mask_num}, the light coming from two adjacent bars shines on two neighbouring channels of the MAPMT 86\,\% of the time. As a result, a signal recorded in a neighbouring channel cannot be attributed with certitude to crosstalk, as it is likely to be a real signal. In addition, the primary signal energy resolution is quite low as it is impossible to guarantee that the whole signal comes from the energy deposited in the scintillating bar or if part of it stems from crosstalk of an adjacent bar.

\begin{figure}[!htb]
  \centering
  \begin{subfigure}[t]{.68\textwidth} 
    \centering
    \includegraphics[width=\textwidth]{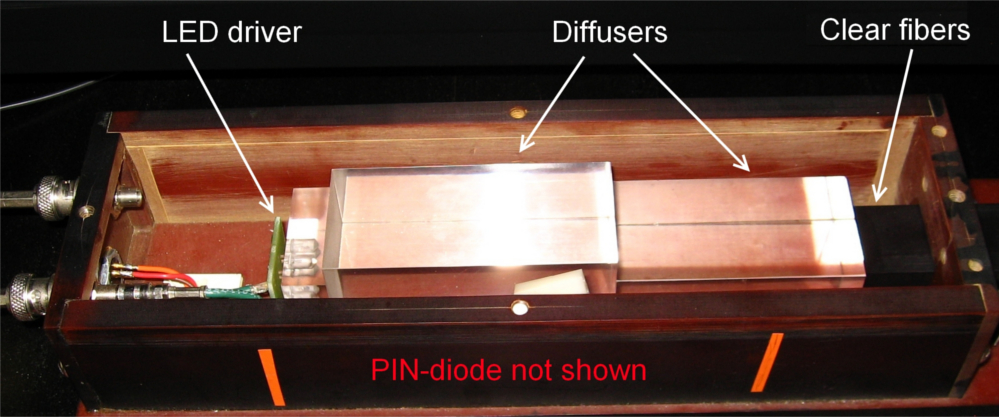}
    \caption{}
  \end{subfigure}
  \hfill
  \begin{subfigure}[t]{.29\textwidth} 
    \centering
    \includegraphics[width=\textwidth]{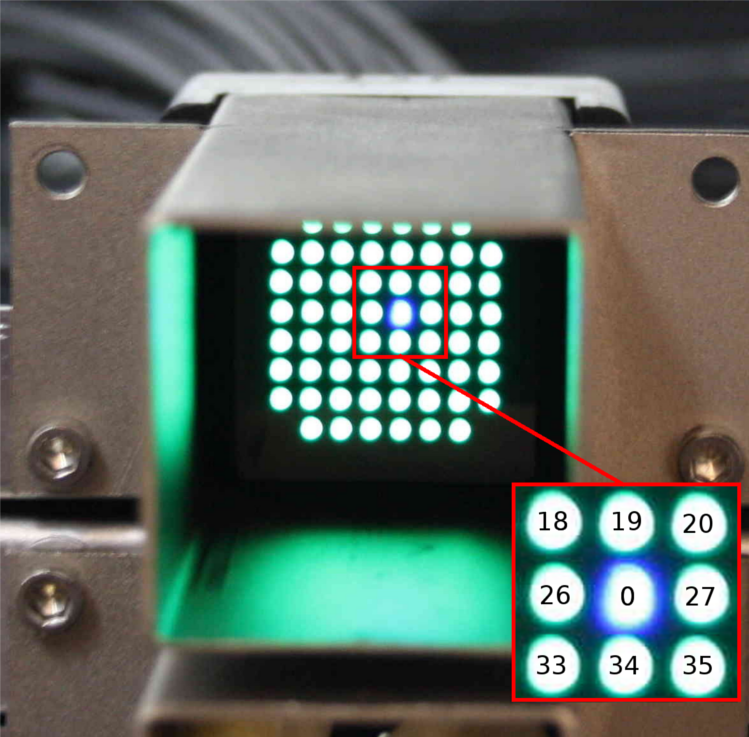}
    \caption{}
  \end{subfigure}
  \caption{Pictures of (a) the LED driver and (b) output light of a single plane fibre bunch.}
  \label{fig:led_driver}
\end{figure}

These constraints motivated the use of an LED light source, pictured in figure\,\ref{fig:led_driver} (a), to perform the analysis. An LED driver pulses light on diffusers that direct the light into a bundle of 48 fibres. Each clear fibre conducts the light towards one specific channel of one of the MAPMTs. The test channel is labelled 0 and is surrounded by channels 18, 19, 20, 26, 27, 33, 34 and 35. For simplicity purposes, the adjacent channels are labelled by cardinal points as represented in figure\,\ref{fig:mask_coord}. With a single fibre lit at each trigger, this guarantees that any signals in the neighbouring photosensors comes from crosstalk. The light coming from the LED driver is blue in contrast with the green light coming from the WLS fibres glued inside of the scintillating bars (figure\,\ref{fig:led_driver} (b)).

When the data sample was recorded, the EMR was completed and located in the MICE hall at RAL. The LED driver was tuned with a variety of voltages ranging from 11.0\,V to 22.0\,V by steps of 0.5\,V. The trigger consisted in the coincidence of an arbitrary spill gate and hits in a channel 0. For each voltage, 100 spills of 100 triggers were recorded to reach a total of $10^4$\,triggers per setting. 

\subsection{Events structure}
A hit in channel 0 can generate hits in the surrounding channels. The four channels directly adjacent to it (N, S, W, E) are the most likely to receive a signal. The four corners neighbouring the test channel (NW, NE, SW, SE) are less likely to be shined on but are included in this analysis for completeness. The channels located two or more compartment away from channel 0 have a negligibly low probability of registering a hit and are not included in the analysis. 

\begin{figure}[!htb]
	\centering
	\includegraphics[width=.7\textwidth]{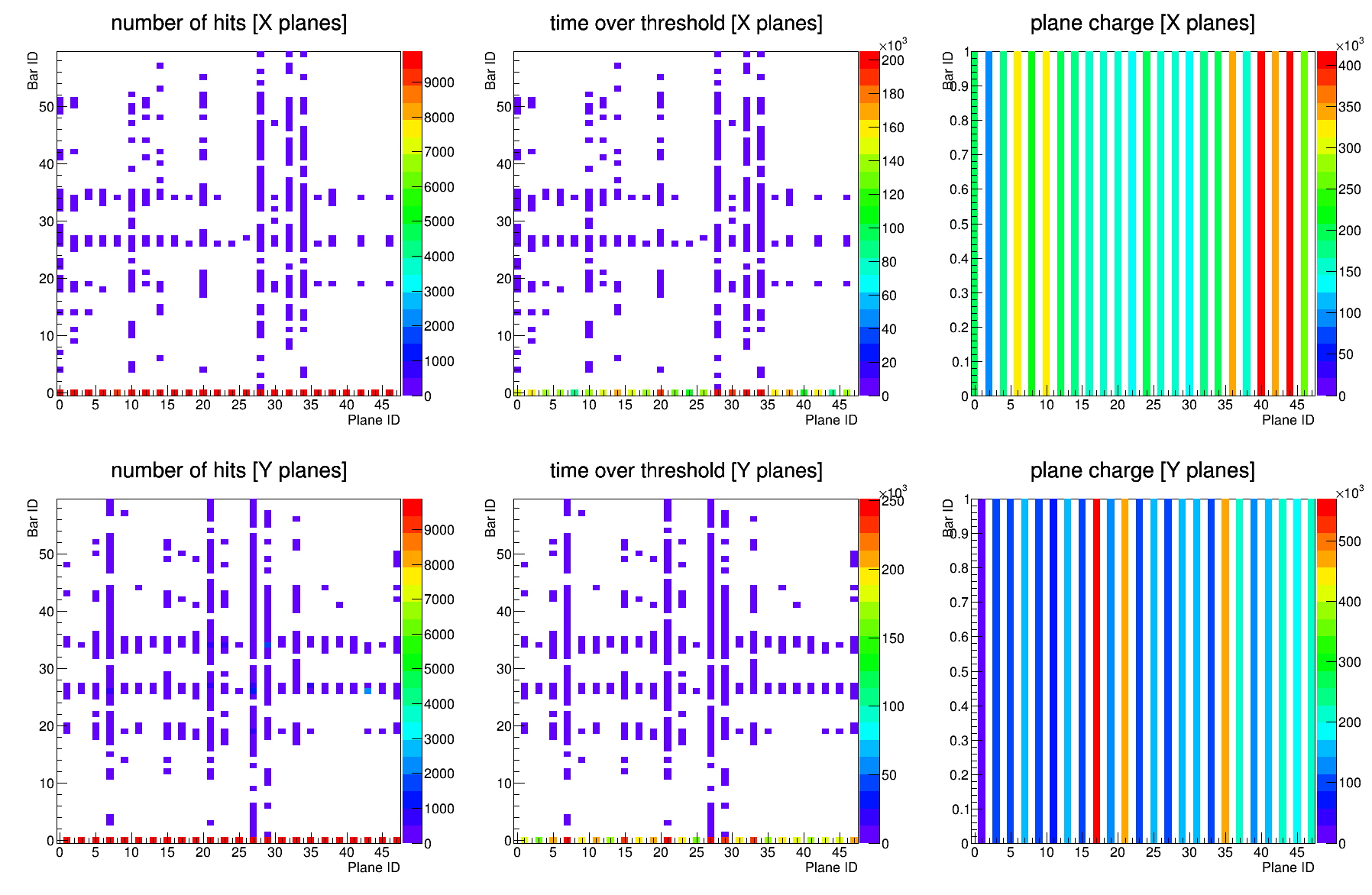}
	\caption{Integrated amount hits, time-over-threshold and charge over $10^4$\,LED pulses in channel 0 of intensity matched to the expected level coming from MIP energy deposition.}
	\label{fig:ledev_le}
\end{figure}

\begin{figure}[!htb]  
	\centering
	\includegraphics[width=.7\textwidth]{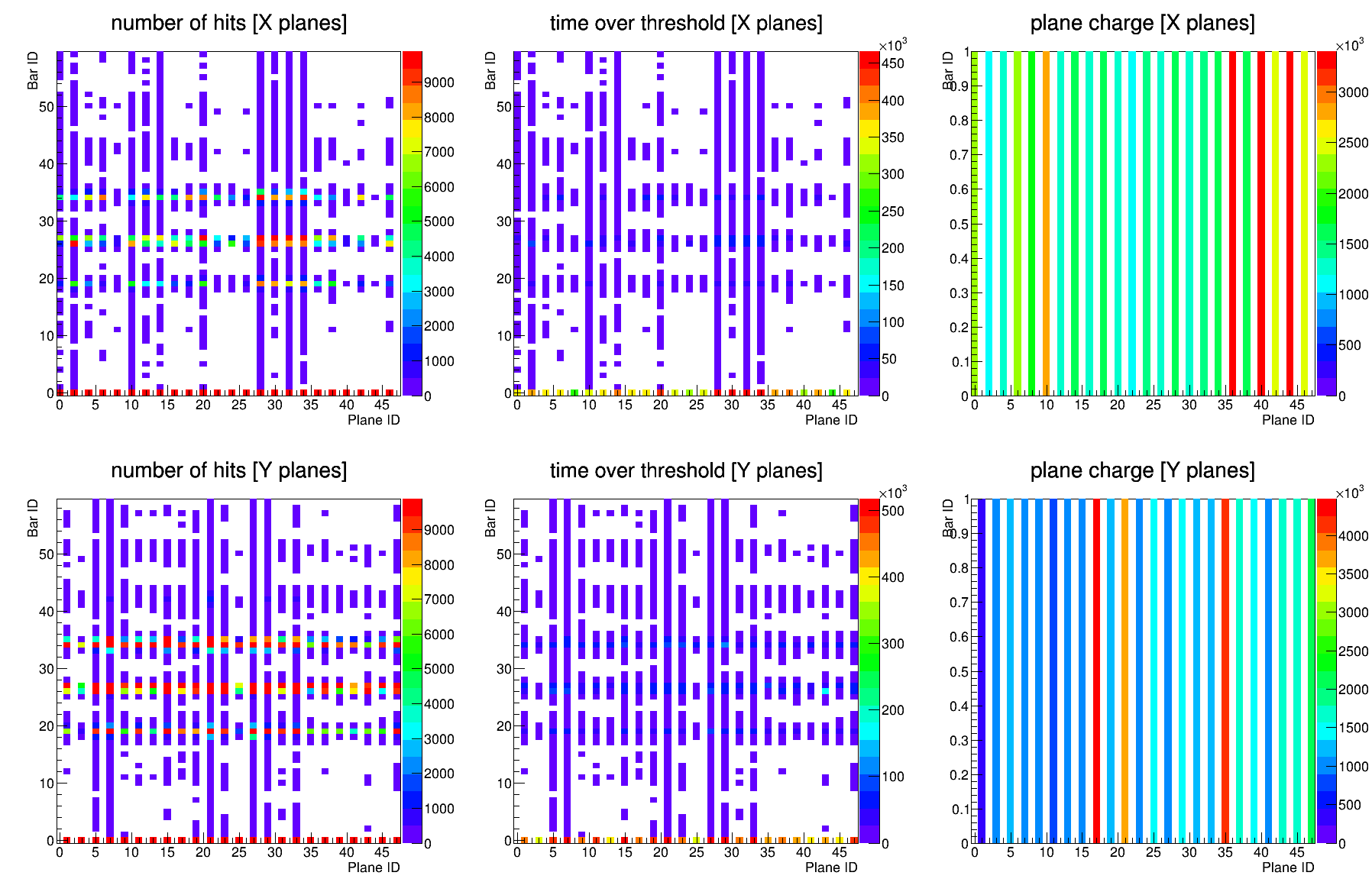}
	\caption{Integrated amount hits, time-over-threshold and charge over $10^4$\,LED pulses in channel 0 at the level expected of very energetic showers or particles stopping in the detector.}
	\label{fig:ledev_he}
\end{figure}

The integrated amount of hits, time-over-threshold and charge of each channel are represented for $10^4$\,LED pulses for two voltage settings: one matched to MIP energy deposition levels (figure\,\ref{fig:ledev_le}) and the other to very high energy depositions due to showers or particles stopping in the EMR (figure\,\ref{fig:ledev_he}). In both cases, four lines appearing at the level of bars 19, 26, 27 and 34, as expected. There is a noticeable difference from plane to plane. Some planes barely experience crosstalk, others record hits in every single bar at least once during a run. Their corresponding integrated charges and time-over-threshold are not identical. At the time of data taking, all PMTs were powered at the same voltage and did not undergo calibration. It does not affect the results as the planes are analysed independently and the energy measurement is consistent within the same plane. It is apparent that the amount of hits, although constant in channel 0, is dramatically higher in the adjacent channels at high voltage. Some of the surrounding bars are hit practically every single time a signal in channel 0 is recorded. 

\begin{figure}[!htb]
	\centering
	\includegraphics[width=.75\textwidth]{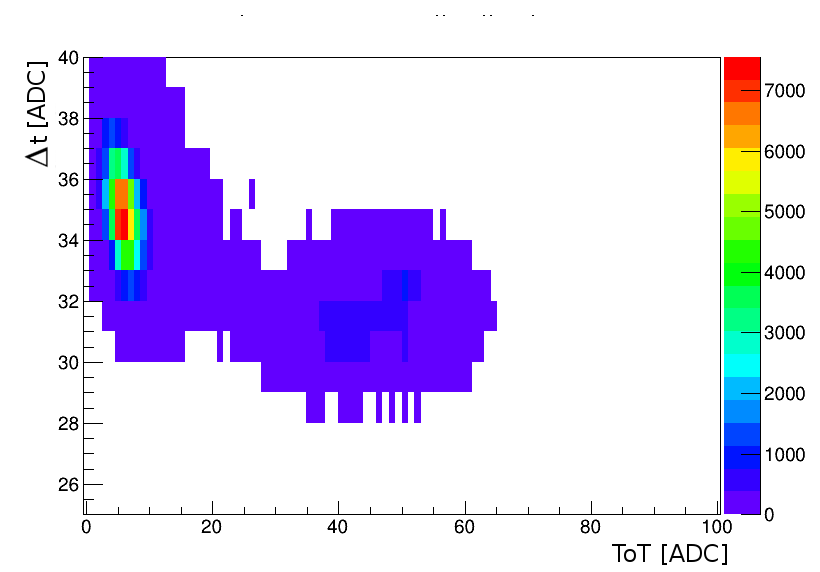}
	\caption{Energy and time distribution of a high voltage LED data sample.}
	\label{fig:LEDdist}
\end{figure}

The energy and time distribution of the high intensity run is represented in figure\,\ref{fig:LEDdist}. The very high energy hits (ToT$\sim45$) are the primary hits in channel 0. The levels of crosstalk for high voltages are very high and create a clear bunch on the distribution; the lower energy hits (ToT$\sim5$) are secondary and correspond to crosstalk in adjacent bars. The crosstalk signals are easily separated from the primary as they are both much lower in energy and shifted in time by typically 10\,ns. The amount of hits is much higher at low energy because a given light pulse only generates one signal in the test channel but may give rise to an array of hits in the surrounding ones. The cuts applied on each data sample are rudimentary. The only requirement to associate a hit with a given LED setting is that its timing coincides with the trigger time. In terms of trigger time minus hit time, $\Delta t$, it corresponds to the interval 25\,ADC counts $<\Delta t<$ 40\,ADC counts.

\subsection{Setting selection}
\begin{figure}[!htb]
  \centering
  \includegraphics[width=.75\textwidth]{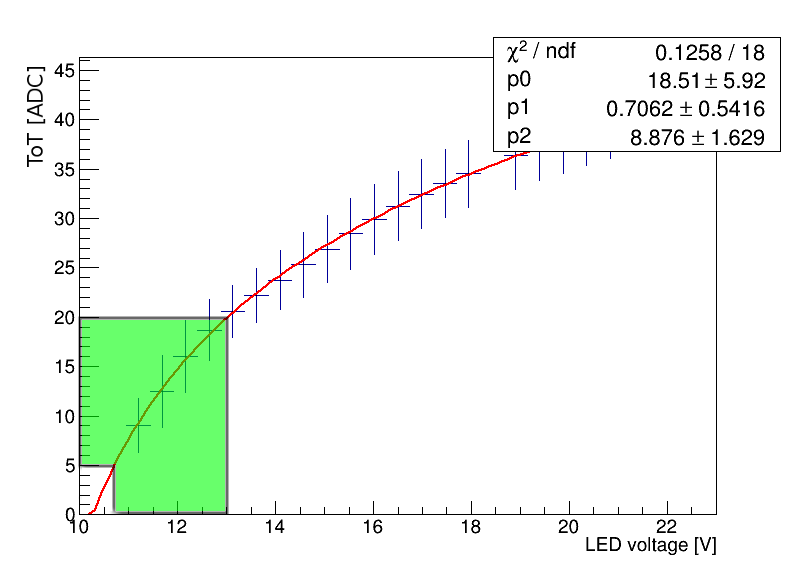}
  \caption{Time-over-threshold (ToT) distribution as a function of the voltage applied to the LED driver. The green area corresponds to a ToT range consistent with MIP energy deposition. The fitted parameters correspond to the logarithmic function of the form $\text{ToT}=p_0\ln\left[p_1(V-p_2)\right]$.}
  \label{fig:tot_vs_voltage}
\end{figure}

For a given setting, a time-over-threshold and a charge distribution are measured in channel 0 in each of the 48 MAPMTs. Two raw data samples (18.5\,V and 21.5\,V) failed to unpack and were excluded from the analysis. The average time-over-threshold (ToT) is represented as a function of the LED voltage for an example MAPMT in figure\,\ref{fig:tot_vs_voltage}. The $y$ error bars are the RMS spread in time-over-threshold. The mean time-over-threshold follows a clear logarithmic trend. A similar graph for the total charge exhibits a linear dependency. The green area represents the voltage region for which the recorded mean time-over-threshold is consistent with a MIP energy deposition. Higher voltages match the energy deposition pattern of a particle stopping in the EMR or a very energetic shower.

\begin{figure}[!htb]
  \centering
  \includegraphics[width=.75\textwidth]{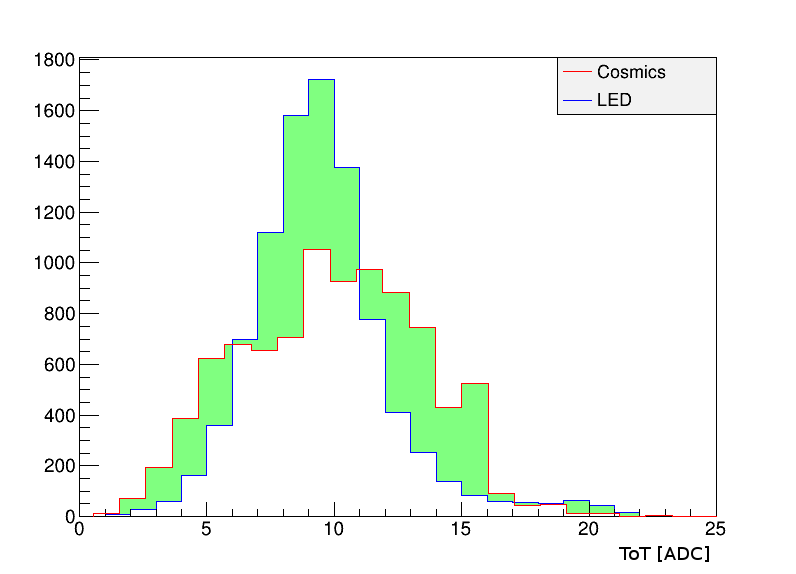}
  \caption{Normalized distributions of time-over-threshold measurements of both cosmic (red) and LED (blue) data in one of the planes. The green area has to be minimized to achieve the most accurate estimation of the crosstalk at MIP energies.}
  \label{fig:cosmics_vs_LED}
\end{figure}

LED data do not straightforwardly correspond to a given particle energy deposition. To assess their correspondence, cosmic data were recorded at RAL in the same experimental condition. A distribution of ToT measurements was binned into a histogram for cosmics, $f_\text{C}$, and for LED, $f_\text{LED}$, for each plane in the EMR. These histograms were used as a tool to identify, for a given MAPMT, the LED voltage that most closely reproduces the cosmic distribution. The method used was to loop over the first ten voltage settings, calculate the area, $\mathcal{A}$, between two normalized distributions and select the setting that gives the lowest value to be the MIP run. For each MAPMT, one has to minimize
\begin{equation}
\mathcal{A}(V)=\int_{0}^{+\infty}\mid f_\text{C}(\text{ToT})-f_\text{LED}(\text{ToT},V)\mid \text{dToT}.
\end{equation}
A comparison of the two types of distribution is provided for the 12\,V setting in figure\,\ref{fig:cosmics_vs_LED}. The LED time-over-threshold distribution is sharper than cosmic data but still give a good estimation of the crosstalk associated with an MIP energy deposition pattern.

\subsection{Crosstalk figures-of-merit}
The crosstalk level is characterized by two main measurements. The first variable is the ratio $R_Q$ of the signal amplitude in a given adjacent channel over the primary amplitude in channel 0. The percentage of light that leaks in a surrounding channel gives an estimate of the significance of the crosstalk and the ability of the detector to discriminate it from true signals. The challenge in making an accurate measurement of this quantity is related to the time-over-threshold (ToT) used as an indirect metric for energy deposition.

Signals are expected to be recorded only above a certain energy deposition. Digitization simulations have shown that a single photoelectron generated at the beginning of an MAPMT accelerating section generates a signal with a ToT of a few ADC counts. As the resolution on this measurement is of order 1\,ADC count, the ratios involving low values of ToT in channel 0 carry a large uncertainty due to this phenomenon. The choice of a higher voltage to measure the value of $R_Q$ removes this limitation as the relative uncertainty decreases proportionally to mean ToT growth. 

\begin{figure}[!htb]
    \centering
    \includegraphics[width=.6\textwidth]{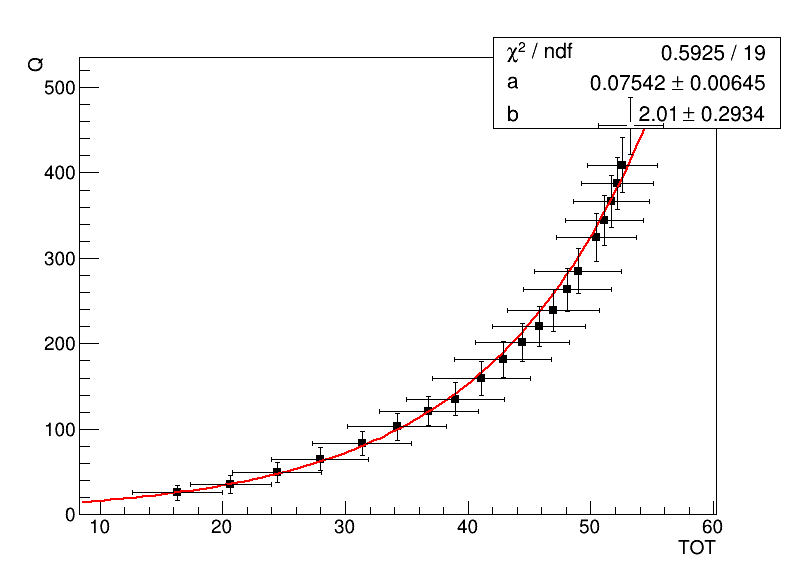}
    \caption{Exponential fit to the charge $Q$ as a function of ToT graph in a single MAPMT.}
    \label{fig:qvstot}
\end{figure}

\begin{figure}[!htb]
  \begin{minipage}[b]{.45\textwidth} 
    \centering
    \includegraphics[width=.85\textwidth]{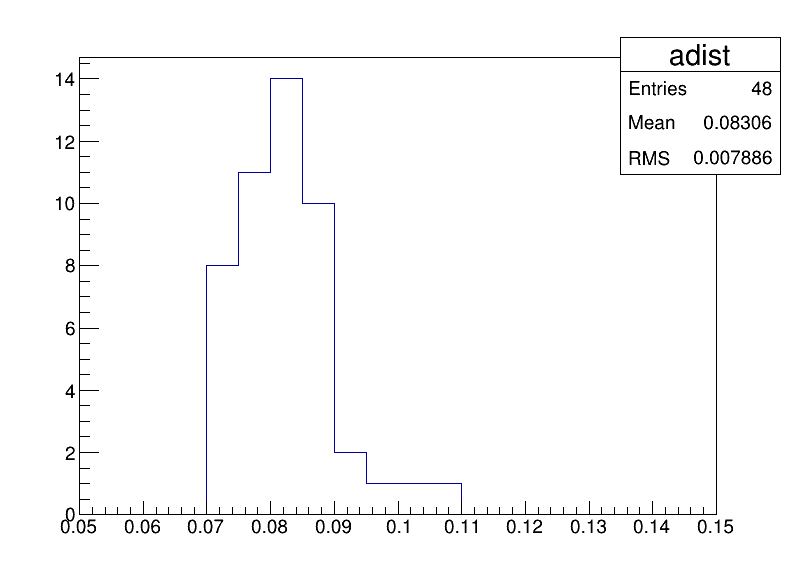}
    \caption{Distributions of the exponential fit parameter $a$ across all MAPMTs.}
    \label{fig:expa}
  \end{minipage}
  \hfill
  \begin{minipage}[b]{.45\textwidth} 
    \centering
    \includegraphics[width=.85\textwidth]{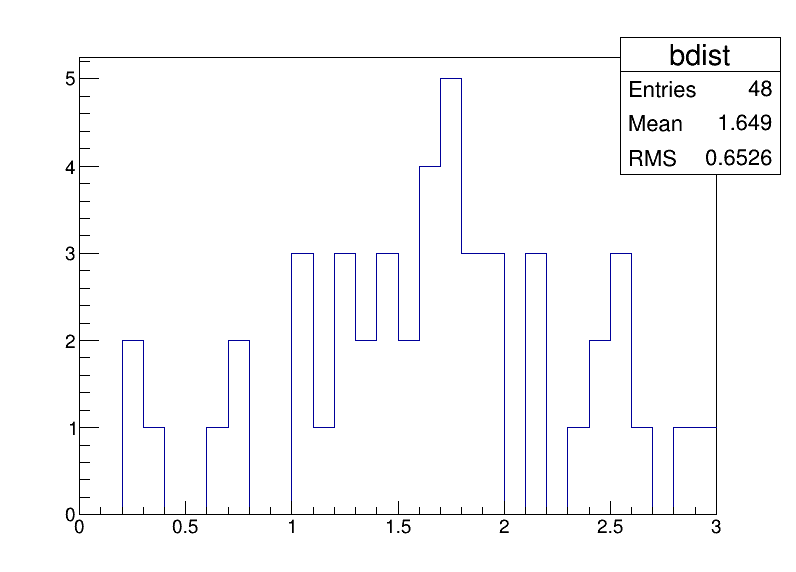}
    \caption{Distributions of the exponential fit parameter $b$ across all MAPMTs.}
    \label{fig:expb}    
  \end{minipage}
\end{figure}

In addition, time-over-threshold measurements are not linearly proportional to the signal amplitude. A ratio of ToT measurements does not correspond to an energy deposition ratio. The exact dependency between ToT and the charge, $Q$, was not thoroughly investigated in the scope of this crosstalk analysis but it is in first approximation related to the ToT through an exponential function of the form
\begin{equation}
Q=e^{a\times \text{ToT}+b}
\label{eq:exp}
\end{equation}
with $a$ and $b$ two unknown parameters. $a$ is the slope of the exponential in log scale and depends on the EMR characteristics such as the scintillation time constant, the FEB shaping function or the threshold level. It is expected to be constant across the detector with small variations. The parameter $b$ depends on the two MAPMT and SAPMT gains and vary significantly from one plane to another, as they are not calibrated. These parameters were measured experimentally by fitting the charge $Q$ as a function of ToT with equation\,\ref{eq:exp} as shown for a single plane in figure\,\ref{fig:qvstot}. The measured distributions of $a$ and $b$ across all MAPMTs are shown in figures\,\ref{fig:expa} and \ref{fig:expb}, respectively.

The fitted parameters, $a$ and $b$, are used to convert the time-over-threshold to charge measurements in order to calculate the correct ratio $R_Q$. Given a measurement $\text{ToT}_i$ in one of the neighbouring channel and $\text{ToT}_0$ in the test channel, the ratio reads
\begin{equation}
R_Q=\frac{Q_i}{Q_0}=\frac{e^{a\times \text{ToT}_i+b}}{e^{a\times \text{ToT}_0+b}}=e^{a(\text{ToT}_i-\text{ToT}_0)}\,.
\end{equation}

The second main parameter used to characterize the crosstalk is the rate at which crosstalk happens. The figure-of-merit that is measured is the ratio, $R_N$, of hits in a given surrounding channel, $N_i$, to the total amount of hits in channel 0, $N_0$, i.e. $R_N=N_i/N_0$. This quantity is measured in each of the 8 surrounding channels for the MIP run, as it is the only one relevant to real data taking.

\subsection{Results}
\subsubsection{Individual MAPMTs}
This analysis produced a plethora of graphs for each MAPMT that cannot be included in the core of this paper but are compiled in \cite{bib:cm_xt}. For each MAPMT, the following measurements are produced:

\begin{enumerate}
\item charge, $Q$, as a function of ToT;
\item ToT as a function of LED voltage;
\item charge, $Q$, as a function of LED voltage;
\item for each of the 8 channels around 0:
\begin{enumerate}
  \item $\text{ToT}_i$ as a function of $\text{ToT}_0$;
  \item $R_Q$ as a function of $Q_0$;
  \item $R_N$ as a function of $\text{ToT}_0$.
\end{enumerate}
\end{enumerate}

\begin{figure}[!htb]
  \centering
  \includegraphics[width=.75\textwidth]{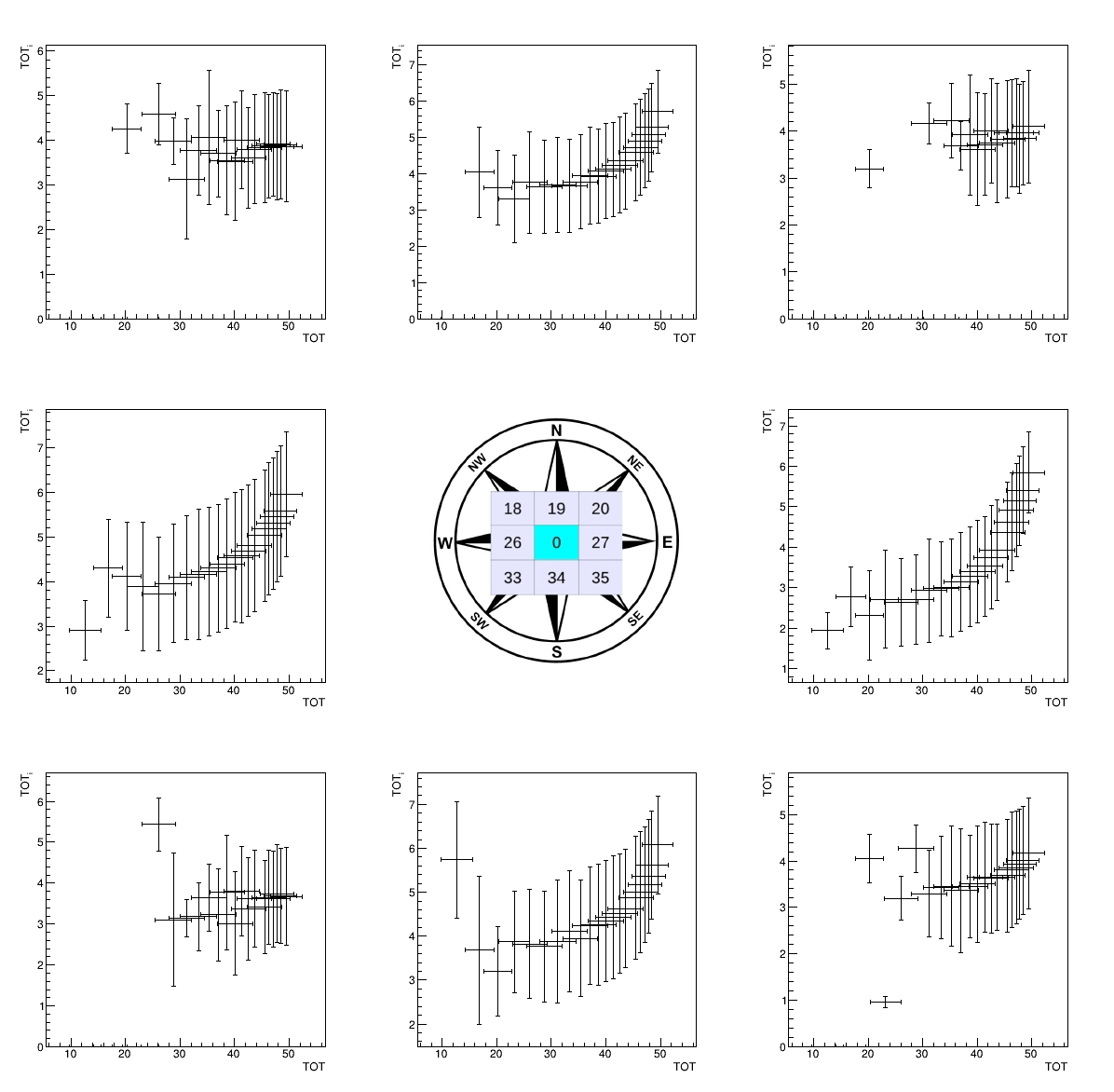}
  \caption{Average time-over-threshold, $\text{ToT}_i$, in the surrounding channels as a function of the average primary signal time-over-threshold, $\text{ToT}_0$, in the 8 channels adjacent to the test channel.}
  \label{fig:toti}
\end{figure}

The eight graphs of series 4.(a) are represented for one MAPMT in figure\,\ref{fig:toti}. The values given for $\text{ToT}_i$ are averaged over one voltage setting and the error bars represent the RMS. 

In the four channels directly adjacent to channel 0 (N, S, W, E), after a flat stretch, the time-over-threshold starts raising as a function of the signal intensity. This is the expected behaviour from the digitization simulations; the observed minimum revolves around 4 ADC counts. In the 4 corners, there is no clear detachment of the curve from the threshold level, as the light leaked is not sufficient to tear off more than a single photoelectron.

\begin{figure}[!htb]
  \centering
  \includegraphics[width=.75\textwidth]{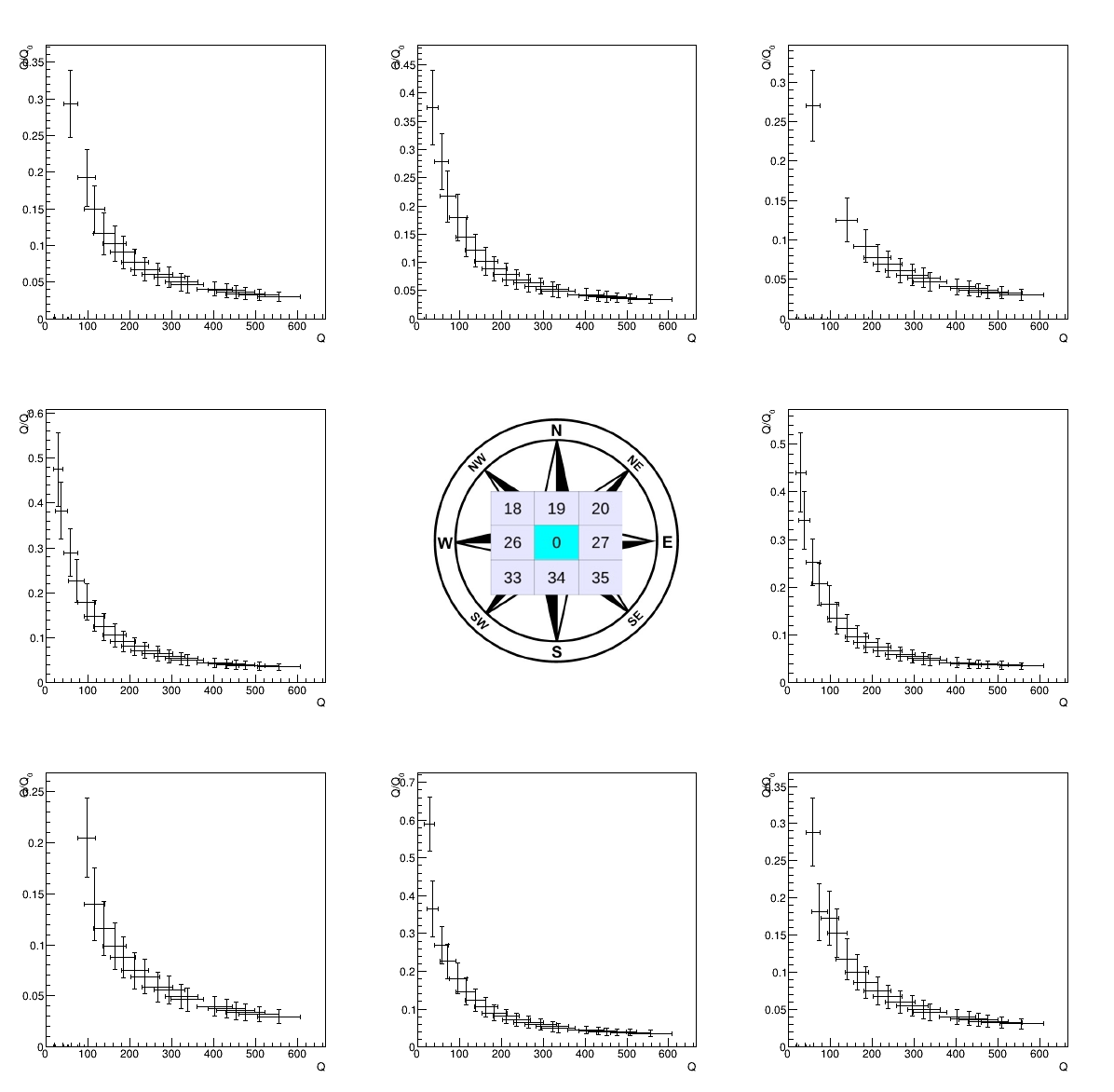}
  \caption{Average crosstalk signal ratio, $R_Q$, as a function of the average charge in channel 0, $Q_0$, in the 8 channels adjacent to the test channel.}
  \label{fig:rq}
\end{figure}

The series of graphs 4.(b) are represented for one MAPMT in figure\,\ref{fig:rq}. The values of $R_Q$ are averaged over one voltage setting and the error bars represent the RMS.

The ratio should be constant as a function of $Q_0$ as the crosstalk signals charge, $Q_i$, are linearly dependant to the primary signal charge. Due to the measurement threshold apparent in figure\,\ref{fig:toti}, it is not. In these graphs there are two different regimes. While $ToT_i$ is close to threshold, $Q_i$ stays constant and the ratio decreases as $1/Q_0$. Above threshold, the function stabilizes and reaches a constant as $Q_i$ starts to increase. The value of $R_Q$ is measured for the highest energy setting, as it yields the smallest uncertainty. In the corners, the stabilization is never achieved and the values only represent an upper boundary. 

\begin{figure}[!htb]
  \centering
  \includegraphics[width=.75\textwidth]{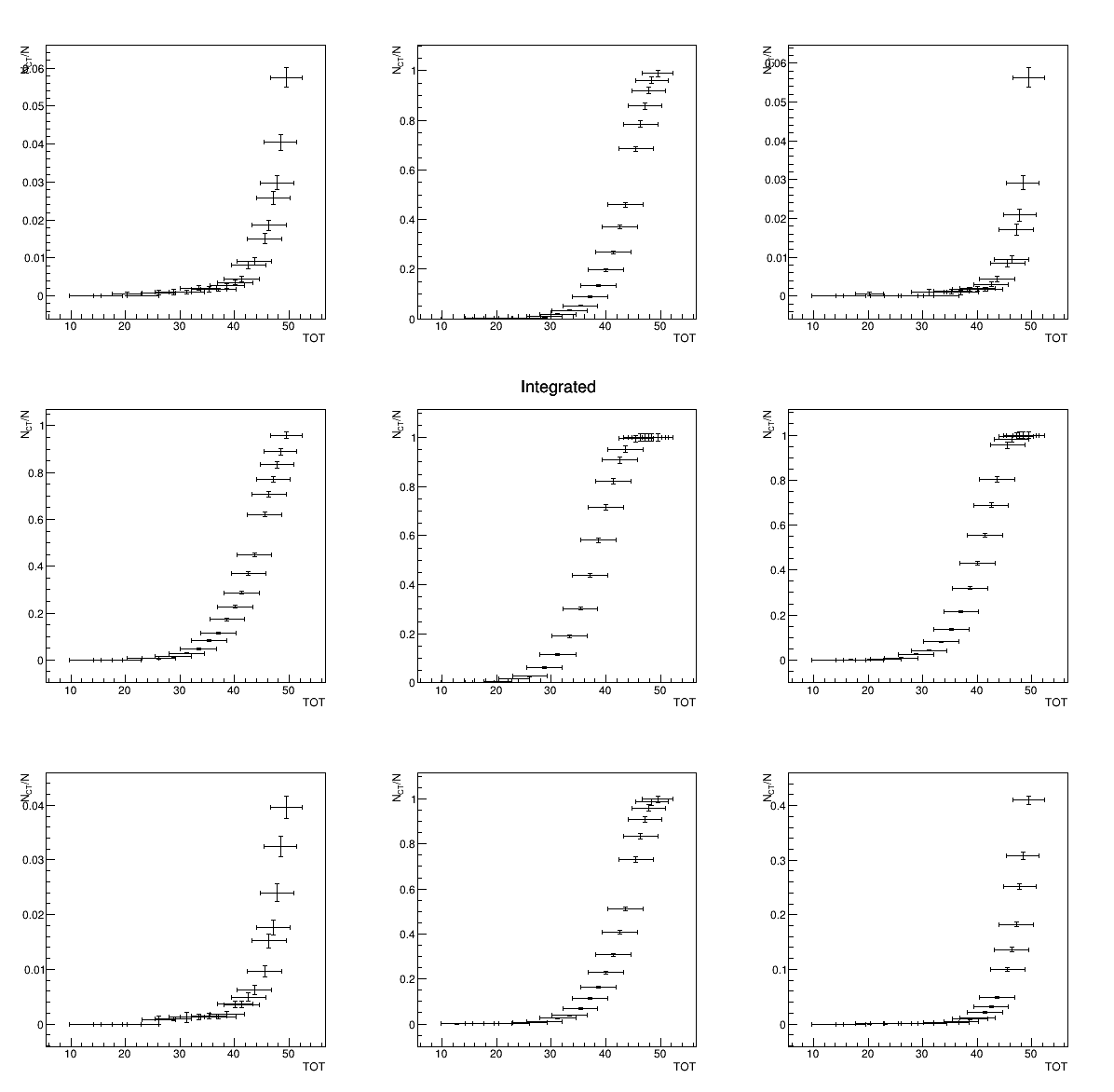}
  \caption{Crosstalk rate, $R_N$, as a function of the average time-over-threshold in channel 0, $\text{ToT}_0$, in the 8 channels adjacent to the test channel. The centre plot represents the integrated rate.}
  \label{fig:rn}
\end{figure}

The series of graphs 4.(c) are represented for one MAPMT in figure\,\ref{fig:rn}. The 8 graphs on the sides of the picture correspond to the eight surrounding channels, as for the previous sets of graphs. The central graph represents the integrated rate of crosstalk, i.e. the probability that at least one of the surrounding channels receives a signal when channel 0 does.

It appears that the probability of crosstalk at low voltage, which corresponds to the MIP run, is very low. It increases as a function of the signal amplitude and reaches 100\,\% at high voltages. This is the expected behaviour of optical crosstalk. 

\subsubsection{Summary}
\begin{figure}[!htb]
 \begin{minipage}[b]{.45\textwidth}
  \centering
  \includegraphics[width=.75\textwidth]{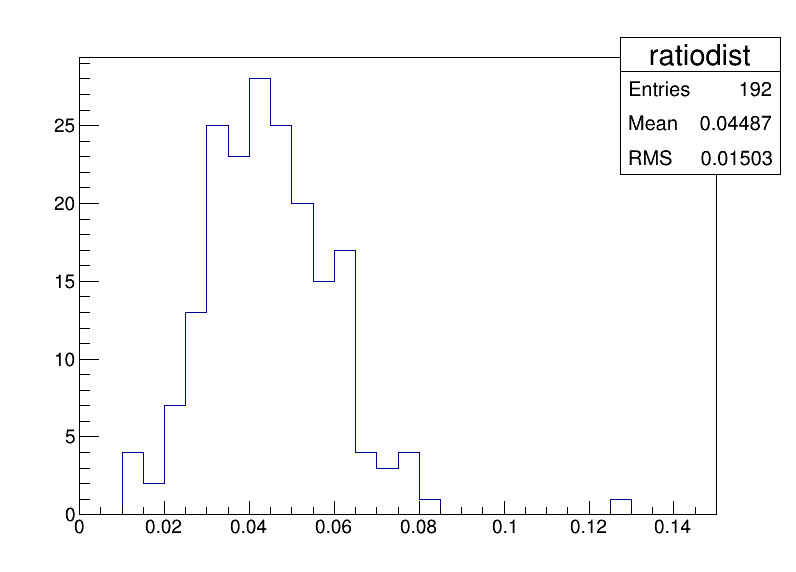}
  \caption{Distribution of crosstalk ratio in the adjacent channels for all MAPMTs.}
  \label{fig:ratiodist}
 \end{minipage}
 \hfill
 \begin{minipage}[b]{.45\textwidth}
  \centering
  \includegraphics[width=.75\textwidth]{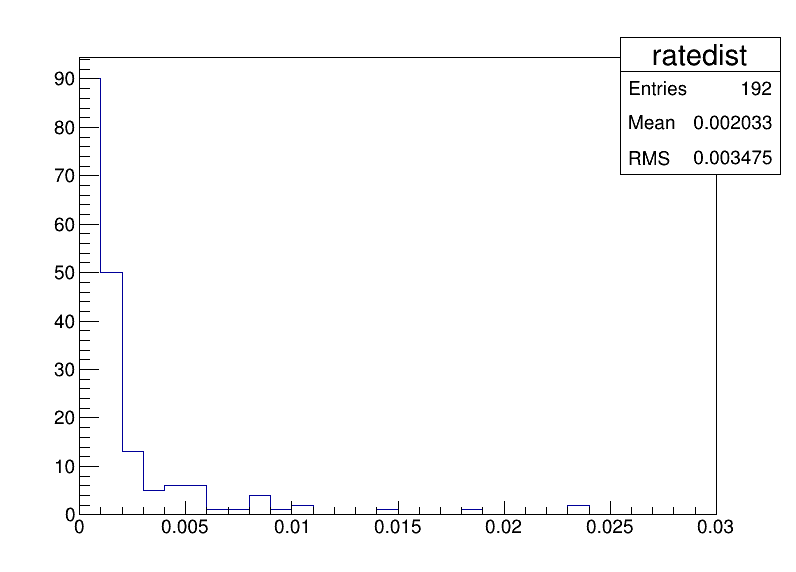}
  \caption{Distribution of crosstalk rate in the adjacent channels for all MAPMTs.}
  \label{fig:ratedist}
 \end{minipage}    
\end{figure}

The measured values of the ratio $R_Q$ and the rate $R_N$ are compiled in a table in \cite{bib:cm_xt} for the four directly adjacent channels in each individual MAPMT. $R_Q$ is the value of the ratio measured for the highest voltage setting and $R_N$ the rate measurement in the MIP tuned LED setting. The distributions are presented in figures \ref{fig:ratiodist} and \ref{fig:ratedist} for all MAPMTs. The average probability of crosstalk is $0.20\pm0.03$\,\% and the average signal recorded in an adjacent channel represent $4.5\pm0.1$\,\% of the initial signal.

\section{Misalignment}
\subsection{Definitions}
The data sample used to measure the misalignment of the MAPMTs is the same LED data that were processed for the crosstalk analysis. 

\begin{figure}[!htb]
  \centering
  \includegraphics[width=.6\textwidth]{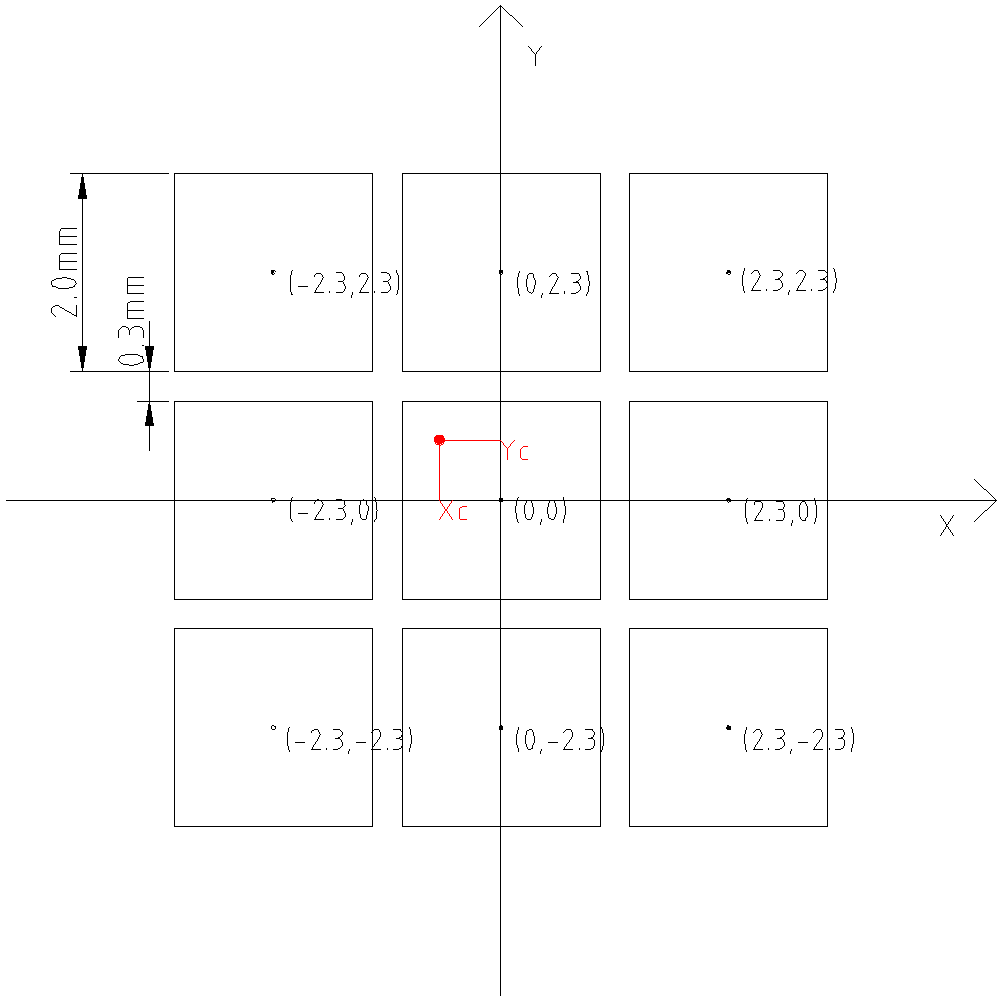}
  \caption{System of coordinates used to determine the misalignment centre. Channel 0 is placed at the origin of the axes and the surrounding channels are set according to their distances from the centre of channel 0 as described in the MAPMT data sheet.}
  \label{fig:misalignment}
\end{figure}

In this section, the coordinates of the MAPMT mask centre $(x_C,y_C)$ with respect to the fibre bundle mask centre are evaluated. To reconstruct the position of this point, the phenomenon of optical crosstalk is an advantage. In the case of a mask shifted with respect to the fibre bundle, light leaks and creates crosstalk signals more often in some channels than others. The centre is computed as the weighted average
\begin{equation}
(x_C,y_C)=\left(\frac{\sum_ix_iw_i}{\sum_iw_i},\frac{\sum_iy_iw_i}{\sum_iw_i}\right),
\end{equation}
with $x_i$, $y_i$ the coordinates of the surrounding channels and $w_i$ the amount of hits recorded in them; the coordinate system used for this analysis is represented in figure\,\ref{fig:misalignment}. The uncertainty on this measurement is related to the amount of hits in the surrounding bars through
\begin{equation}
\Delta x_C=x_C\left(\frac{\underset{i}{\bigoplus}x_i\sqrt{w_i}}{\sum_ix_iw_i}\oplus\frac{\underset{i}{\bigoplus}\sqrt{w_i}}{\sum_iw_i}\right).
\end{equation}

\subsection{Results}
\begin{figure}[!htb]
	\begin{minipage}[b]{.45\textwidth}
		\centering
  \includegraphics[width=1.1\textwidth]{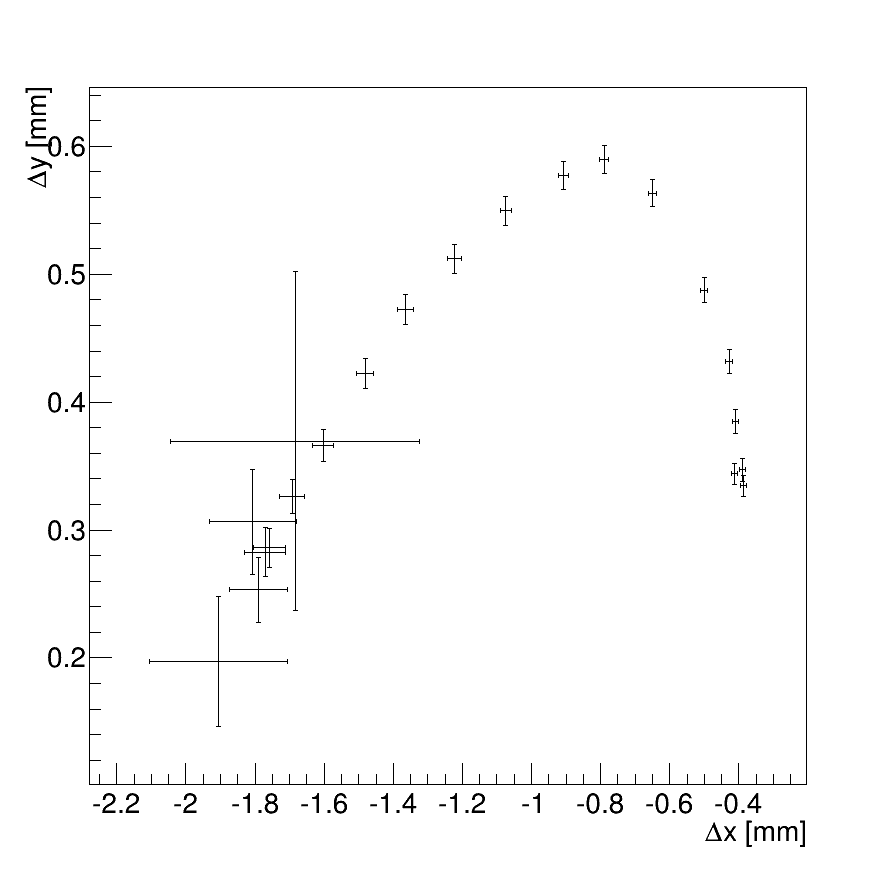}
		\caption{Misalignment centre of an MAPMT for different LED settings. The uncertainties are broad for low voltage settings and are reduced as the voltage is increased.}
		\label{fig:comvssetting}
	\end{minipage}
	\hfill
	\begin{minipage}[b]{.45\textwidth}
		\centering
  \includegraphics[width=1.1\textwidth]{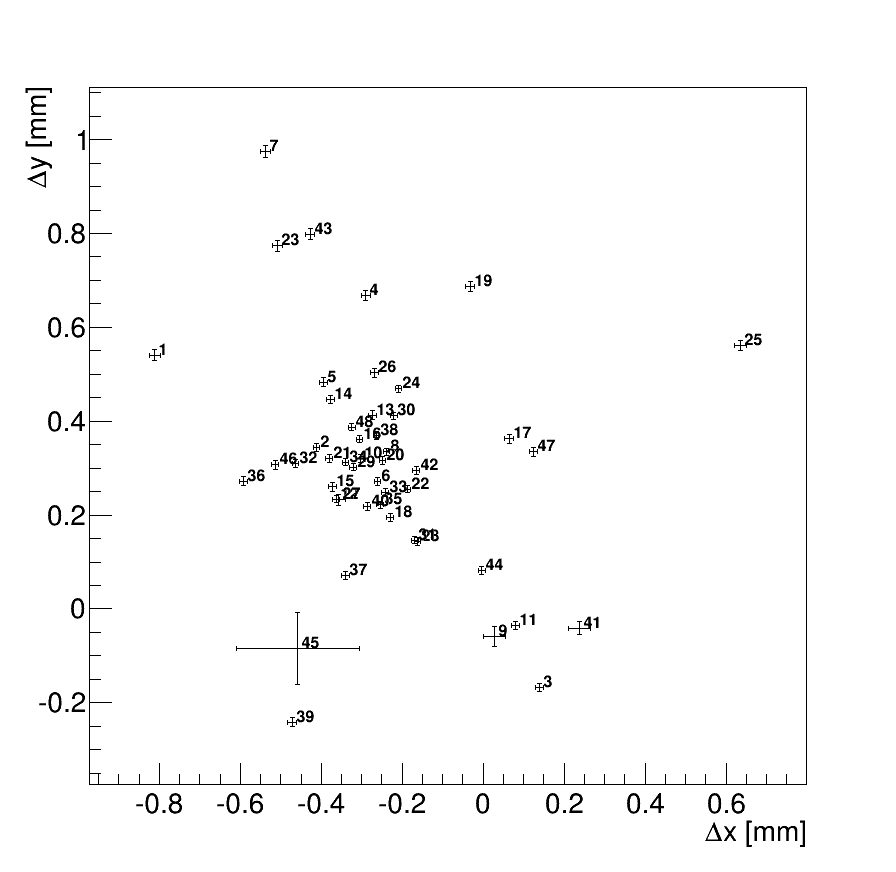}
		\caption{Distribution of the misalignment centres of the 48 MAPMTs. \\ \\}
		\label{fig:comvsplane}
	\end{minipage}    
\end{figure}

The resolution on the misalignment is driven by the amount of hits in the adjacent channels. The setting with the brightest primary signals provides the largest statistical sample in the adjacent bars. This is illustrated in the graph of figure\,\ref{fig:comvssetting}. The points with broader uncertainties correspond to low voltage settings and a significant precision improvement is observed as higher voltages are reached. The points cluster at the end of the trail, indicating convergence of the estimate.

The misalignment centres of all MAPMTs have been calculated and are represented in figure\,\ref{fig:comvsplane} . The tags on each correspond to the plane ID. There is a noticeable cluster of points around $(-0.3,0.3)$, part of which is explained by the fact that the mask were systematically shifted by $-0.5$\,mm along the $x$ axis during construction. The misalignments are small and, although they impact the crosstalk geometry, they do not influence the performance of the EMR.

An exhaustive list of the misalignment centres is provided for all the planes in \cite{bib:cm_xt}.

\section{Signal acquisition efficiency}
\label{section:eff}
\subsection{Description}
This final analysis aims at determining the probability of a signal to be lost in the EMR. The critical requirement for this detector is to have at least one hit in each plane. A missed plane does not provide one of the coordinate necessary to reconstruct a space point in the module, reducing the resolution on range reconstruction. It is necessary to determine if signal loss is a predominant factor that needs to be improved in the EMR. The signal acquisition efficiency, $E_\text{SA}$, is defined as the probability that an MIP muon leaves at least a hit in a plane that it crosses.

More specific quantities have been measured in the context of this analysis. The efficiency of individual planes depend on each MAPMT's characteristics and selected voltage. An MAPMT that does not respond as well as the others has a lower efficiency and its voltage needs to be tuned. In addition, the distribution of the amount of bars hit per plane was measured. The correlation between the production of a signal and the energy deposited in a scintillating bar was studied along with a digitized Monte Carlo simulation to make predictions on the expected average amount of bars hit per plane.

\subsection{Data acquisition}
An array of data samples were used for this analysis to assess the influence of energy deposition on the amount of bars hit in one plane. Using muons that stop in the detector allows for the measurement of the influence of energy deposition on the amount of bars hit, e.g. due to crosstalk. It does not allow for the analysis of the most downstream planes behaviour as the tracks do not reach them. Two additional settings with much higher energy were selected to probe every plane of the detector. Monte Carlo studies of the EMR have shown that muons with longitudinal momenta, $p_z$, above 280\,MeV/$c$ leave the detector without stopping\,\cite{bib:emr_simulation}. The chosen settings include two well under that predicted value and two far above. The data samples processed in this analysis are summarised in table\,\ref{tab:beam_settings} and have been extracted from MICE Step I data recorded in October 2013.

\begin{table}[!htb]
  \centering
  \begin{tabular}{c|c|c|c|c}
    Run ID & Beam setting & TOF1 triggers & $p_z^{\text{Q9}}$ [MeV/$c$] & $p_z^{\text{EMR}}$ [MeV/$c$] \\
    \hline
    5428 & $e^+$ & 60511 & 300.38 & 239.64 \\
    5429 & $e^+$ & 21860 & 300.38 & 239.64 \\
    \hline
    5439 & $\pi^+$ & 17265 & 293.83 & 232.87 \\
    5450 & $\pi^+$ & 55823 & 293.83 & 232.87 \\
    \hline
    5401 & $\pi^+$ & 90757 & 424.37 & 365.29 \\
    5403 & $\pi^+$ & 37969 & 424.37 & 365.29 \\
    5405 & $\pi^+$ & 30007 & 424.37 & 365.29 \\
    \hline
    5410 & $\pi^+$ & 50670 & 450.17 & 388.77 \\
    5414 & $\pi^+$ & 37969 & 450.17 & 388.77 \\
 \end{tabular}
 \caption{Set of beam settings selected for the signal efficiency analysis. $p_z^\text{Q9}$ is the mean theoretical momentum upon exiting the last beam line quadrupole and $p_z^\text{EMR}$ is the estimated mean momentum at the entrance of the EMR.}
 \label{tab:beam_settings}
\end{table}

\begin{figure}[!htb]
 \centering
 \includegraphics[width=.75\textwidth]{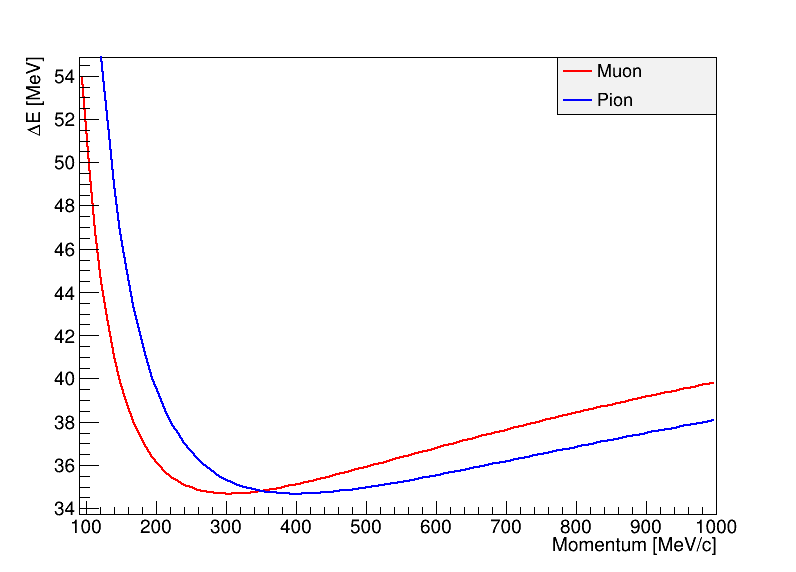}
 \caption{Mean energy loss in the KL through ionization by muons (red) and pions (blue) as a function of impinging momentum. The energy lost by a MIP is $\sim35$\,MeV.}
 \label{fig:kl_energy_loss}
\end{figure}

Each setting is provided with the value of the selected momentum at target from which the momentum upon exiting Q9 (ninth quadrupole of the MICE beam line) has been computed using Monte Carlo simulations of the MICE experiment. After Q9, the beam goes through two time-of-flight detectors (TOF1 and TOF2), 9.48\,m of air between the two TOFs and a lead-based preshower calorimeter (KL). The mean energy loss in the TOFs and the air has been previously evaluated and estimated at 10.12\,MeV in each TOF and 1.6\,MeV in the air\,\cite{bib:flight_counters}. A simplified estimation of the mean energy loss in the KL was developed specifically for this analysis and is represented for muons and pions as a function of the momentum in \ref{fig:kl_energy_loss}. These functions have been computed using the relativistic Bethe-Bloch formula\,\cite{bib:sigmund} corresponding to the KL composition. Applying these losses to a given momentum at Q9 provides a good estimation of the momentum upon entering the EMR.

For the Monte Carlo simulation, an ideal 250\,MeV/$c$ negative muon beam was fired at the detector with a typical angular distribution. $10^4$ events were generated for this analysis.

\subsection{Hit pre-selection}
\begin{figure}[!htb]
 \centering
 \includegraphics[width=.75\textwidth]{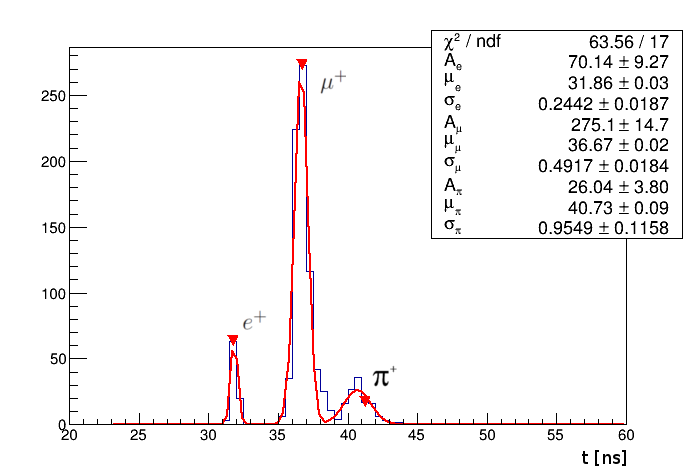}
 \caption{Time-of-flight between TOF1 and TOF2. The leftmost peak corresponds to particles that travel very close to the speed of light ($e^+,e^-$) and the two others are the muon and pion peaks. The three peaks are fitted with a trimodal Gaussian mixture which scales, $A_\alpha$, means, $\mu_\alpha$, and spreads, $\sigma_\alpha$, are summarised in the legend for $\alpha=e,\,\mu,\,\pi$.}
 \label{fig:tof12}
\end{figure}

Only muons are used to calculate the efficiency, as they are the primary focus of the EMR detector and the electron showers are not expected to hit every plane on their path. A filter has to be used to discriminate the other particles and select the muon events. The time-of-flight between TOF1 and TOF2 provide an excellent statistic to separate the particle species. Events that produced a single hit (i.e. a pair of slabs) in TOF1 and TOF2 are selected and the time-of-flight is computed. The distribution of time-of-flight is represented in figure\,\ref{fig:tof12} and exhibits three distinct peaks. The three modes are identified using the ROOT TSpectrum class and fitted with a trimodal Gaussian mixture. The leftmost Gaussian function corresponds to ultrarelativistic particles, i.e. electrons or positrons. The two others are the muon and the pion peaks. They are clearly separated due to the momentum selection at the second dipole and the difference of mass between the two particles ($m_\mu = 105.66$\,MeV/$c^2$, $m_{\pi^\pm} = 139.57$\,MeV/$c^2$). The pion peak is in fact a mixture of pions and muons as the decay of the former produces an array of possible time-of-flights. The probability of belonging to peak $\alpha$ is calculated for each event of time-of-flight $t$ as
\begin{equation}
p_\alpha(t) = \frac{1}{\sqrt{2\pi}\sigma_\alpha}\exp\left[-\frac{(t-\mu_\alpha)^2}{2\sigma_\alpha^2}\right],\,\alpha=e,\,\mu,\,\pi.
\end{equation} The events for which $p_\mu$ is the largest of the three are selected for this analysis. Different particle types would lead to different energy deposition pattern.

Only the primary particle tracks are used. The decay products of the muons ($e^+,\,e^-$) have different energy and have a much broader angular distribution, as they cover the entire range of solid angles. This would result in a lot of planes being hit in more than 2 bars. 

A minimum of 10 hits is requested for the track to be included. The trigger used in MICE is based on a single hit in TOF1, which does not always give a full track in the EMR. This cut rejects noise events or showers that only produce scarce hits in the EMR.

\subsection{Processing}
The raw data samples are reconstructed, i.e. the primary tracks are separated from the noise by associating them in time with a trigger and the secondary tracks are isolated in an array for further processing, as described in section\,\ref{section:recon}.

A distribution of the amount of hits recorded is stored in a histogram for each plane. Each time a plane is missed by a track, i.e no signal is produced in one plane even though it is on the path of a particle, its plane ID is registered as missed.

The two settings with energies below the threshold for muons to cross the whole detector have been combined into one to increase the level of statistics. To improve the accuracy of the efficiency measurement, it is critical to have a lot of events, as the probability of missing a plane is very small. The same thing has been done with the two settings above threshold. The amount of times each plane was hit reached $\sim3000$ after combining the samples above threshold. 
\subsection{Results}

\subsubsection{Signal reproduction}
In real data, the MAPMT measures a time-over-threshold for individual bars and a the SAPMT records an integrate charge for the whole plane. The MAPMT is the focus of this analysis as it allows for the spacial reconstruction of tracks. The SAPMT efficiency was not included as the photomultipliers will be replaced by the end of 2014 and an analysis will be performed then.

These quantities depend on a lot of digitization parameters that translate the energy deposited in the scintillator to a measurable signal. To estimate the probability that a given energy deposition is recorded at the level of the photomultipliers, the Monte Carlo sample described above is used. It provides the exact energy deposition for each bar, value that is digitized by following these steps:

\begin{enumerate}
\item convert the energy deposited into a mean number of scintillating photons $\overline{n}_\text{sph}$ (2000 $\gamma$/MeV);

\item sample $n_\text{sph}$ from a Poisson distribution of mean $\overline{n}_{\text{sph}}$;

\item convert $n_\text{sph}$ into a mean number of trapped photons $\overline{n}_\text{tph}=E_Tn_\text{sph}$ ($E_T$ = 2\%);

\item sample $n_\text{tph}$ from a Poisson distribution of mean $\overline{n}_\text{tph}$;

\item attenuate $n_\text{tph}$ through the length of the WLS and clear fibres to get the mean number of attenuated photons $\overline{n}_\text{aph} = 10^{\alpha_\text{WLS}L_\text{WLS}+\alpha_\text{CL}L_\text{CL}}n_\text{tph}$ ($\alpha_\text{WLS}=-2.0$\,dB/m, $\alpha_\text{CL}=-0.2$\,dB/m);

\item apply the connector attenuation map (up to 30\,\%);

\item sample $n_\text{aph}$ from a Poisson distribution of mean $\overline{n}_\text{aph}$;

\item convert $n_\text{aph}$ to the mean number of photoelectrons $\overline{n}_\text{pe}=QE\times n_\text{aph}$ (QE = 20\,\%);

\item sample $n_\text{pe}$ from a Poisson distribution of mean $\overline{n}_\text{pe}$;

\item correct $n_\text{pe}$ for cathode non-uniformity (up to 40\%);

\item convert $n_\text{pe}$ to the number of ADC counts $n_\text{ADC}$ (8 ADC counts/$n_\text{pe}$);

\item simulate electronics response using Gaussian smearing (width $\sigma = 10$\,ADC counts);

\item convert $n_\text{ADC}$ to ToT through $n_\text{ADC} = a+b\times\ln(\text{ToT}/c+d)$ with fixed parameters;

\item convert GEANT4 time to mean ADC counts $\overline{\Delta t}$ (2.5\,ns/ADC count);

\item sample $\Delta t$ from a Gaussian distribution of mean $\overline{\Delta t}$ (width $\sigma = 2$\,ADC counts).
\end{enumerate}

If the energy initially deposited in the detector is not high enough, the light produced and transferred could be too dim to extract a photoelectron from the MAPMT cathode. In that case, no hit is recorded by the DAQ system.

\begin{figure}[!htb]
 \centering
 \includegraphics[width=.75\textwidth]{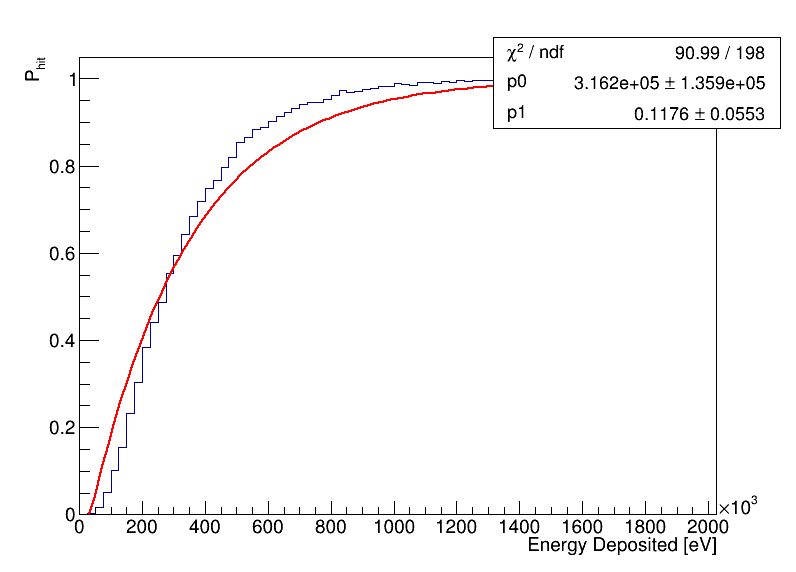}
 \caption{Probability of a hit creating a digitized signal as function of the energy deposited in a bar.}
 \label{fig:digi_edep}
\end{figure}

For each bar in the EMR, the total energy deposited is reconstructed and digitized to a ToT measurement. If the latter is zero, the energy deposited is lost. The probability that a signal is recorded as a function of energy deposition in represented in figure\,\ref{fig:digi_edep}. Above one MeV, approximatively no signal is lost through the digitization process.

\begin{figure}[!htb]
 \centering
 \begin{subfigure}[t]{.49\textwidth}
  \centering
  \includegraphics[width=\textwidth]{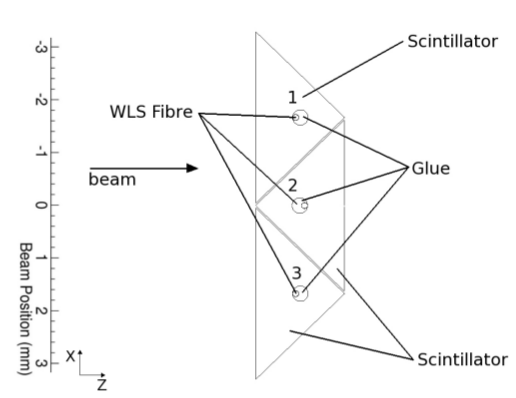}
 \end{subfigure}
 \hfill
 \begin{subfigure}[t]{.49\textwidth}
  \centering
  \includegraphics[width=\textwidth]{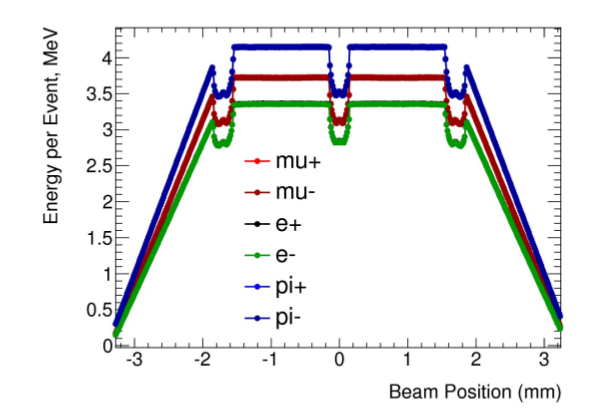}
 \end{subfigure}
 \caption{Simulation of the energy deposition as a function of particle position for a group of three triangular scintillating bars of the EMR.}
 \label{fig:edep_pos}
\end{figure}

In the EMR, due to the triangular shape of the scintillating bars, a particle always goes through at least two of them in each plane. In the case of beam data, as the trajectories are generally perpendicular to the planes, it is safe to consider that they hit exactly two bars per plane. In an optimal situation, the distribution of bars hit per plane should be a single bin corresponding to two bars. As demonstrated by figure\,\ref{fig:digi_edep}, not every hit produces a signal recorded at the level of the MAPMT. A hit can be lost depending on where the particle crosses with respect to the centre of a bar. The energy deposition as a function of the position of the trajectory with respect to the bar has been estimated for MIPs using a GEANT4 simulation in a previous analysis\,\cite{bib:emr_simulation} and is depicted in figure\,\ref{fig:edep_pos}.

\begin{figure}[!htb]
 \centering
 \includegraphics[width=.75\textwidth]{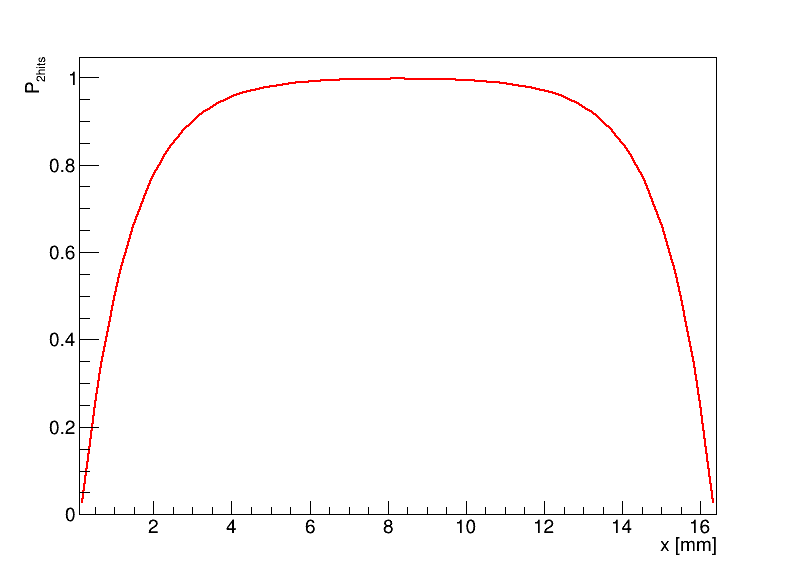}
 \caption{Probability to have two hits in one plane as a function of track position.}
 \label{fig:prob_pos}
\end{figure}

The previous simulation results are combined with the probability of producing a hit as a function of deposited energy to estimate the probability of producing a hit as a function of position. The figure of interest is the probability of having two bars hit as a function of the position of the beam with respect to a set of two bars. The digitized ratio measured in figure\,\ref{fig:digi_edep} is first fitted with a function of the form
\begin{equation}
f_\text{hit}(E) = 1 - \exp(-E/p_0+p_1),
\label{eq:elossfit}
\end{equation}
with $p_0$ and $p_1$ the fitting parameters. The base of the triangular section of a bar measures 33\,mm; the energy deposition increases linearly from 0 to 16.5\,mm away from the edge of the bar. In the Monte Carlo simulation, the situation is simplified by neglecting the presence of a hole and estimate that $E = p_2x$ with $p_2 \simeq 4.2/16.5 \simeq 0.25$\,MeV/mm. Implementing this estimate in equation\,\ref{eq:elossfit} yields
\begin{equation}
f_{\text{hit}}(x) = 1 - \exp(-p_2x/p_0+p_1),
\label{eq:bar_hit}
\end{equation}
the probability of having a hit in one bar as a function of the distance from the edge of it. To compute the probability of having two bars hit, the function in equation \ref{eq:bar_hit} is combined with its symmetrical equivalent (maximal at 0, minimal at 16.5\,mm), i.e. 
\begin{equation}
f_{\text{2hits}}(x) = (1 - \exp(-p_2x/p_0+p_1))(1 - \exp(p_2(x-16.5)/p_0+p_1)).
\end{equation}
This probability function is represented in figure\,\ref{fig:prob_pos}. The probability of having two hits drops as the particle approaches the edge of one of the bars, as the energy is asymmetrically deposited. This pattern is repeated all the way to the edges of the plane where a bar is isolated on one end. As the beam is unlikely to stray that far, the edges are ignored and averaging $f_{\text{2hits}}$ over 16.5 mm is equivalent to averaging the probability over the whole plane. The probability of producing two hits in one plane hence reads
\begin{equation}
P_{2hits} = \frac1{16.5}\int_{0}^{16.5}f_\text{2hits}(x)dx=84.18\%.
\label{eq:p2hits}
\end{equation}

The low energy events are in reality much less represented than the simulation would suggest. This obviously results in a higher proportion of single bar hits. Three or more bar hits are theoretically impossible but not with real data. Crosstalk and noisy channels are the main causes of additional hits. In this simple estimation, missed planes cannot theoretically exist but it does not account for irregularities that might cause such a situation.

\subsubsection{Global Efficiency}
To compute the global efficiency of the detector, $E_\text{SA}$, only the data samples above crossing threshold are used. The samples under threshold provide a biased estimate as the energy deposited is a function of the depth. The plane in which the muon stops yields more light and is less likely to be missed. More energy also increases the likelihood of crosstalk as demonstrated in section\,\ref{section:xt}. A muon that goes through the whole detector deposits a more or less constant amount of energy in each plane and is expected to produce a signal in each of them.

\begin{figure}[!htb]
  \centering
  \includegraphics[width=.75\textwidth]{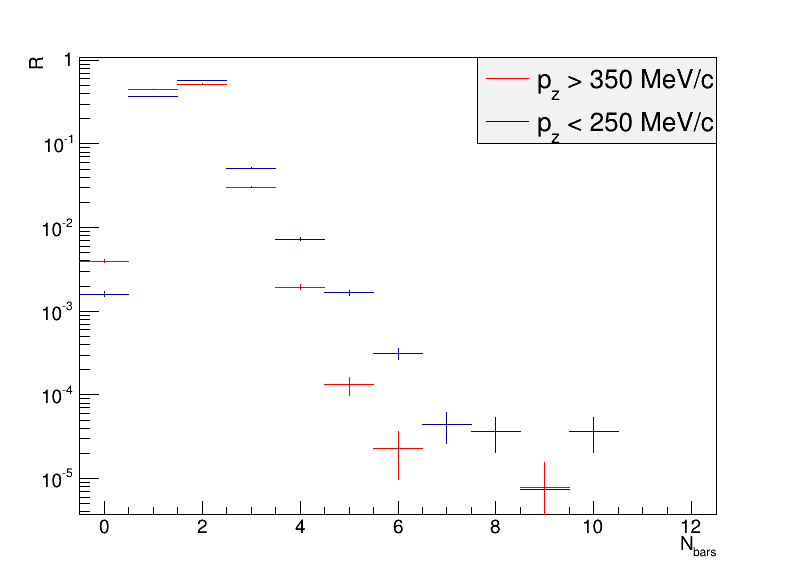}
  \caption{Fraction of the sample that hit a given number of bars per plane for a beam of particles below (blue) and above (red) the crossing threshold of the EMR detector.}
  \label{fig:bars_hit}
\end{figure}

The distribution of the amount of bars hit per plane is given in terms of fraction of the sample, $R$, in figure\,\ref{fig:bars_hit} for the samples below and above threshold. The exact fractions are summarised in table\,\ref{tab:bars_hit}. A significantly larger fraction of planes record high numbers of bars hit and the probability of missing a plane decreases when the muon stops inside of the detector.

\begin{table}[!htb]
	\centering
  \begin{tabular}{c|c|c}
    Number of bars & $p_z>350$ MeV/c & $p_z<250$ MeV/c\\
    \hline
    0 & \textcolor{red}{$0.43\pm0.02$ \%} & $0.16\pm0.01$ \%\\
    1 & $45.25\pm0.25$ \% & $36.62\pm0.19$ \% \\
    2 & $51.22\pm0.27$ \% & $57.16\pm0.26$ \% \\
    3 & $2.90\pm0.05$ \% & $5.13\pm0.06$ \% \\
    4 & $0.18\pm0.01$ \% & $0.73\pm0.02$ \% \\
    5 & $0.011\pm 0.003$ \% & $0.17\pm0.01$ \% \\
    $\geq6$ & $<0.01$ \% & $<0.1$ \% \\
  \end{tabular}
  \caption{Values and uncertainty of the bins in figure\,\ref{fig:bars_hit}.}
  \label{tab:bars_hit}  
\end{table}

As long as at least one hit is recorded in each plane on the path of the particle, it is sufficient to have a pair of coordinates per module. The efficiency of the detector is summarised by
\begin{equation}
E_\text{SA} = 1-R_0 = 99.57\pm0.02\,\%,
\end{equation}
with $R_0$ the fraction of planes missed.

Additional information on the detector is obtained by observing the distribution of bars hit for the two types of settings. At lower beam momentum, muons deposit more energy, stop in the detector and the proportion of planes hit twice is close to 6\,\% higher. Higher bar counts are more likely for low momentum muons, not only because they stop in the detector, but also because they produce more crosstalk on their path. The rate of crosstalk increases as a function of the time-over-threshold as represented in figure\,\ref{fig:rn}. The percentage of missed bars is lower for the same set of reasons.

The percentage of single bars measured is translated in the simulation by applying a cut off at 800\,keV of energy deposition. Anything below this value is not observed in practice at the end of the digitization chain. Including this postulate in the Monte Carlo digitization yields a revised probability in equation\,\ref{eq:p2hits} of $P_{\text{2hits}} \simeq 61.3$\,\%, much closer to the measured value.

The percentage of planes hit only once is quite high. It does not influence the resolution on the range but does affect the energy reconstruction as some of the energy deposited is lost. The probability of a plane being missed is very low as is expected from a detector with no dead areas.

\subsubsection{Single plane Efficiency}
In the current set-up of the EMR, all of the MAPMTs are set to the same 700\,V voltage. No high voltage scan has been performed yet. An MAPMT which voltage needs to be raised corresponds to a plane that has a higher percentage of single hit and misses particles more frequently. If the voltage needs to be lowered, the probability of having more than two hits rises. In an ideal situation, the efficiency needs to be maximized without producing an excessive amount of crosstalk, i.e. limiting the fraction hit more than twice.

\begin{figure}[!htb]
	\centering
	\includegraphics[width=.75\textwidth]{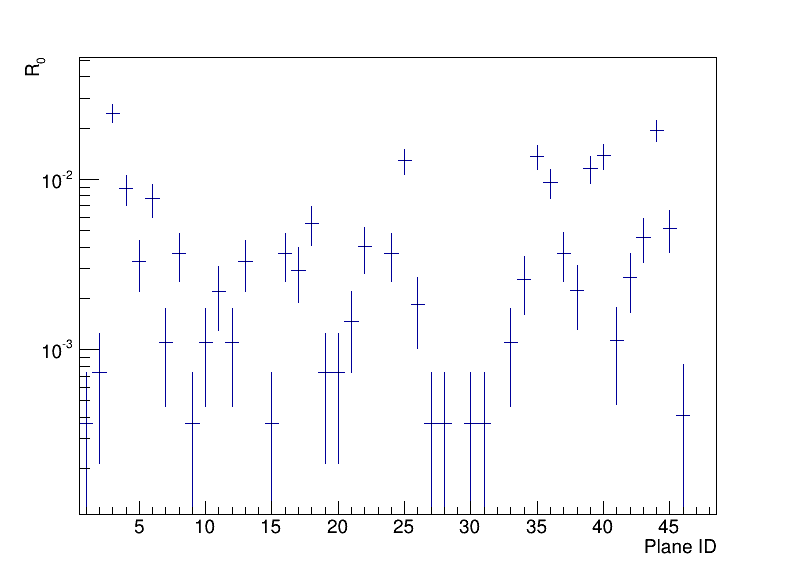}
	\caption{Probability of a single plane to not record a signal, $R_0$.}
	\label{fig:plane_missed}  
\end{figure}

The probability of missing a plane, $R_0$ is represented as a function of the plane ID in figure\,\ref{fig:plane_missed}. In this case, only the sample above crossing threshold, $p_z>350$\,MeV/$c$, is included to have a coherent energy deposition across the whole depth of the detector. The planes are categorized into four ranges of inefficiency in table\,\ref{tab:plane_missed}.

\begin{table}[!htb]
	\centering
	\begin{tabular}{c|c}
		Probability & Number of planes \\
		\hline
		$R_0>1$ \% & \textcolor{red}{6} \\
		$0.1$ \% $<R_0<1$ \% & 25 \\
		$0<R_0<0.1$ \% & 11 \\
		$R_0=0$ & \textcolor{green}{6} \\
	\end{tabular}
	\caption{Number of planes sorted in four ranges of inefficiency.}
	\label{tab:plane_missed}  
\end{table} 

It is apparent that the measured ratio $R_0$ is strongly plane dependant as is expected from an uncalibrated detector. Some planes always produce at least a hit on track whereas some of them are lost as frequently as 2.5\,\% of the time. This indicates that some the MAPMT do not have a perfectly suited voltage supply. The efficiency, however, never falls under 97.5\,\%, which is satisfactory. A high voltage scan will be performed by the end of 2014 to optimize the power supply to fit the specifics of each MAPMT.

\subsubsection{Energy dependency}
The last analysis studies the influence of energy deposition on bar multiplicity. In the following, all the energy settings of table\,\ref{tab:beam_settings} are used as they represent different energy deposition patterns.

\begin{figure}[!htb]
 \centering
 \begin{minipage}[b]{.45\textwidth}
  \centering
  \includegraphics[width=\textwidth]{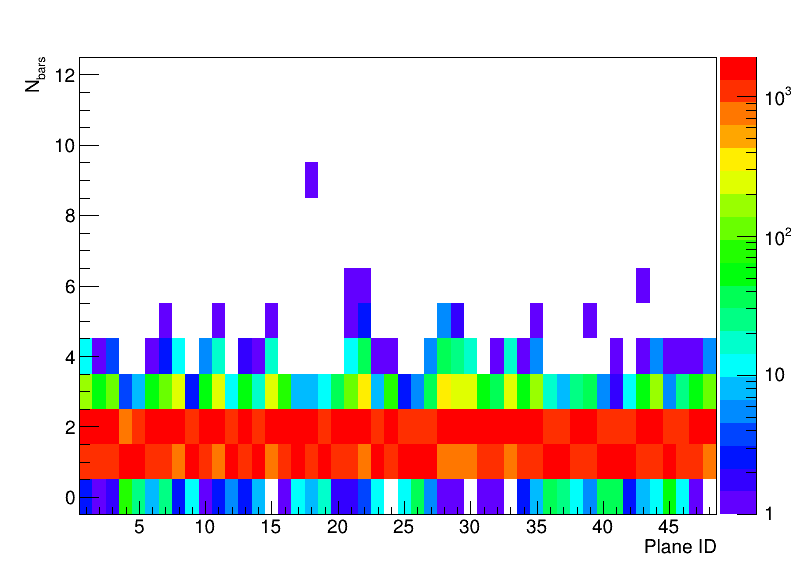}
 \caption{Bar multiplicity distributions in each plane in the high momentum setting.}
 \label{fig:bars_high}
 \end{minipage}
 \hfill
 \begin{minipage}[b]{.45\textwidth}
  \centering
  \includegraphics[width=\textwidth]{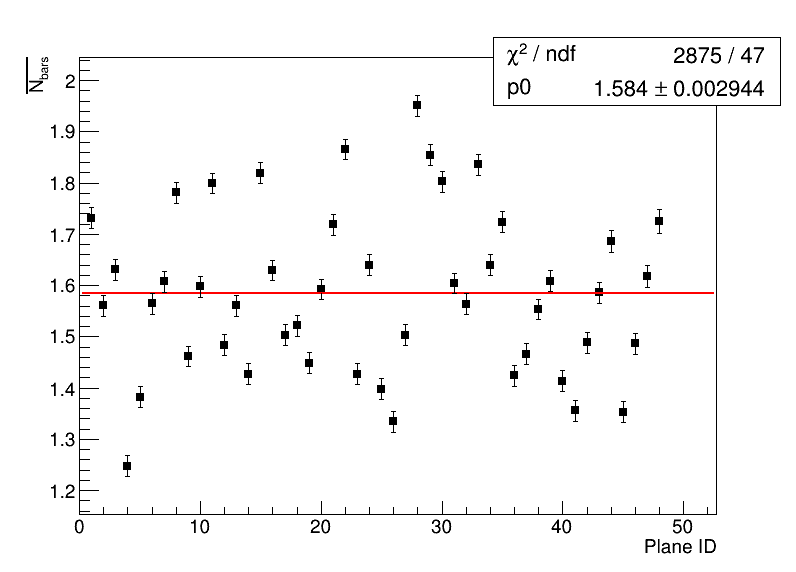}
 \caption{Average bar multiplicity in each plane in the high momentum setting.}
 \label{fig:avbars_high}
 \end{minipage}
\end{figure}

The behaviour of the different planes when subjected to a high energy beam is examined first. The energy deposition in each plane is practically uniform as the particles are minimum ionizing for this setting. The distribution of bars hit is represented for each plane in figure\,\ref{fig:bars_high} and the average number of bars hit in figure\,\ref{fig:avbars_high}. 

Small variations are observed between the different planes which correspond to the relative efficiency of each MAPMT, the voltage chosen, fibre mask misalignment, etc. There is a strong correlation between the planes that have a high number of bars on average, i.e. $\overline{N_{bars}}> 1.75$, and the crosstalk level. This is expected as they are both directly proportional to the MAPMT sensitivity and the misalignment. Bar multiplicity could be used to tune the MAPMTs voltage and achieve a central value. Overly sensitive photosensors are less likely to miss a track entirely and not record a single hit.  

\begin{figure}[!htb]
	\centering
	\begin{minipage}[b]{.45\textwidth}
		\centering
		\includegraphics[width=\textwidth]{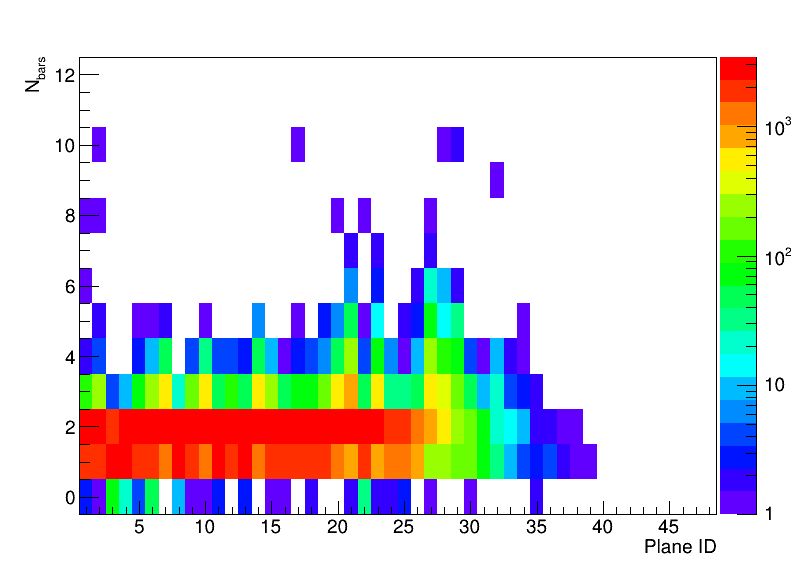}
		\caption{Bar multiplicity distributions in each plane in the low momentum setting.}
		\label{fig:bars_low}
	\end{minipage}
	\hfill
	\begin{minipage}[b]{.45\textwidth}
		\centering
		\includegraphics[width=\textwidth]{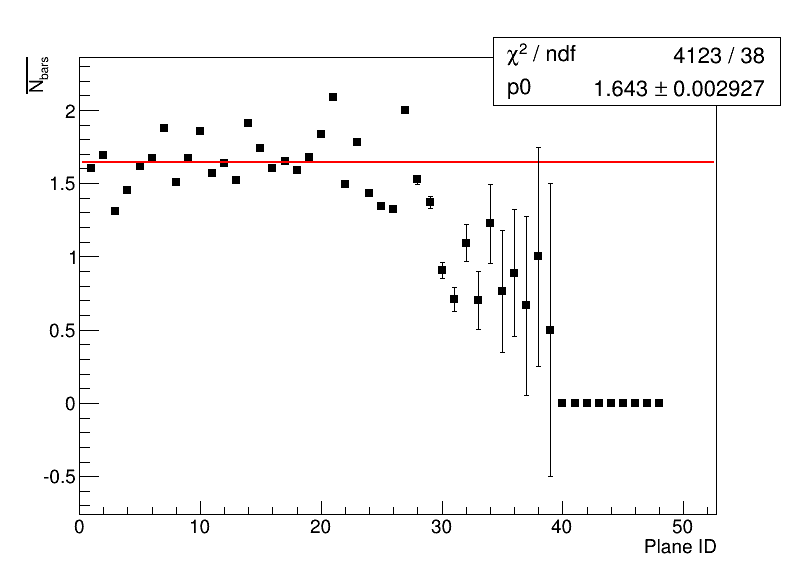}
		\caption{Average bar multiplicity in each plane in the low momentum setting.}
		\label{fig:avbars_low}
	\end{minipage}
\end{figure}

For a below crossing threshold setting, the muons are stopped in the detector and deposit more energy in the plane where they come to a stand still. The distribution of bars hit is represented for each plane in figure\,\ref{fig:bars_low} and the average number of bars hit in figure\,\ref{fig:avbars_low}. 

The average of all bars is shifted higher, as shown by the constant fit to the sample, due to the beam higher energy deposition. There is a structural modification in the right-hand side of the distribution of bars hit for the different planes. Where the particle stops, it produces a burst of light that leaks into a lot of channels surrounding the stopping point. In plane 26, for instance, the probability to have three or four bars is higher than to have one. Hitting an even bigger amount is not a rare phenomenon and a peak is clearly obsrved in the distribution at the level of this plane and the surrounding ones.

This bar multiplicity pattern is predictable. It smears the energy of a single particle in several channels which makes it less straight forward to reconstruct a single point. The scintillation light in the main channel is always the brightest and the average plane coordinate should be weighted in terms of energy deposition to achieve maximum accuracy on the particle range.

\section{Conclusions}
The analyses performed in the context of this master thesis have covered a wide array of potential sources of issues and have demonstrated that the EMR hardware performs as designed.

Only 4 of the 2832 channels of the EMR (0.14\,\%) are not functioning. They are on the edge of the fiducial volume and in one the deepest planes, which does not influence the efficiency of the detector in a realistic beam line. The 48 sets of front-end electronics integrated in each plane are all operating as expected and were shown to be reliable in the context of the MICE run cycle in October 2013. The deviation from the average luminosity that was measured is within the expected range, which means that none of the fibres are suspected to be broken and that the implemented connector system works. 

A closer look at the construction of the EMR did not unveil any major misconceptions. The mismatch channel analysis only revealed two swapped fibres, a considerably low number considering that a little under 6000 cables had to be plugged in manually during the assembly of the detector. The identification of the two channels in this analysis allows for the correction of the DAQ channel map and guarantees an entirely exact cabling from signal to data.

The investigation of the crosstalk in the MAPMTs and their misalignment with respect to the fibre bundle revealed low rates of crosstalk and the 48 photomultipliers misalignments were within an acceptable range. The probability of crosstalk at MIP energy depositions is on average of $0.20\pm0.03$\,\% while the mean crosstalk signal intensity is $4.5\pm0.1$\,\% of the primary signal.

The signal acquisition efficiency analysis of the detector as a whole showed that $99.73\pm0.02$\,\% of the $1.3\times 10^5$ planes hit in the analysis produced a recorded signal in the MAPMT. The procedure revealed a non-negligible asymmetry between the different photomultipliers that will be subjected to a high voltage scan in the months to come in order to regularize their efficiency.

The EMR hardware works to specification and has the necessary efficiency to perform as foreseen in the MICE cooling channel. On a personal note, this master thesis taught me a great deal about the work of a particle physicist: the challenges of the constant technical and programming hurdles to overcome, the satisfaction of finding what is sought after and the thrilling wonder of what is yet to be discovered. Special thanks to Prof. Alain Blondel for the opportunity of being part of a fantastic international effort to develop the particle physics instruments of tomorrow and to Ruslan Asfandiyarov and Yordan Karadzhov for chaperoning me along in the last year and teaching me how to be an experimental physicist.


\newpage
\begin{thebibliography}{99}

  \bibitem{bib:nu_mass}
  V. M. Lobashev \emph{et al.} \emph{Direct Search for the Mass of Neutrino and Anomaly in the Tritium Beta-Spectrum: Status of ’Troitsk Neutrino Mass’ Experiment.} Nucl. Instr. and Meth. in Phys. Res. B (Proc. Suppl.) 91: 280--286. 2001.
  
  \bibitem{bib:nu_sun_flux}
  R. Davis. \emph{Nobel Lecture: A half-century with solar neutrinos.} Review of Modern Physics 75: 985--994. 2003.
  
  \bibitem{bib:nu_sn_flux}
  R. M. Bionta \emph{et al.} \emph{Observation of a neutrino burst in coincidence with Supernova 1987A in the Large Magellanic Cloud.} Phys. Rev. Lett. 58: 1494--1496. 1987.
  
  \bibitem{bib:cnub}
  B. Eberle. \emph{Big Bang Relic Neutrinos and Their Detection.}
  PhD Thesis, University of Hamburg. 2005.
 
  \bibitem{bib:nu_atm}
  M. C. Gonzalez-Garcia and Y. Nir. \emph{Neutrino masses and mixing: evidence and implications.} Review of Modern Physics 75: 345--402. 2003.
 
  \bibitem{bib:nu_geo}
  T. Araki \emph{et al.} \emph{Experimental investigation of geologically produced antineutrinos with KAMLAND.} Nature 436: 499--503. 2005.
  
  \bibitem{bib:nu_reactor}
  C. Bemporad \emph{et al.} \emph{Reactor-based neutrino oscillation experiments.} Review of Modern Physics 74: 297--328. 2002.

  \bibitem{bib:nu_acc}  
  K. Abe \emph{et al.} \emph{The T2K Neutrino Flux Prediction}. Physical Review D 87: 012001. 2013.
  
  \bibitem{bib:beta_spectrum}
  O. v. Baeyer, O. Hahn, L. Meitner, \emph{Magnetische Spektren der $\beta$-Strahlen des Radiums}, Phys. Z. 12: 1099--1101. 1911.
  
  \bibitem{bib:beta_decay}
  H. R. Crane. \emph{The Energy and Momentum Relations in the Beta-Decay, and
  the Search of the Neutrino.} Rev. of Mod. Phys. 20, 1: 278--295. 1948.
  
  \bibitem{bib:pauli_letter}
  W. Pauli. \emph{Letter to a physicist’s gathering at Tubingen, December 4, 1930.} Reprinted in Wolfgang Pauli, Collected Scientific Papers vol. 2, ed. R. Kronig and V. Weisskopf: 1313. 1964.
  
  \bibitem{bib:fermi_theory}
  F. L. Wilson \emph{Fermi's Theory of Beta Decay}. American Journal of Physics, Volume 36, Number 12. 1968.
  
  \bibitem{bib:nu_discovery}
  C. L. Cowan Jr., F. Reines, F. B. Harrison, H. W. Kruse, A. D. McGuire. \emph{Detection of the Free Neutrino: a Confirmation.} Science 124 (3212): 103--104. 1956.
  
  \bibitem{bib:nu_helicity}
  M. Goldhaber \emph{et al.} \emph{Helicity of neutrino.} Phys. Rev. 109: 1015--1017. 1958.
  
  \bibitem{bib:nu_generations}
  G. Danby \emph{et al.} \emph{Observation of high-energy neutrino reactions and the existence of two kinds of neutrinos.} Phys. Rev. Lett. 9: 36--44. 1962.
  
  \bibitem{bib:nu_tau}
  K. Kodama \emph{et al.} \emph{Observation of tau-neutrino interactions.} Phys. Lett. B 504: 218--224. 2001.
  
  \bibitem{bib:homestake}
  B. T. Cleveland \emph{et al.} \emph{Measurement of the solar electron neutrino flux with the Homestake chlorine detector.} Astrophys. Journal 496: 505--526. 1998.
  
  \bibitem{bib:kamland}
  K. Eguchi \emph{et al.} \emph{First results from KAMLAND: evidence for reactor antineutrino disappearance.} Phys. Rev. Lett. 90, 2: 021802. 2003.
  
  \bibitem{bib:nu_osc}
  S. Bilenky, B. Pontecorvo. \emph{Lepton mixing and neutrino oscillations} Zh. Eksp. Teor. Fiz., 34: 247. 1957.
  
  \bibitem{bib:nu_osc_mns}
  Z. Maki \emph{et al.} \emph{Remarks on the Unified Model of Elementary Particles.} Prog. Theor. Phys. 28: 870--880. 1962.
  
  \bibitem{bib:nu_bahcall}
  J. N. Bahcall \emph{et al.} \emph{Present Status of the Theoretical Predictions for the $\prescript{36}{}{Cl}$ Solar-Neutrino Experiment.} Phys. Rev. Lett. 20: 1209--1212. 1968.
  
  \bibitem{bib:sage}
  SAGE Collaboration. \emph{Measurement of the solar neutrino capture rate with gallium metal.} Phys. Rev. C 60: 055801. 1999.
  
  \bibitem{bib:gallex}
  W. Hampel \emph{et al.} \emph{GALLEX solar neutrino observations: results for GALLEX IV.} Phys. Lett. B 447: 127--133. 1999.
  
  \bibitem{bib:sno}
  The SNO Collaboration. \emph{Measurements of the rate of $\nu_e+d\rightarrow p+p+e^-$ interactions produced by $\prescript{8}{}{B}$ solar neutrinos at the Sudbury Neutrino Observatory.} Phys. Rev. Lett. 87: 071301. 2001.
  
  \bibitem{bib:nu_atm_osc}
  Y. Fukuda \emph{et al.} \emph{Evidence for Oscillation of Atmospheric Neutrinos.} Phys. Rev. Lett. 81: 1562--1567. 1998.
  
  \bibitem{bib:t2k}
  T2K Collaboration. \emph{Measurement of Neutrino Oscillation Parameters from Muon Neutrino Disappearance with an Off-axis Beam.} Phys. Rev. Lett. 111: 211803. 2013.
  
  \bibitem{bib:double_chooz}
  Y. Abe \emph{et al.} \emph{Reactor electron antineutrino disappearance in the Double Chooz experiment} Phys. Rev. D 86: 052008. 2012.
  
  \bibitem{bib:global_fits}
  D. V. Forero, M. Tórtola, J. W. F. Valle \emph{Global status of neutrino oscillation parameters after Neutrino-2012}. Phys. Rev. D 86: 073012. 2012.
  
  \bibitem{bib:PDG}
  J. Beringer \emph{et al.} [Particle Data Group]. \emph{Review of Particle Physics}. Phys. Rev. D 86: 010001. 2012.
  
  \bibitem{bib:nu_mix}
  V. N. Gribov, B. Pontecorvo. \emph{Neutrino astronomy and lepton charge.} Physics Letters B28: 493. 1969
  
  \bibitem{bib:nu_physics}
  F. J. P. Soler, C. D. Froggatt, F. Muheim. \emph{Neutrinos in Particle Physics, Astrophysics and Cosmology.} CRC Press, Taylor and Francis Group. 2009.
  
  
  \bibitem{bib:nu_fact}
  C. Albright \emph{et al.} \emph{Physics at a Neutrino Factory.} Fermilab-FN-692. 2000. 
  
  \bibitem{bib:mu_collider}
  G.I. Budker, in Proc. of the 7th Int. Conf. on High Energy Accelerators, Yerevan, Armenia: 33. 1969.
  
  \bibitem{bib:nu_fact_feas}
  S. Ozaki, R. Palmer, M. Zisman, J. Gallardo. \emph{Feasibility Study-II of a Muon-Based Neutrino Source.} Eds. BNL-52623. 2001.
  
  \bibitem{bib:icool_perf}
  K. Hanke. \emph{Muon Front-End without Cooling.} NuFact Note 59. 2000.
  
  \bibitem{bib:nu_fact_cp}
  A. De Rujula, M. B. Gavela, P. Hernandez. \emph{Neutrino oscillation physics with a neutrino factory.} Nucl. Phys. B 547: 21. 1999.
  
  \bibitem{bib:msw}
  A. Yu. Smirnov. \emph{The MSW effect and Solar Neutrinos.} arXiv:hep-ph/0305106. 2003.
  
  \bibitem{bib:nu_lbl_exp}
  M. Lindner. \emph{The Physics Potential of Future Long Baseline Neutrino Oscillation Experiments.} Springer Tracts Mod. Phys. 190: 209--242. 2003.
  
  \bibitem{bib:jhf_sk}
  P. Huber \emph{et al.} \emph{Synergies between the first-generation JHF-SK and NuMI superbeam experiments.} Nucl. Phys. B 654: 3--29. 2003
  
  \bibitem{bib:jhf_sk2}
  Y. Itow \emph{et al.} \emph{The JHF-Kamioka neutrino project.} KEK report 2001-4: ICRR--report--477--2001--7. 2001.
  
  \bibitem{bib:nu_fact_status}
  G. Prior. \emph{Status of the Neutrino Factory accelerator design studies.} J. Phys.: Conf. Ser. 408: 012013. 2013.
  
  \bibitem{bib:merit}
  K.T. McDonald et al. \emph{The MERIT high-power target experiment at the CERN PS,} in Proc. of IPAC10, Kyoto, Japan. 2010.
  
  \bibitem{bib:beam_cool}
  A. N. Skrinskii and V. V. Parkhomchuk. \emph{Method of cooling beams of
  charged particles.} Sov. J. Part. Nucl. 12(3): 223--247. 1981.
  
  \bibitem{bib:sigmund}
  P. Sigmund. \emph{Particle Radiation and Radiation Effects.} Springer Series in Solid State Sciences, 151. Berlin Heidelberg: Springer-Verlag. ISBN 3-540-31713-9. 2006.
  
  \bibitem{bib:icool}
  R. P. Johnson. \emph{Ionization Cooling and Muon Colliders,} in Proc. of EPAC08, Genova, Italy. 2008.
  
  \bibitem{bib:mu_collider_feas}
  The $\mu^+\mu^-$ Collaboration. \emph{Muon Muon Collider: a feasibility study.} BNL 52503. 1997.
  
  \bibitem{bib:mice_proposal}
  MICE Collaboration. \emph{An International Muon Ionization Cooling Experiment (MICE).} Proposal to the Rutherford Appleton Laboratory. MICE-NOTE-21. 2003.  
  
  \bibitem{bib:mice_diff}
  M. Apollonio \emph{et al.} \emph{The MICE diffuser system,} in Proc. of EPAC08, Genova, Italy. 2008.
  
  \bibitem{bib:mice_rf}
  D. Li \emph{et al.} \emph{A 201-MHz normal conductig RF cavity for the International MICE Experiment,} in Proc. of EPAC08, Genova, Italy. 2008.
  
  \bibitem{bib:mice_tofs}
  M. Bonesini \emph{et al.} \emph{The design and commissioning of the MICE upstream Time-Of-Flight system.} Nucl. Instr. and Meth. in Phys. Res. A 615: 14--26. 2010.
  
  \bibitem{bib:mice_ckov}
  L. Cremaldi  \emph{et al.} \emph{A Cherenkov Radiation Detector with High Density Aerogels.} IEEE Transactions on Nuclear Science 56: 1475--1478. 2009.
  
  \bibitem{bib:mice_pid}
  M. Bonesini and M. Rayner. \emph{The MICE PID Detector System.} MICE-NOTE-304. 2010.
  
  \bibitem{bib:mice_tracker}
  M. Ellis \emph{et al.} \emph{The design, construction and performance of the MICE scintillating fibre trackers.} MICE-NOTE-254. 2010.
  
  \bibitem{bib:mice_kl}
  J. Lee-Franzini \emph{et al.} \emph{The KLOE electromagnetic calorimeter.} Nucl. Instr. and Meth. in Phys. Res. A 360: 201--205. 1995.
  
  \bibitem{bib:mice_instrumentation}
  A. R. Sandstr\" om. \emph{Background and Instrumentations in MICE.} PhD Thesis, University of Geneva. 2007.
  
  \bibitem{bib:emr_philips}
  R. Asfandiyarov \emph{et al.} \emph{Selecting Philips XP 2972 Photomultiplier Tubes for the Electron-Muon Ranger (EMR)}. MICE-NOTE-383. 2012.
  
  \bibitem{bib:mice_status}
  A. Alekou. \emph{MICE Status.} MICE-NOTE-297. 2010.
  
  \bibitem{bib:emr_scintillator}
  A. Pla-Dalmau \emph{et al.} \emph{Low-cost extruded plastic scintillator.} Nucl. Instr. and Meth. in Phys. Res. A 466: 482--491. 2001.
  
  \bibitem{bib:emr_wlsfibres}
  Saint-Gobain Crystals. \emph{Scintillating Products: Scintillating Optical Fibres.} Saint-Gobain Ceramics and Plastics, Inc. 2005.
  
  \bibitem{bib:emr_clfibres}
  Kuraray. \emph{Kuraray Scintillation Materials}. \url{http://www.phenix.bnl.gov/WWW/publish/donlynch/RXNP/Safety\%20Review\%206_22_06/Kuraray-PSF-Y11.pdf}. 2007.
  
  \bibitem{bib:emr_design_change}
  R. Asfandiyarov \emph{et al.} \emph{Modifications to EMR Design}. MICE-NOTE-357. 2011.
  
  \bibitem{bib:EMR}
  R. Asfandiyarov \emph{et al.}. \emph{Electron-Muon Ranger (EMR) Construction and Tests}. MICE Collaboration Meeting 29. 2011.
  
  \bibitem{bib:hamamatsu_pmt}
  Hamamatsu. \emph{Multianode Photomultiplier Tube Assembly H7546A, H7546B}. \url{https://www.hamamatsu.com/resources/pdf/etd/H7546A_H7546B_TPMH1240E.pdf}. 2008.
  
  \bibitem{bib:emr_fADC}
  R. Asfandiyarov \emph{et al.}. \emph{Features of CAEN V1731 Flash ADC Waveform Digitizer}. MICE-NOTE-358. 2011.
  
  
  
  \bibitem{bib:electronics_qt}
  F. Drielsma \emph{et al.} \emph{Electron-Muon Ranger (EMR) Electronics Quality Tests.} MICE-NOTE-441. 2013.
  
  \bibitem{bib:hamamatsu_handbook}
  Hamamatsu. \emph{Hamamatsu Handbook, Chapter 9 : Position Sensitive Photomultiplier Tubes}. \url{http://www.hamamatsu.com/resources/pdf/etd/PMT_handbook_v3aE-Chapter9.pdf}. 2006.
  
  \bibitem{bib:cm_xt}
  F. Drielsma \emph{et al.} \emph{Crosstalk and Misalignment in the Electron-Muon Ranger (EMR)} MICE-NOTE-440. 2013.
  
  \bibitem{bib:emr_simulation}
  R. Asfandiyarov \emph{et al.} \emph{Geant4 Simulation of the EMR Detector Response Study}. MICE-NOTE-388. 2012.
  
  \bibitem{bib:flight_counters}
  M. Rayner. \emph{The MICE Experiment: Momentum Measurements Using the Time of Flight Counters}. Joint UKNF, INO, UKIERI meeting. 2008.
  
\end{thebibliography}
\end{document}